\documentclass[12pt,epsf]{article}
\usepackage{amssymb,amscd}
\usepackage{graphics}
\usepackage{epsfig}

\DeclareFontFamily{U}{rsf}{}
\DeclareFontShape{U}{rsf}{m}{n}{
  <5> <6> rsfs5 <7> <8> <9> rsfs7 <10-> rsfs10}{}
\DeclareMathAlphabet\Scr{U}{rsf}{m}{n}

\catcode`\@=11
\def\@citex[#1]#2{%
\if@filesw \immediate \write \@auxout {\string \citation {#2}}\fi
\@tempcntb\m@ne \let\@h@ld\relax \def\@citea{}%
\@cite{%
  \@for \@citeb:=#2\do {%
    \@ifundefined {b@\@citeb}%
      {\@h@ld\@citea\@tempcntb\m@ne{\bf ?}%
      \@warning {Citation `\@citeb ' on page \thepage \space undefined}}%
      {\@tempcnta\@tempcntb \advance\@tempcnta\@ne%
      \@tempcntb\number\csname b@\@citeb \endcsname \relax%
      \ifnum\@tempcnta=\@tempcntb 
        \ifx\@h@ld\relax%
          \edef \@h@ld{\@citea\csname b@\@citeb\endcsname}%
        \else%
          \edef\@h@ld{\ifmmode{-}\else--\fi\csname b@\@citeb\endcsname}%
        \fi%
      \else
        \@h@ld\@citea\csname b@\@citeb \endcsname%
        \let\@h@ld\relax%
      \fi}%
    \def\@citea{,\penalty\@highpenalty\,}%
  }\@h@ld
}{#1}}

\def\@citeb#1#2{{[#1]\if@tempswa , #2\fi}}
\def\@citeu#1#2{{$^{#1}$\if@tempswa , #2\fi }}
\def\@citep#1#2{{#1\if@tempswa , #2\fi}}

\def\bcites{         
        \catcode`\@=11
        \let\@cite=\@citeb
        \catcode`\@=12
}

\def\upcites{         
        \catcode`\@=11
        \let\@cite=\@citeu
        \catcode`\@=12
}

\def\plaincites{      
        \catcode`\@=11
        \let\@cite=\@citep
        \catcode`\@=12
}

\newcount\hour
\newcount\minute
\newtoks\amorpm
\hour=\time\divide\hour by 60
\minute=\time{\multiply\hour by 60 \global\advance\minute by-\hour}
\edef\standardtime{{\ifnum\hour<12 \global\amorpm={am}%
        \else\global\amorpm={pm}\advance\hour by-12 \fi
        \ifnum\hour=0 \hour=12 \fi
        \number\hour:\ifnum\minute<10 0\fi\number\minute\the\amorpm}}
\edef\militarytime{\number\hour:\ifnum\minute<10 0\fi\number\minute}

\def\draftlabel#1{{\@bsphack\if@filesw {\let\thepage\relax
   \xdef\@gtempa{\write\@auxout{\string
      \newlabel{#1}{{\@currentlabel}{\thepage}}}}}\@gtempa
   \if@nobreak \ifvmode\nobreak\fi\fi\fi\@esphack}
        \gdef\@eqnlabel{#1}}
\def\@eqnlabel{}
\def\@vacuum{}
\def\marginnote#1{}
\def\draftmarginnote#1{\marginpar{\raggedright\scriptsize\tt#1}}
\def\draft{
        \pagestyle{plain}
        \overfullrule=2pt
        \oddsidemargin -.5truein
        \def\@oddhead{\sl \phantom{\today\quad\militarytime} \hfil
        \smash{\Large\sl DRAFT} \hfil \today\quad\militarytime}
        \let\@evenhead\@oddhead
        \let\label=\draftlabel
        \let\marginnote=\draftmarginnote
        \def\ps@empty{\let\@mkboth\@gobbletwo
        \def\@oddfoot{\hfil \smash{\Large\sl DRAFT} \hfil}
        \let\@evenfoot\@oddhead}
        \def\@eqnnum{(\theequation)\rlap{\kern\marginparsep\tt\@eqnlabel}%
        \global\let\@eqnlabel\@vacuum}  }

\def\section{\@startsection {section}{1}{\z@}{3.ex plus 1ex minus
 .2ex}{2.ex plus .2ex}{\Large\bf}}
\def\subsection{\@startsection{subsection}{2}{\z@}{2.75ex plus 1ex minus
 .2ex}{1.5ex plus .2ex}{\large\bf}}        

\def\appendix{{\newpage\section*{Appendix}}\let\appendix\section%
        {\setcounter{section}{0}
        \gdef\thesection{\Alph{section}}}\section}

\def\abstract{\if@twocolumn
\section*{Abstract}
\else 
\begin{center}
{\bf Abstract\vspace{-.5em}\vspace{0pt}}
\end{center}
\quotation
\fi}

\catcode`\@=12

\newcommand{\beq}{\begin{equation}}
\newcommand{\eeq}{\end{equation}}
\newcommand{\beqa}{\begin{eqnarray}}
\newcommand{\eeqa}{\end{eqnarray}}
\newcommand{\dd}{{\rm d}}

\newcommand{\Z}{{\bf Z}}

\newcommand{\R}{{\bf R}}

\newcommand{\C}{{\bf C}}

\newcommand{\e}{\,{\rm e}}

\newcommand{\ket}{\rangle}
\newcommand{\no}{\nonumber}

\newcommand{\be}{\begin{equation}}
\newcommand{\ee}{\end{equation}}
\newcommand{\bea}{\begin{eqnarray}}
\newcommand{\eea}{\end{eqnarray}}

\def\to{\rightarrow}

\def\lae{\mathrel{\mathop{\smash{\lower .5 ex \hbox{$\stackrel<\sim$}}}}}
\def\lae{\mathrel{\mathop{\smash{\lower .5 ex \hbox{$\stackrel>\sim$}}}}}

\def\ket#1{\left| #1 \right\rangle}

\def\Tr{{\rm Tr}}
\def\l:{\mathopen{:}\,}
\def\r:{\,\mathclose{:}}

\catcode`\@=11
\def\theequation{\arabic{equation}}
\catcode`\@=12

\bcites

\catcode`\@=11
\def\theequation{\thesection.\arabic{equation}}
\@addtoreset{equation}{section}
\@addtoreset{footnote}{section}
\@addtoreset{footnote}{subsection}
\catcode`\@=12

\newcommand{\bi}{\overline{\imath}}
\newcommand{\bj}{\overline{\jmath}}
\newcommand{\bz}{\overline{z}}

\newcommand{\tr}{{\rm tr}}
\newcommand{\nn}{\nonumber}

\newcommand{\fg}{\mathfrak{g}}
\newcommand{\fh}{\mathfrak{h}}

\newcommand{\whfg}{\widehat{\fg}}
\newcommand{\whfh}{\widehat{\fh}}
\newcommand{\wtJ}{\widetilde{J}}

\newcommand{\V}{\widehat{V}}
\newcommand{\wtV}{\widehat{V}}
\newcommand{\B}{\Scr{H}}
\newcommand{\wtB}{\Scr{H}}

\newcommand{\half}{{1\over 2}}

\newcommand{\s}{\sigma}
\newcommand{\bk}{\overline{k}}

\newcommand{\kket}[1]{\vert \Scr{B}, #1\rangle\!\rangle}
\newcommand{\bbra}[1]{\langle\!\langle \Scr{B},#1 \vert}
\newcommand{\cket}[1]{\vert \Scr{C},#1\rangle\!\rangle}
\newcommand{\cbra}[1]{\langle\!\langle \Scr{C},#1 \vert}

\begin{document}

\begin{titlepage}

\begin{center}

\hfill
CERN-TH/2002-196\\
\hfill
hep-th/0208141\\
\hfill

\vskip 1.5 cm
{\large \bf Notes on Orientifolds of Rational Conformal Field Theories}
\vskip 2 cm 
{Ilka Brunner${}^*$ and Kentaro Hori${}^{\dag}$}\\
\vskip 0.6cm
{\it ${}^*$Theory Division, CERN,
CH-1211 Geneva 23, Switzerland}\\[0.3cm]
{\it ${}^{\dag}$Institute for Advanced Study,
Princeton, New Jersey 08540, U.S.A.\\
and\\
University of Toronto,
Toronto, Ontario M5S 1A7, Canada}

\end{center}

\vskip 2 cm
\begin{abstract}
We review and develop the construction of crosscap states
associated with parity symmetries
in rational conformal field theories.
A general method to construct crosscap states
in abelian orbifold models is presented.
It is then applied to rational $U(1)$ and parafermion
systems, where in addition we study  the geometrical interpretation of
the corresponding parities.
\end{abstract}

\vskip 3.5 cm
CERN-TH/2002-196

August 2002

\end{titlepage}

\section{Introduction}\label{sec:intro}

Four-dimensional
string compactifications
with ${\mathcal N}=1$ supersymmetry
allowing non-abelian gauge symmetries and chiral matter contents
are phenomenologically appealing.
Recently, D-branes on Calabi--Yau threefolds
were studied extensively, partly because one may obtain ${\mathcal N}=1$
supersymmetry in four dimensions if the D-branes extend in the
dimensions transverse to the Calabi-Yau.
However, consistency conditions require either
the Calabi--Yau to be non-compact or the tadpoles
to be cancelled by some other objects.
In the former case,
we are left with four-dimensional theories where
gravity is essentially decoupled.
Although such systems are interesting in their own right,
our main concern is the theory in four dimensions
with a {\it finite} Newton's constant.
Thus, we need to consider compact internal spaces.
The only known candidates to cancel the tadpoles while maintaining
${\mathcal N}=1$ supersymmetry are orientifold planes.

Orientifolding means to gauge a parity symmetry of the worldsheet theory.
The basic example is the gauging of the worldsheet
orientation reversal $\Omega$ of the Type IIB superstring, 
resulting in the Type I superstring.
One is free, however, to consider more general parity actions
where $\Omega$ is combined with some action on 
space-time with a necessary consistency condition that this combination is
a symmetry of the underlying theory
\cite{Stalk,BSI,BSII,HOI,HOII} (see \cite{ASreview} for
a recent review).
The fixed-point sets of the space-time action, called orientifold planes,
can carry tension and RR charges
opposite to those of D-branes
and can be used for tadpole cancellation.
A crosscap state is associated to each parity symmetry, just as a
boundary state is associated with a boundary condition or D-brane
\cite{CLNY,PC}. These crosscap states encode the physical data, such as  
tension and RR charges of the orientifold planes.

One approach to study orientifolds of Calabi--Yau manifolds is to consider
special points in the moduli space
where the worldsheet theory is exactly solvable.
One class of such models are toroidal orbifolds, which have been
extensively studied \cite{GP,PS-GP,torus}.
Another important class of systems are Gepner models whose basic building
blocks are rational ${\mathcal N}=2$ superconformal field theories
(see \cite{BlWi,ABPSS} for earlier work on orientifolds of Gepner models).
However, general methods to
study parity symmetries and orientifolds of such models
are not developed to the same extent as in the case of D-branes.

The purpose of the present paper is to collect and
review the known techniques to study orientifolds
of rational conformal field theories (RCFTs) and further develop
them.
We present a coherent method to construct
parity symmetries and the corresponding crosscap states in RCFTs 
and their orbifolds.
The method is then applied to two simple examples,
the rational $U(1)$ and the parafermions $SU(2)/U(1)$.
This serves as a warm-up
to the ${\mathcal N}=2$ models, which will be reported in a forthcoming paper
\cite{BH2}.
Along the way, we also find the geometrical interpretation
of the parity symmetries of these examples.

In Section \ref{general} we describe boundary
and crosscap states and the corresponding parities in RCFTs. 
We review the construction of a universal crosscap state by
Pradisi--Sagnotti--Stanev (PSS) \cite{PSS},
which applies in any RCFT with the charge conjugation modular invariant.
The corresponding parities can be combined with the discrete symmetries
of the system, giving rise to the class of crosscap states considered 
in \cite{HSS}. These crosscap states, as well as the rational boundary
states constructed by Cardy, preserve the diagonal subalgebra ${\cal A}$
of the full symmetry algebra ${\cal A} \otimes {\cal A}$. 

We then proceed to study parity symmetries of orbifold models
and provide a general new method to construct the crosscap states.
The emerging picture is much cleaner than the construction
of boundary states, which suffers from the fixed-point resolution problem.
Our method can be used to explain the result of an earlier paper
\cite{FHSSW}, which also studied the same subject
(see \cite{HuSc,HSSI} for earlier work concerning various
special cases). 

In the last subsection, we consider D-branes and parities preserving
the subalgebra ${\cal A}$ embedded into the symmetry algebra ${\cal A}
\otimes {\cal A}$ through an automorphism $\omega$ of ${\cal A}$:
$W \mapsto A\otimes 1 + 1\otimes \omega(W)$, in particular
the mirror automorphism that acts as charge conjugation.
Extending the terminology
of the ${\mathcal N}=2$ supersymmetry algebra \cite{OOY,BH2},
we call them B-branes/B-parities while the ones preserving the ordinary
diagonal subalgebra shall be called A-branes/A-parities.
Sometimes,
an orbifold model is the mirror of the original model. In such cases,
B-branes and B-parities can be obtained by applying the mirror map
to A-branes/A-parities of the mirror, which are constructed using
orbifold techniques.

Sections \ref{sec:arbR}, \ref{sec:ratR}, \ref{sec:gWZW} and 
\ref{sec:paraf}
are devoted to examples. 
In Section~\ref{sec:arbR} we revisit
the free boson compactified
on a circle of arbitrary radius, including an extension of
the standard construction \cite{CLNY,PC} to non-involutive parities.
We prove that the orientifold corresponding to the parity
where the target space action consists of a half-period shift
of the circle is T-dual to the orientifold associated to the reflection
$X\to -X$ of the circle coordinate with one 
$SO$- and one $Sp$-type orientifold plane (see \cite{ASreview} for a 
recent related discussion).
Section~\ref{sec:ratR} is concerned with the special case 
that the radius of the circle
is $R=\sqrt{k}$, $k$ a positive integer.
In this case, the theory becomes rational and one can apply the
method of Section \ref{general}. It is instructive to see how the
RCFT data encodes geometrical and physical information of the
orientifolds in this simple case.

Our second example is the parafermion system, which is
discussed in Sections~\ref{sec:gWZW} and \ref{sec:paraf}. This model
has a lagrangian description in terms of a $SU(2)_k$ mod
$U(1)_k$  gauged WZW model that is 
particularly well adapted to a study of the geometrical interpretation of 
parity actions. Geometrically, the parafermion theory can be understood
as a sigma-model with a disk target space parametrized by a
complex coordinate $z$ with $|z|\leq 1$. D-branes in this model
have been studied in \cite{MMS}.
A-type parities act as antiholomorphic involutions of the target
geometry, the basic example being $z\to \bar{z}$.  It is
possible to combine this with
an element of the $\Z_k$ symmetry of the theory,
which acts as a phase multiplication on the target space coordinate. 
Accordingly, the corresponding
orientifold planes are  located along diameters of the disk. B-type parities
act holomorphically on the target space, the fundamental B-type parity
being $z\to z$. This involution leaves the whole disk fixed
and therefore corresponds to an orientifold 2-plane.
Combining this with phase rotations leads generically
to non-involutive parities,
which we also consider. In the case where the level $k$ of the
parafermion theory is even, there is a second involutive parity,
$z\to -z$, which leaves only the center of the disk fixed and hence
describes an orientifold 0-plane. Finally we discuss
the same model purely in terms of rational conformal field theory
and give a detailed map of RCFT results to geometrical properties.
The questions discussed in Section~\ref{sec:paraf}
have been partially
addressed in \cite{hikida}, but we disagree with some of the 
results in that paper. In particular, we find that the geometric 
interpretation of the PSS crosscap states is different.

\section{Crosscaps in RCFT}\label{general}

\newcommand{\tilW}{\widetilde{W}}
\newcommand{\kke}{\rangle\!\rangle}
\newcommand{\bbr}{\langle\!\langle}
\newcommand{\ICk}{\kke^{C}}
\newcommand{\IBk}{\kke}
\newcommand{\ICb}{{}^{C}\bbr}
\newcommand{\IBb}{\bbr}
\newcommand{\shalf}{\mbox{$1\over 2$}}
\newcommand{\whchi}{\widehat{\chi}}
\newcommand{\bg}{\overline{g}}

We begin by describing the construction of boundary and crosscap states
of rational conformal field theories. A review and extension of 
previous work in
\cite{Cardy,Ishibashi,PSS,HSS,FHSSW,PSSII,bantay,bantay2,FSbranes} 
is followed by
developing new techniques to construct crosscap states in orbifolds.

We consider a quantum field theory in $1+1$ dimensions.
Let ${\mathcal H}_g$ be the space of states
of the system formulated on a circle
with the $g$-twisted periodic
boundary condition, where $g$ is an internal symmetry.
Let ${\mathcal H}_{\alpha_1,\alpha_2}$ be
the space of states on a segment
with the boundary conditions $\alpha_1$ and $\alpha_2$
at the left and the right ends.
We denote by $|\Scr{B}_{\alpha}\rangle$ and
$|\Scr{C}_{P}\rangle$
the boundary and crosscap states corresponding to
a boundary condition $\alpha$ and a parity symmetry $P=\tau\Omega$
\footnote{$\tau$ is an internal transform and $\Omega$ is the space
coordinate inversion. If the system has fermions,
$\Omega$ is assumed to include the exchange of left and right components.}.
The cylinder, Klein bottle (KB), and M\"obius strip (MS)
amplitudes are expressed in two ways
\beqa
&&
\Tr\!\!\mathop{}_{{\mathcal H}_{\alpha_1,\alpha_2}}g\e^{-\beta H_o(L)}
=\mathop{}_{g}\!\langle \Scr{B}_{1}|
\e^{-LH_c(\beta)}|\Scr{B}_{2}\rangle\!\!\mathop{}_{g},
\label{BB}\\
&&
\Tr\!\!\mathop{}_{{\mathcal H}_{P_1P_2^{-1}}}\!\!
P_2\e^{-\beta H_c(L)}
=\langle \Scr{C}_1|\e^{-{L\over 2}H_c(2\beta)}|\Scr{C}_2\rangle,
\label{CC}\\
&&
\Tr\!\!\mathop{}_{{\mathcal H}_{\alpha, P(\alpha)}}\!\!
P\e^{-\beta H_o(L)}
=\mathop{}_{P^{2}}\!\langle \Scr{B}_{\alpha} |\e^{-{L\over 2}H_c(2\beta)}
|\Scr{C}_P\rangle.
\label{BC}
\eeqa
Here $H_c(\ell)$ and $H_o(\ell)$ are the Hamiltonians of the system
on a circle of circumference $\ell$ and
segment of length $\ell$ respectively.
(The Hilbert spaces and boundary/crosscap states also depend
 on the lengths which are omitted for notational simplicity.)
The subscripts of the boundary states
show the periodicity of the boundary circle.
For instance, $|\Scr{B}_{2}\rangle_g$ consists of elements
in ${\mathcal H}_g$.
Note that $|\Scr{C}_P\rangle$ has a periodicity determined by $P^2$.
(Eq~(\ref{CC}) makes sense only if $P_1^2=P_2^2$.)
In (\ref{BC}), $P(\alpha)$ stands for the $P$-image of
the boundary condition $\alpha$.
The left and the right hand sides of (\ref{BB})--(\ref{BC})
may be referred to as {\it loop channel}
and
{\it tree channel} expressions respectively.

\begin{figure}[tb]
\centerline{\includegraphics{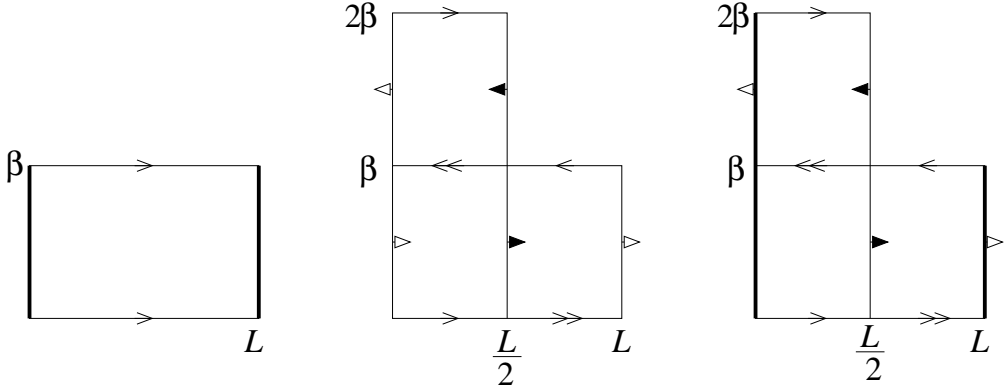}}
\caption{Cylinder, Klein Bottle and M\"obius Strip}
\label{LT}
\end{figure}

In what follows, we will consider
{\it conformally invariant} quantum field theories
and study boundary conditions and parity symmetries
that preserve the conformal invariance.
In such a theory, one can rescale the lengths
$(L,\beta)\to (\lambda L,\lambda \beta)$
without changing the amplitudes.
It is customary to choose the circumference of the closed string
to be $2\pi$ and the length of the open string to be $\pi$.
Suppose we choose $(L,\beta)=(\pi, -2\pi i\tau),
(2\pi,-2\pi i\tau), (\pi,-2\pi i\tau)$ in the loop channel expressions of
(\ref{BB}), (\ref{CC}), (\ref{BC}) respectively, where $\tau$
is a complex number on the positive imaginary axis.
Then in the tree-channel expressions, we take
$(L,\beta)=(-\pi i/\tau,2\pi),
(-\pi i/\tau,\pi), (-\pi i/2\tau,\pi)$.
In string theory, Eqs.~(\ref{BB})-(\ref{BC}) are referred to
as {\it loop/tree channel duality} and have played an important role
(see e.g. \cite{GP}).

\subsubsection*{\it The System We Consider}

We consider an RCFT based on a chiral algebra ${\mathcal A}=\{W_n^{(r)}\}$
with a set of representations $\{\Scr{H}_i\}$.
We primarily consider
the model ${\mathcal C}$
with the Hilbert space of states
\beq
{\mathcal H}^{\mathcal C}=\bigoplus_i\Scr{H}_i\otimes\Scr{H}_{\bi}.
\label{CCM}
\eeq
where $\bi$ is the BPZ conjugate of $i$. 
This model has ${\mathcal A}\otimes{\mathcal A}$ symmetry algebra
generated by $W^{(r)}=W^{(r)}\otimes 1$ and
$\tilW^{(r)}=1\otimes W^{(r)}$.
For each representation $i$, we fix an antiunitary operator
$U:\Scr{H}_i\to \Scr{H}_{\bi}$
such that
\beq
UW_n^{(r)}U^{-1}=(-1)^{s_r}{W_{-n}^{(r)}}^{\dag} \quad\forall r,
\eeq
where $s_r$ is a spin of the current $W^{(r)}$.

\subsection{Symmetry-Preserving D-branes/Orientifolds}

We first study D-branes and orientifolds that preserve
a diagonal subalgebra of the ${\mathcal A}\otimes{\mathcal A}$
symmetry.
On the Minkowski worldsheet with time and space coordinates
$(t,x)$, ``symmetry-preserving'' means the following:
for D-branes the associated boundary conditions (say, at $x=0$)
are such that $W^{(r)}(t,0)=\tilW^{(r)}(t,0)$,
while orientifolds should be associated with parity symmetries
that map $W^{(r)}(t,x)$ to $\tilW^{(r)}(t,-x)$.

\subsubsection{Constraints on Boundary/Crosscap Coefficients}

A Wick rotation followed by a 90${}^{\circ}$ rotation show
that the boundary and crosscap states obey
\beqa
(W^{(r)}_n-(-1)^{s_r}\tilW^{(r)}_{-n})|\Scr{B}\rangle=0,
\label{Bcond}\\
(W^{(r)}_n-(-1)^{s_r+n}\tilW^{(r)}_{-n})|\Scr{C}\rangle=0.
\label{Ccond}
\eeqa
The basic set of solutions to these equations
was found by Ishibashi \cite{Ishibashi}.
Let us denote by $\{|i,N\rangle\}_N$ an orthonormal basis
of the representation $\Scr{H}_i$.
Equations
(\ref{Bcond}) and (\ref{Ccond}) are solved respectively by
\beqa
&&\kket{i}:=\sum_N |i,N\rangle\otimes U|i,N\rangle,\\
&&\cket{i}:=\e^{\pi i(L_0-h_i)}\kket{i}.
\eeqa
It follows from the definition that
\beqa
&&\bbra{i}\e^{2\pi i \tau H_c}\kket{j}=\delta_{i,j}\chi_i(2\tau),
\label{bbchi}\\
&&\cbra{i}\e^{2\pi i \tau H_c}\cket{j}=\delta_{i,j}\chi_i(2\tau),
\label{ccchi}\\
&&\bbra{i}\e^{2\pi i \tau H_c}\cket{j}=\delta_{i,j}
\whchi_i(2\tau),
\label{bcchi}
\eeqa
where $H_c=L_0+\widetilde{L}_0-c/12$
and $\whchi_i(\tau)=\e^{-\pi i (h_i-c/24)}\chi_i(\tau+{1\over 2})$.
The actual boundary and crosscap states are linear
combinations of these Ishibashi-states:
\beqa
&&|\Scr{B}_{\alpha}\rangle=\sum_in_{\alpha i}\kket{i},\\
&&|\Scr{C}_{\mu}\rangle=\sum_i \gamma_{\mu i}\cket{i}.
\eeqa
Here $\alpha$ and $\mu$ are the labels for the boundary conditions
and parity symmetries.
We first assume that the parity symmetries are involutive
 $P_{\mu}^2=1$.
(We will later treat those that are not involutive.)
A set of constraints on the coefficients $n_{\alpha b}$ and $\gamma_{\mu b}$
are found by using the loop/tree channel duality (\ref{BB})--(\ref{BC}):
\beqa
&&
\Tr\!\!\mathop{}_{{\mathcal H}_{\alpha,\beta}}\!\e^{2\pi i\tau H_o}
=\langle \Scr{B}_{\alpha}|\e^{-{\pi i\over\tau} H_c}|\Scr{B}_{\beta}\rangle,
\label{TrBB}
\\
&&
\Tr\!\!\mathop{}_{{\mathcal H}_{g_{\mu\nu}}}\!\!
P_{\nu}\e^{2\pi i\tau H_c}
=\langle \Scr{C}_{\mu}|\e^{-{\pi i\over 2\tau}H_c}|\Scr{C}_{\nu}\rangle,
\label{TrCC}
\\
&&
\Tr\!\!\mathop{}_{{\mathcal H}_{\alpha,\mu(\alpha)}}\!\!
P_{\mu}\e^{2\pi i\tau H_o}
=\langle \Scr{B}_{\alpha}|\e^{-{\pi i\over 4\tau} H_c}|\Scr{C}_{\mu}\rangle,
\label{TrBC}
\eeqa
where
$g_{\mu\nu}$ is the internal symmetry $P_{\mu}P_{\nu}^{-1}$
that commutes with the chiral algebra ${\mathcal A}\otimes {\mathcal A}$,
and $\mu(\alpha)$ is the $P_{\mu}$-image of the boundary condition $\alpha$.

Since the diagonal symmetry ${\mathcal A}$ is preserved by
the boundary conditions $\alpha$ and $\beta$,
open string states fall into a sum of irreducible representations
\beq
{\mathcal H}_{\alpha,\beta}=\bigoplus_{i}
n_{\alpha\beta}^i\Scr{H}_i
\eeq
on which $H_o$ acts as $L_0-{c\over 24}$, where
$n_{\alpha\beta}^i$ are non-negative integers.
Using (\ref{bbchi}), (\ref{TrBB}) is expressed as
\beq
\sum_in_{\alpha\beta}^i\chi_i(\tau)=\sum_in_{\alpha i}^*n_{\beta i}
\chi_i(-1/\tau).
\label{Retrbb}
\eeq

For a symmetry $g$ that commutes with the chiral algebra
${\mathcal A}\otimes {\mathcal A}$,
the space ${\mathcal H}_{g}$ of $g$-twisted closed string states
can be decomposed into the representations of ${\mathcal A}\otimes
{\mathcal A}$,
\beq
{\mathcal H}_g=\bigoplus_{ij}
h^{ij}_g \Scr{H}_i\otimes\Scr{H}_j,
\eeq
where $h^{ij}_g$ are non-negative integers.
Note that $P_{\nu}$ transforms the $g$-twisted boundary condition
into the $\tau_{\nu}g^{-1}\tau_{\nu}^{-1}$-twisted boundary condition.
Since $P_{\nu}$ is a symmetry-preserving parity,
$P_{\nu}W_n^{(r)}=\tilW_n^{(r)}P_{\nu}$,
it acts on the closed string states essentially by the exchange
of the left and right factors.
To be more precise, it maps the subspace
$h^{ij}_g\Scr{H}_i\otimes\Scr{H}_j$
of ${\mathcal H}_g$ to a subspace of
${\mathcal H}_{\tau_{\nu}g^{-1}\tau_{\nu}^{-1}}$ as
\beq
P_{\nu}:\xi\otimes u\otimes v\in \C^{h^{ij}_g}\otimes
\Scr{H}_i\otimes \Scr{H}_{j}\mapsto
K_{\nu}^{ij}(g)\xi\otimes  v\otimes u\in
\C^{h^{ji}_{\tau_{\nu}g^{-1}\tau_{\nu}^{-1}}}
\otimes\Scr{H}_{j}\otimes \Scr{H}_{i},
\eeq
where $K_{\nu}^{ij}(g)$ is a matrix acting on the multiplicity space
$\C^{h^{ij}_g}\cong \C^{h^{ji}_{\tau_{\nu}g^{-1}\tau_{\nu}^{-1}}}$.
(Note that it has to be the case that
$h^{ij}_{g}=h^{ji}_{\tau_{\nu}g^{-1}\tau_{\nu}^{-1}}$.)
$P_{\nu}^2=1$ requires
$K_{\nu}^{ji}(\tau_{\nu}g^{-1}\tau_{\nu}^{-1})K_{\nu}^{ij}(g)=1$.
In particular, $K_{\nu}^{ii}(g_{\mu\nu})$ is a matrix that squares to 1
and therefore its eigenvalues must be $\pm 1$.
Thus, using (\ref{ccchi}), (\ref{TrCC})
is expressed as
\beq
\sum_{i}k_{\mu\nu}^{i}
\chi_i(2\tau)=\sum_i\gamma_{\mu i}^{*}\gamma_{\nu i}
\chi_i(-1/2\tau),
\label{Retrcc}
\eeq
where $k_{\mu\nu}^i=\tr K_{\nu}^{ii}(g_{\mu\nu})$.
Since $K_{\nu}^{ii}(g_{\mu\nu})$ squares to $1$,
the number $k_{\mu\nu}^i$ must be an integer such that
$|k_{\mu\nu}^i|\leq h^{ii}_{g_{\mu\nu}}$ and
$k_{\mu\nu}^i\equiv h^{ii}_{g_{\mu\nu}}$ mod 2.

Let us next consider the action of $P_{\mu}$
on open string states. Since it exchanges the left and right
boundaries of the string,
the symmetry-preserving condition becomes
$P_{\mu} W_n^{(r)}=(-1)^n W_n^{(r)} P_{\mu}$.
It therefore has to transform the subspace
$\Scr{H}_i^{\oplus n_{\alpha\beta}^i}$ of ${\mathcal H}_{\alpha,\beta}$
to a subspace of ${\mathcal H}_{\mu(\beta),\mu(\alpha)}$
as
$$
P_{\mu}:\eta\otimes u\in \C^{n_{\alpha\beta}^i}\otimes \Scr{H}_i\mapsto
M_{\alpha\beta,\mu}^i\eta\otimes
\e^{\pi i (L_0-h_a)}u\in \C^{n_{\mu(\beta)\mu(\alpha)}^i}\otimes
\Scr{H}_i,
$$
where $M_{\alpha\beta,\mu}^i$
is a matrix acting on $\C^{n_{\alpha\beta}^i}$.
(Note that $n_{\alpha\beta}^i$ must be equal to
$n_{\mu(\beta)\mu(\alpha)}^i$ for any $i$.)
$P_{\mu}^2=1$ requires
$M_{\mu(\beta)\mu(\alpha)}^iM_{\alpha\beta,\mu}^i=1$.
In particular, the eigenvalues of the matrix $M_{\alpha\mu(\alpha)}^i$
have to be $\pm 1$.
Using this, we can rewrite (\ref{TrBC}) as
\beq
\sum_i m_{\alpha,\mu}^i\whchi_i(\tau)
=\sum_i n_{\alpha i}^*\gamma_{\mu i}\whchi_i(-1/4\tau),
\label{Retrbc}
\eeq
where $m_{\alpha\mu}^i=\tr M_{\alpha\mu(\alpha),\mu}^i$.
For $P_{\mu}^2=1$, the number
$m_{\alpha\mu}^i$ must be an integer such that
$\pm m_{\alpha\mu}^i\leq n_{\alpha\mu(\alpha)}^i$ and
$m_{\alpha\mu}^i\equiv n_{\alpha\mu(\alpha)}^i$ mod 2.

We have found the constraints (\ref{Retrbb}), (\ref{Retrcc})
and (\ref{Retrbc}). At this stage, we use the modular transformation property
of characters
\beqa
&&\chi_j(-1/\tau)=\sum_i\chi_i(\tau)S_{ij},
\nn\\
&&
\chi_j(\tau+1)=\sum_{i}\chi_i(\tau)T_{ij},
\nn\\
&&\widehat{\chi}_j(-1/4\tau)
=\sum_i\widehat{\chi}_i(\tau)P_{ij},
\nn
\eeqa
where
\beq
P=\sqrt{T}ST^2S\sqrt{T},
\eeq
in which $\sqrt{T}_{ij}=\delta_{i,j}\e^{\pi i(h_i-{c\over 24})}$.
Here we are in the standard convention, $STS=T^{-1}ST^{-1}$.
The three constraints can then be rewritten as
\beqa
&&n_{\alpha\beta}^i=\sum_jn_{\alpha j}^*n_{\beta j}S_{ij},
\label{Ca1}
\\
&&k_{\mu\nu}^i
=\sum_j\gamma_{\mu j}^{*}\gamma_{\nu j}S_{ij},
\label{Ca2}
\\
&&m_{\alpha\mu}^i
=\sum_jn_{\alpha j}^{*}\gamma_{\mu j}P_{ij},
\label{Ca3}
\eeqa
where $n_{\alpha\beta}^i$,
$k_{\mu\nu}^i$ and
$m_{\alpha\mu}^i$ are integers such that
\beq
\begin{array}{l}
\bullet\,\,\,\,n_{\alpha\beta}^i\geq 0,\\
\bullet\,\,\,\,-h^{ii}_{g_{\mu\nu}}\leq
k_{\mu\nu}^i\leq h^{ii}_{g_{\mu\nu}},\quad
k_{\mu\nu}^i\equiv h^{ii}_{g_{\mu\nu}}\,\,(\mbox{mod 2}),\\
\bullet\,\,\,\,-n_{\alpha\mu(\alpha)}^i\leq
m_{\alpha\mu}^i\leq n_{\alpha\mu(\alpha)}^i,\quad
m_{\alpha\mu}^i\equiv n_{\alpha\mu(\alpha)}^i\,\,(\mbox{mod 2}),\\
\end{array}
\label{Ca4}
\eeq
where $h_g^{ij}$ is the multiplicity of $\Scr{H}_i\otimes\Scr{H}_j$
in the space ${\mathcal H}_g$ of the $g$-twisted closed string states.

\subsubsection{Cardy--PSS Solution}

A simple solution to the above constraints that applies to all RCFTs
has been found by Cardy \cite{Cardy} for boundary
states and by Pradisi--Sagnotti--Stanev \cite{PSS} for a crosscap state.

Cardy's boundary
conditions carry the same labels as the representations $\{i\}$.
The coefficients $n_{ij}$ and the multiplicities $n_{ij}^k$
are given in terms of the modular $S$-matrix and the fusion coefficients:
\beqa
&&n_{ij}={S_{ij}\over \sqrt{S_{0j}}},\\
&&n_{ij}^k=N_{\bi j}^{\bk}.
\eeqa
The constraints (\ref{Ca1}) translate into the Verlinde formula
\cite{verlinde}.

PSS have found a crosscap state that corresponds to a parity symmetry
transforming the Cardy boundary states as
\beq
P_0:i\mapsto \bi.
\label{P0mapC}
\eeq
We label this state by ``0'' for a reason that will become clear shortly.
The coefficients $\gamma_{0i}$ and the numbers $k_{00}^i$, $m_{i0}^{j}$
are given by
\beqa
&&\gamma_{0i}={P_{0i}\over\sqrt{S_{0i}}},\\
&&k_{00}^i=Y_{i0}^0,\label{Ce2}\\
&&m_{i0}^j=Y_{\bi 0}^{\bj}\label{Ce3},
\eeqa
where $Y_{ij}^k$ is defined by
\beq
Y_{ij}^k=\sum_l{S_{il}P_{jl}P_{kl}^*\over S_{0l}}.
\eeq
It is straightforward to show that the constraints (\ref{Ca2}) and (\ref{Ca3})
are satisfied for the PSS crosscap state. 
To see that they obey (\ref{Ca4}), we first note that
$g_{00}=1$ and therefore $h^{ii}_{g_{00}}=h^{ii}=\delta_{i,\bi}$
because ${\mathcal H}_1={\mathcal H}=\oplus_i\Scr{H}_i\otimes \Scr{H}_{\bi}$.
The conditions (\ref{Ca4}) therefore require $k_{00}^i=\pm \delta_{i,\bi}$.
Also, since $n_{iP(i)}^j=n_{i\bi}^j=N_{\bi\bi}^{\bj}$,
the number $m_{i0}^j$ must be an integer such that
$|m_{i0}^j|\leq N_{\bi\bi}^{\bj}$ and $m_{i0}^j\equiv N_{\bi\bi}^{\bj}$
(mod 2).
These constraints are obeyed by the above solution because of

\noindent
{\bf Bantay's Relation} \cite{bantay}
\beq
\begin{array}{l}
|Y_{i0}^j|\leq N_{ii}^j,\\
Y_{i0}^j\equiv N_{ii}^j\quad \mbox{mod 2}.
\end{array}
\label{bantays}
\eeq

\noindent
Indeed the condition on $m_{i0}^j=Y_{\bi 0}^{\bj}$ is nothing
but Bantay's relation.
The condition on $k_{00}^i=Y_{i0}^0$ also follows from this since
$N_{ii}^0=\delta_{i,\bi}$.

The number $Y_{i0}^0$ (which is $\pm 1$ if $i=\bi$ and 0 otherwise)
is a CFT analog of the Frobenius--Schur indicator for the theory
of group representations
\footnote{The Frobenius-Schur indicator of an irreducible representation $R$
of a finite group $G$
is defined to be $+1$ in the case when $R$ is real, $0$ when $R$ is complex
and $-1$ when $R$ is pseudo-real.}.

\subsubsection{Dressing by Discrete Symmetries}\label{sec:dressing}

If an RCFT has a discrete symmetry of a certain type,
one can find additional parity symmetries together with
the corresponding crosscap states.

Let $\Scr{G}$ be the finite abelian group generated by the simple currents $\{g\}$.
We recall that a simple current $g$ is a representation such that
the fusion product of
$g$ with any representation $i$ contains only one representation, which we
denote by $g(i)$.
Let us introduce the number
\beq
Q_g(i)=h_i+h_g-h_{g(i)} \quad\mbox{mod 1}.
\label{defQ}
\eeq
The map $g\mapsto \e^{2\pi iQ_g(i)}$ defines for each $i$ a homomorphism
$\Scr{G}\to U(1)$
(some of the properties of $Q_g(i)$ are summarized in Appendix \ref{app:Q}).
We can thus define a representation of $\Scr{G}$
on the Hilbert space 
${\mathcal H}$ in such a way that $g$ acts by the
phase multiplication
$\e^{2\pi i Q_g(i)}\times$ on the subspace $\Scr{H}_i\otimes\Scr{H}_{\bi}$.
The Cardy state
$|\Scr{B}_i\rangle$ is mapped by $g$ as
\beq
|\Scr{B}_i\rangle=\sum_{j}{S_{ij}\over\sqrt{S_{0j}}}\kket{j}
\stackrel{g}{\mapsto}
\sum_j{S_{ij}\over\sqrt{S_{0j}}}\e^{2\pi i Q_g(j)}\kket{j}
=\sum_j{S_{g(i)j}\over \sqrt{S_{0j}}}\kket{j}=|\Scr{B}_{g(i)}\rangle,
\label{mapofC}
\eeq
where we have used $S_{g(i)j}=\e^{2\pi i Q_g(j)}S_{ij}$.

We now find new parity symmetries $P_g$ that act on
${\mathcal H}$ as $g\circ P_0$.
It follows from (\ref{P0mapC}) and (\ref{mapofC}),
that these parities should map the Cardy branes as
\beq
P_g:i\mapsto g(\bi).
\label{Pgi}
\eeq
The crosscap coefficients and the numbers
$k_{g_1g_2}^i$, $m_{ig}^j$ are given by
\beqa
&&\gamma_{gi}={P_{gi}\over\sqrt{S_{0i}}},\\
&&k_{g_1g_2}^i=Y_{i g_2}^{g_1}
\label{zg1g2}\\
&&m_{ig}^j=Y_{\bi g}^{\bj}.
\eeqa
It is easy to show that this solves the constraints (\ref{Ca2}) and (\ref{Ca3})
\cite{PSS,HSS}.
The Y-tensors can be rewritten as
\beq
Y_{i g_2}^{g_1}=\e^{\pi i (\hat{Q}_{g_2}(g_2^{-1}g_1)
-2Q_{g_2}(i))}Y_{i0}^{g_2^{-1}g_1},\quad
Y_{\bi g}^{\bj}
=\e^{\pi i (\hat{Q}_g(g^{-1}(\bj))-2Q_g(\bi))}Y_{\bi 0}^{g^{-1}(\bj)},
\eeq
where $\hat{Q}_g(i):=h_i+h_g-h_{g(i)}$ (not just modulo integers).
Since
$\e^{\pi i (\hat{Q}_g(b)-2Q_g(a))}=\pm 1$ if
$Y_{a 0}^{b}\ne 0$, the integrality of $k_{g_1g_2}^i$
and $m_{ig}^j$ is also satisfied.

In order to show that the last constraint (\ref{Ca4}) is satisfied, we need to 
find the multiplicities
$h_{g_{12}}^{ii}$ and $n_{iP_g(i)}^j$, where
$g_{12}=P_1P_2^{-1}=g_1g_2^{-1}$ and $P_g(i)=g(\bi)$ by (\ref{Pgi}).
The space of $g$-twisted closed string states is given by
${\mathcal H}_g=\oplus_i\Scr{H}_i\otimes \Scr{H}_{g(\bi)}$ 
(see Appendix \ref{subap:gtw}),
and therefore $h_g^{ii}=\delta_{i,g(\bi)}$.
Thus, (\ref{Ca4}) requires $k_{g_1g_2}^i=\pm \delta_{i,g_1g_2^{-1}(\bi)}$.
We note that $N_{ii}^{g_2^{-1}g_1}=N^0_{i g_1^{-1}g_2(i)}
=\delta_{\bi,g_1^{-1}g_2(i)}=\delta_{i,g_1g_2^{-1}(\bi)}$, and hence
$Y_{i 0}^{g_2^{-1}g_1}=\pm\delta_{i,g_1g_2^{-1}(\bi)}$
using Bantay's relation
(\ref{bantays}).
This indeed shows that
$k_{g_1g_2}^i=
Y_{i g_2}^{g_1}=\pm Y_{i 0}^{g_2^{-1}g_1}
=\pm \delta_{i,g_1g_2^{-1}(\bi)}$.
On the other hand, the open string multiplicity is
$n_{ig(\bi)}^j=N_{\bi g(\bi)}^{\bj}=N_{\bi\bi}^{g^{-1}(\bj)}$.
Then the claimed solution
$m_{ig}^i=\pm Y_{\bi 0}^{g^{-1}(\bj)}$
obeys the last condition of (\ref{Ca4}),
again due to Bantay's relation (\ref{bantays}).

\subsubsection{Summary}

We have found D-branes $\Scr{B}_i$ and parity symmetries $P_g$
($g\in\Scr{G}$),
which preserve the diagonal symmetry ${\mathcal A}\subset
{\mathcal A}\otimes{\mathcal A}$.
The corresponding boundary and crosscap states are given by
\beqa
&&|\Scr{B}_i\rangle
=\sum_j{S_{ij}\over\sqrt{S_{0j}}}\kket{j},\\
&&|\Scr{C}_{P_g}\rangle
=\sum_i{P_{gj}\over\sqrt{S_{0j}}}\cket{j}.
\eeqa
The cylinder, Klein bottle and M\"obius strip amplitudes are given by
\beqa
&&\Tr\!\mathop{}_{{\mathcal H}_{i,i'}}\!\e^{2\pi i\tau H}
=\langle \Scr{B}_i|\e^{-{\pi i\over \tau}H}|\Scr{B}_{i'}\rangle
=\sum_j N_{\bi i'}^{\bj}\chi_j(\tau),
\label{trr1}\\
&&\Tr\!\mathop{}_{{\mathcal H}_{gh^{-1}}}\!P_h\e^{2\pi i\tau H}
=\langle \Scr{C}_g|\e^{-{\pi i\over 2\tau}H}|\Scr{C}_h\rangle
=\sum_j Y_{jh}^g\chi_j(2\tau),
\label{trr2}\\
&&\Tr\!\mathop{}_{{\mathcal H}_{i,g(\bi)}}\!P_g\e^{2\pi i\tau H}
=\langle \Scr{B}_i|\e^{-{\pi i\over 4\tau}H}|\Scr{C}_g\rangle
=\sum_j Y_{\bi g}^{\bj}\widehat{\chi}_j(\tau),
\label{trr3}
\eeqa
where a shorthand notation $|\Scr{C}_g\rangle$ for
$|\Scr{C}_{P_g}\rangle$ has been used.
We can simplify  expression (\ref{trr1}) by using
$N_{\bi i'}^{\bj}=N_{i'j}^i$.
Also, taking the complex conjugate of (\ref{trr3})
and using $(Y_{\bi g}^j)^*=Y_{ij}^g$, we have
\beq
\Tr\!\mathop{}_{{\mathcal H}_{g(\bi),i}}\!P_g\e^{2\pi i\tau H}
=\langle \Scr{C}_g|\e^{-{\pi i\over 4\tau}H}|\Scr{B}_i\rangle
=\sum_j Y_{ij}^g\widehat{\chi}_j(\tau).
\eeq
This can also be obtained from (\ref{trr3}) by replacing
$i\to g(\bi)$ and using $Y_{\overline{g(\bi)} g}^{\bj}=Y_{ij}^g$,
which can be derived by using $P_{\bj k}^*=P_{jk}$ and
$\e^{-2\pi i Q_g(k)}P_{gk}=P_{gk}^*$.

\subsection{Crosscaps in Orbifolds}\label{subsec:orb}

We have constructed parity symmetries together with the crosscap states
for the charge-conjugate modular invariant ${\mathcal C}$.
We now turn  to the orbifold model ${\mathcal C}/G$
where $G$ is a group of simple currents, $G\subset \Scr{G}$.
To define a consistent orbifold theory, 
$G$ must have a symmetric bilinear form
$q(g_1,g_2)$ with values in $\R/\Z$ such that
\beq
Q_{g_1}(g_2)=2q(g_1,g_2)\qquad\mbox{mod 1},
\eeq
and $q(g,g)=-h_g$.
The orbifold model we consider has modular invariant partition function
\beq
Z^{{\mathcal C}/G}
={1\over |G|}\sum_{i,g_1,g_2}
\e^{2\pi i(Q_{g_2}(i)-q(g_2,g_1))}
\chi_i(\tau)\overline{\chi_{g_1^{-1}(i)}(\tau)}.
\label{orbPF}
\eeq
We will find $|\Scr{G}|$ symmetry preserving
parities/crosscaps in 1-to-1 correspondence to
the number $|\Scr{G}|$ of simple currents, just as in the original model
${\mathcal C}$.

Let us first present the basic idea behind the construction of
a crosscap state in the orbifold model ${\mathcal C}/G$
for the parity symmetry that is induced by a parity
$P$ of the original model ${\mathcal C}$.
The twisted partition function with respect to the induced parity,
denoted again by $P$,
is expressed as
\beq
\Tr\!\!\mathop{}_{{\mathcal H}^{{\mathcal C}/G}}Pq^H
=\sum_{g_1\in G}\Tr\!\!\mathop{}_{{\mathcal H}_{g_1}}\left(
\Bigl({1\over |G|}\sum_{g_2\in G}g_2\Bigr)\,Pq^H\right),
\eeq
where the first sum is over the $G$-twisted spaces and
${1\over |G|}\sum_{g_2\in G}g_2$ is the projection onto
the subspace of $G$-invariant states.
Let us rearrange the sum as $(1/|G|)\sum_{g_1,g_2}\Tr_{{\mathcal H}_{g_1}}
g_2Pq^H$, and make a replacement $g_1\to g_1g_2^{-1}$.
At this point we recall from (\ref{CC}) that
$$
\Tr\!\!\mathop{}_{{\mathcal H}_{g_1g_2^{-1}}}(
g_2Pq^H)
={}^{\mathcal C}\!\langle \Scr{C}_{g_1P}|
q_t^H|\Scr{C}_{g_2P}\rangle^{\!{\mathcal C}},
$$
where the superscript shows that the state
$|-\rangle^{\mathcal C}$ is in the theory ${\mathcal C}$.
Then, we find
\beq
\Tr\!\!\mathop{}_{{\mathcal H}^{{\mathcal C}/G}}Pq^H
={1\over |G|}\sum_{g_1,g_2}{}^{\mathcal C}\!\langle \Scr{C}_{g_1P}|
q_t^H|\Scr{C}_{g_2P}\rangle^{\!{\mathcal C}}.
\eeq
This implies that the crosscap state
for the induced parity $P$
is given by
\beq
|\Scr{C}_P\rangle^{\!{\mathcal C}/G}
={1\over\sqrt{|G|}}\sum_{g\in G}|\Scr{C}_{gP}\rangle^{\!{\mathcal C}}.
\label{cand}
\eeq

\subsubsection{PSS Parities Induced on Orbifold Models}

We would like to apply this construction by identifying
$P$ as one of the parity symmetries
of ${\mathcal C}$ obtained in the previous section,
say the original PSS parity $P_0$.
We know that $gP_0$ is equal to $P_g$ at least in the action on
the untwisted states ${\mathcal H}^{\mathcal C}$ and we know the
crosscap states for all $P_g$.
Thus, (\ref{cand}) appears to be an ideal formula
for constructing a crosscap
state in the orbifold theory.
However, one has to be careful when identifying $P_0$ as $P$
and $P_g$ as $gP_0$.
The subtleties are:\\
(i) $P$ may differ from $P_0$ in the action
on the twisted Hilbert space
${\mathcal H}_g$ by a $g$ dependent phase.\\
(ii) $P_g$ and $g\circ P_0$ may differ in the
action on the twisted Hilbert space
${\mathcal H}_{g'}$ by a $g$ and $g'$ dependent phase.\\
Let us first examine (ii).
Using $k_{g_1g_2}^i$ in (\ref{zg1g2})
and the $g_2$-action on ${\mathcal H}_{g_1}$ in (\ref{TTT}),
it is straightforward to find
\beq
P_{g_2}=\e^{\pi i(\hat{Q}_{g_2}(g_1)-2q(g_2,g_1))}g_2P_0\qquad
\mbox{on ${\mathcal H}_{g_1}$}.
\eeq
To accommodate the possibility (i) we suppose that $P$ and $P_0$ are related
as
\beq
P=\e^{\pi i \theta(g)}P_0\qquad
\mbox{on ${\mathcal H}_{g}$}.
\eeq
Then the above procedure is modified as follows
\beqa
\Tr\!\!\mathop{}_{{\mathcal H}^{{\mathcal C}/G}}Pq^H
&=&{1\over |G|}
\sum_{g_1,g_2}\Tr\!\!\mathop{}_{{\mathcal H}_{g_1}} g_2Pq^H
\nn\\
&=&{1\over |G|}\sum_{g_1,g_2}
\e^{\pi i \theta(g_1)}
\e^{-\pi i (\hat{Q}_{g_2}(g_1)-2q(g_2,g_1))}
\Tr\!\!\mathop{}_{{\mathcal H}_{g_1}} P_{g_2}q^H
\nn\\
&=&{1\over |G|}\sum_{g_1,g_2}
\e^{\pi i \theta(g_1)}
\e^{-\pi i (\hat{Q}_{g_2}(g_1)-2q(g_2,g_1))}
{}^{\mathcal C}\!\langle\Scr{C}_{P_{g_1g_2}}|q_t^H
|\Scr{C}_{P_{g_2}}\rangle^{\!{\mathcal C}}.
\eeqa
We would like the phase on the RHS to be of the form
$\e^{-i\omega_{g_1g_2}+i\omega_{g_2}}$,
so that the partition function can be expressed
as $\langle -|q_t^H|-\rangle$, where
$$|-\rangle={1\over
\sqrt{|G|}}
\sum_g\e^{i\omega_g}|\Scr{C}_{P_{g}}\rangle^{\mathcal C}.$$
Thus, we need to have
$$
\e^{\pi i(\theta(g_1)-\hat{Q}_{g_2}(g_1)+2q(g_2,g_1))}
=\e^{-i\omega_{g_1g_2}+i\omega_{g_2}}.
$$
Setting $g_2=1$ we find
$\e^{\pi i \theta(g)}=\e^{-i\omega_g+i\omega_1}$.
Inserting this relation,
we find the constraint on $\theta(g)$:
\beq
\theta(g_1g_2)=\theta(g_1)+\theta(g_2)-\hat{Q}_{g_2}(g_1)+2q(g_2,g_1)\qquad
\mbox{mod 2}.
\label{thetaEq}
\eeq
For each solution $\theta(g)$ to this constraint,
we find the crosscap state
\beq
|\Scr{C}_{P^{\theta}}\rangle^{\!{\mathcal C}/G}
={\e^{i\omega_1}\over\sqrt{|G|}}
\sum_{g\in G}
\e^{-\pi i\theta(g)}|\Scr{C}_{P_g}\rangle^{\!{\mathcal C}}.
\label{orbcross1}
\eeq
Let us count the number of solutions to (\ref{thetaEq}).
If we find one solution, $\theta_*(g)$, the other solutions take the form
$\theta_*(g)+{\mit\Delta}\theta(g)$,
where ${\mit\Delta}\theta(g)$ obey the homogeneous equation
${\mit\Delta}\theta(g_1g_2)
={\mit\Delta}\theta(g_1)+{\mit\Delta}\theta(g_2)$ mod 2.
Note that $g\to \e^{i\pi {\mit\Delta}\theta(g)}$
defines a representation of the group $G$ into $U(1)$.
Since there are $|G|$ such representations, we find that 
Eq.~(\ref{thetaEq}) has $|G|$ solutions.

We could have chosen another $P$ in this construction.
In the above, $P$ was equal to $P_0$ when acting
on the untwisted Hilbert space.
Replacing $P_0$ here by $P_{g_1}$ does nothing new if $g_1\in G$,
since the average over $G$ will be taken.
However, replacing $P_0$ by $P_{g'}$ with $g'\in \Scr{G}\setminus G$
will make a difference.
Repeating the above procedure, we find
parity symmetries of the orbifold theory
induced from such a $P$.
There are $|G|$ of them:
one for each solution $\theta(g)$ of (\ref{thetaEq})
which acts on the states as
\beq
P^{\theta}_{g'}=\e^{\pi i (\theta(g)-\hat{Q}_{g'}(g))}P_{g'}
\qquad
\mbox{on ${\mathcal H}_{g}$},
\eeq
and has the crosscap state
\beq
|\Scr{C}_{P^{\theta}_{g'}}\rangle^{\!{\mathcal C}/G}
={\e^{i\omega_{g'}}\over\sqrt{|G|}}
\sum_{g\in G}
\e^{-\pi i(\theta(g)-\hat{Q}_{g'}(g))}
|\Scr{C}_{P_{gg'}}\rangle^{\!{\mathcal C}}.
\label{orbcross2}
\eeq
Again, replacing $g'$ by $g'g_1$ with $g_1\in G$ makes no difference.

To summarize,
for each $P$ we find $|G|$ parities
from the choice of solutions to (\ref{thetaEq}),
and there are $|\Scr{G}/G|$ choices for $P$ itself.
Thus, we have found as many parity symmetries as
$$
|G|\times |\Scr{G}/G|=|\Scr{G}|.
$$ 
 
\subsubsection*{\it The square of $P_{g'}^{\theta}$}

The parity symmetries obtained this way are not necessarily
involutive. Since $P_{g'}$ is involutive,
the square of $P^{\theta}_{g'}$ is given by
\beq
(P^{\theta}_{g'})^2=\e^{2\pi i (\theta(g)-Q_{g'}(g))}\times 
\qquad
\mbox{on ${\mathcal H}_{g}$}.
\label{SqP}
\eeq
We note that
 $\theta(g)$ obeys
$\theta(g_1g_2)=\theta(g_1)+\theta(g_2)$ modulo 1,
since $\hat{Q}_{g_1}(g_2)-2q(g_1,g_2)$ is an integer.
Note also that
$Q_{g'}(g_1g_2)=Q_{g'}(g_1)+Q_{g'}(g_2)$ modulo 1.
Thus,
we find that $g\mapsto
\e^{2\pi i (\theta(g)-Q_{g'}(g))}$ is a homomorphism of
$G$ to $U(1)$, namely a character of $G$.
Therefore, $(P^{\theta}_{g'})^2$ is a quantum symmetry
of the orbifold model. In particular, the crosscap state $|\Scr{C}_{P_{g'}^{\theta}}\rangle$
must be a state on the circle with the boundary condition
twisted by this quantum symmetry.
This means, as shown in Appendix~\ref{subap:quantums},
that the state must transform under the action of $g$ as
\beq
|\Scr{C}_{P_{g'}^{\theta}}\rangle
\,\stackrel{g}{\longmapsto}\,
\e^{2\pi i (\theta(g)-Q_{g'}(g))}
|\Scr{C}_{P_{g'}^{\theta}}\rangle,
\eeq
which can also be confirmed by a direct computation.

\subsubsection{Boundary States in Orbifolds}\label{subsubsec:Borb}

An idea to obtain D-branes in the orbifold model
is to pick a D-brane $i$ in the original system and
to take the ``average'' over the image branes $g(i)$, $g\in G$.
The corresponding boundary states are given by
\beq
|\Scr{B}_{[i]}\rangle^{{\mathcal C}/G}
={1\over \sqrt{|G|}}
\sum_{g\in G}|\Scr{B}_{g(i)}\rangle^{\mathcal C}.
\label{bi0}
\eeq
The normalization factor
$1/\sqrt{|G|}$ is required for the open string partition function
to count the $i$-$g(i)$ string just once.
(A more careful treatment is required if $g(i)=i$
for some $g\ne {\rm id}$, see below.)
Since $g|\Scr{B}_i\rangle=|\Scr{B}_{g(i)}\rangle$ (\ref{mapofC}),
the state $|\Scr{B}_{[i]}\rangle^{{\mathcal C}/G}$ is $G$-invariant
and belongs to the Hilbert space ${\mathcal H}^{{\mathcal C}/G}$.
Obviously the brane $\Scr{B}_{[i]}$ is the same as
$\Scr{B}_{[g(i)]}$.

Since the parities
$P_{g'}^{\theta}$ are not involutive, but
square to quantum symmetries (\ref{SqP}),
one is motivated to consider the boundary states on the circle with twisted
boundary condition.
Let $g_{\rho}$ be the quantum symmetry associated with the character
$g\mapsto \e^{2\pi i \rho(g)}$.
We claim that the $g_{\rho}$-twisted boundary state for the brane
$\Scr{B}_{[i]}$ takes the form
\beq
|\Scr{B}_{[i]}\rangle_{g_{\rho}}^{{\mathcal C}/G}
={\e^{i\lambda}\over \sqrt{|G|}}
\sum_{g\in G}\e^{-2\pi i\rho(g)}|\Scr{B}_{g(i)}\rangle^{\mathcal C}.
\label{bir}
\eeq
Indeed, $g\in G$ transforms it as
\beq
|\Scr{B}_{[i]}\rangle_{g_{\rho}}^{{\mathcal C}/G}
\stackrel{g}{\longmapsto} 
{\e^{i\lambda}\over \sqrt{|G|}}
\sum_{g'\in G}\e^{-2\pi i\rho(g')}|\Scr{B}_{gg'(i)}\rangle^{\mathcal C}
=\e^{2\pi i \rho(g)}|\Scr{B}_{[i]}\rangle_{g_{\rho}}^{{\mathcal C}/G}.
\eeq
As shown in Appendix~\ref{subap:quantums},
this means that $|\Scr{B}_{[i]}\rangle_{g_{\rho}}^{{\mathcal C}/G}$
is a state on the $g_{\rho}$-twisted circle.

As mentioned above,
when $g(i)\ne i$ if $g\ne 1$ for some $i$, the argument has to be further refined.
This is known as ``the fixed-point problem''.
Resolved boundary states have been constructed in 
\cite{FHSSW,FSbreakI,FSbreakII,BFS}.
In this paper, we do not try to reproduce a general solution, but will revisit
the resolutions
in the concrete models we consider later.

\subsubsection{Constraints on Discrete Torsion}

In general, there can be more than one models of orbifold
${\mathcal C}/G$. We have chosen a particular one
with the partition function (\ref{orbPF}), but one could change
the model by turning on a ``discrete torsion'' \cite{Vafa:DT}.
This means adding an extra phase factor $\e^{2\pi i e(g_2,g_1)}$
for each summand of (\ref{orbPF}), where $e(g_2,g_1)$ is an
antisymmetric bilinear form of $G$ with values in $\R/\Z$
such that $e(g,g)=0$.
Let us see how this modifies the above story.

The discrete torsion shifts the bilinear form $q$
as 
$$q(g_2,g_1)\to q(g_2,g_1)-e(g_2,g_1).$$
The argument above goes through
without modification until (\ref{thetaEq}),
at which point one has to be careful.
We note that $\hat{Q}_{g_2}(g_1)=h_{g_2}+h_{g_1}-h_{g_1g_2}$
is symmetric under the exchange $g_1\leftrightarrow g_2$.
Thus, (\ref{thetaEq}) is possible only if $2(q(g_2,g_1)-e(g_2,g_1))$
is symmetric (mod 2). Since $q(g_2,g_1)$ is already symmetric, we find that
$2e(g_2,g_1)$ has to be symmetric modulo 2, or
$e(g_2,g_1)$ has to be symmetric modulo 1.
Since $e(g_2,g_1)$ is antisymmetric at the same time,
it may appear that no discrete torsion is allowed.
However, since $-{1\over 2}
\equiv {1\over 2}$ mod $\Z$,
a symmetric form with
$0$ or ${1\over 2}$ entries is at the same time antisymmetric modulo 1.
Thus, special types of discrete torsion are indeed allowed.
We shall call them {\it $\Z_2$ discrete torsions}.

\noindent
{\bf Remark 1.}
In general, choices are involved in finding a symmetric bilinear form
$q(g_1,g_2)$ such that $2q(g_1,g_2)=Q_{g_1}(g_2)$ mod 1
and $q(g,g)=-h_g$.
A different choice corresponds exactly to the
modification by a $\Z_2$ discrete torsion.
\\
{\bf Remark 2.}
The restriction on the discrete torsion in orientifold models 
is not new, if one recalls that the discrete torsion is a kind of
$B$-field: Type I string theory projects out the NS-NS B-field modes
\cite{GSW}.
Furthermore, it is also known that special kinds of B-fields
(with period $\pi$) are allowed \cite{SS}. 
(See also \cite{AAADS})
\\
{\bf Remark 3.}
It may appear natural to relate the $\Z_2$ discrete torsions to
the group cohomology classes $\alpha\in H^2(G,\Z_2)$ in the standard way:
$\e^{2\pi i e(g,h)}=\alpha(g,h)\alpha(h,g)^{-1}$.
However, unlike the ordinary case where both
$\e^{2\pi i e(g,h)}$ and $\alpha(g,h)$ take values in $U(1)$,
it is not always true that the map
$\alpha\mapsto \e^{2\pi i e}$ is one-to-one
\footnote{
Let $A$ be an abelian group.
$Ext(G,A)=\{\alpha\in H^2(G,A)|{\rm symmetric}\}$, the kernel of the map
$\alpha(g,h)\to\epsilon(g,h)=\alpha(g,h)\alpha(h,g)^{-1}$, is the set of
{\it abelian} extensions of $G$ by $A$.
It is trivial for $A=U(1)$ but not always for other $A$.
For example $\Z_4$ and $\Z_2\times \Z_2$ are inequivalent
$\Z_2$ extensions of $\Z_2$. 
(For a product $G=G_1\times G_2\times \cdots\times G_s$, 
$Ext(G,A)\cong\prod_{i=1}^s Ext(G_i,A)$ (Cor 3.17 of \cite{GCbook}).
Also $Ext(\Z_n,A)=H^2(\Z_n,A)=A/A^n$ (Theorem 3.1 of \cite{GCbook}).
Thus, if $G$ has a $\Z_n$ factor with $n$ even,
$Ext(G,\Z_2)$ cannot be trivial.)}.
Thus, just from the above consideration, one cannot conclude
that $H^2(G,\Z_2)$ characterizes the $\Z_2$ discrete torsion.
However, there is a claim \cite{KlRa} that this is indeed the
case in certain models.

\subsection{New Crosscaps from Mirror Symmetry}

\subsubsection{Twisting the Symmetry by
Automorphisms}

Let $\omega$ be an automorphism of the chiral algebra ${\mathcal A}$
that acts trivially on the Virasoro subalgebra $\{L_n\}$.
The space $\Scr{H}_i$ acted on by ${\mathcal A}$ through $\omega$,
$W:v\to \omega(W)v$, can be viewed as another representation
$\Scr{H}_{\omega(i)}$ of ${\mathcal A}$.
In other words, there is a unitary isomorphism
\beq
V_{\omega}:\Scr{H}_i\to \Scr{H}_{\omega(i)},
\eeq
such that $\omega(W)=V_{\omega}^{-1}WV_{\omega}$.

The algebra ${\mathcal A}$ can be embedded into the symmetry algebra
${\mathcal A}\otimes {\mathcal A}$ as
$W\mapsto W\otimes 1+1\otimes \omega(W)$.
We can then consider D-branes and orientifolds that preserve
such ``$\omega$-diagonal'' subalgebras \cite{FSbranes,RSGep}.
They are associated with  boundary conditions such that
$W^{(r)}(t,0) 
=\omega \widetilde{W}^{(r)}(t,0)$
and parity symmetries that map
$W^{(r)}(t,x)$ to $\omega\widetilde{W}^{(r)}(t,-x)$.
The conditions on the corresponding boundary and crosscap states
are twisted accordingly: $\widetilde{W}_{-n}^{(r)}$
in (\ref{Bcond}) and (\ref{Ccond})
are replaced by $\omega(\widetilde{W}_{-n}^{(r)})$.
The linear basis of solutions is given by the ``$\omega$-type
Ishibashi states''
\beqa
&&
\kket{i}_{\omega}=(V_{\omega}\otimes {\rm id})\kket{\omega^{-1}(i)},\\
&&
\cket{i}_{\omega}
=(V_{\omega}\otimes {\rm id})\cket{\omega^{-1}(i)}=\e^{\pi i
(L_0-h_i)}\kket{i}_{\omega},
\eeqa
which are sums of elements in
$\Scr{H}_i\otimes\Scr{H}_{\overline{\omega^{-1}(i)}}$.
These states have the same mutual inner-products
as the ordinary Ishibashi states (\ref{bbchi})-(\ref{bcchi}).
Inner products of states with different $\omega$'s
(say $\omega=1$ and $\omega\ne 1$)
are given in terms of so-called ``twining characters''.
For instance,
\beqa
\bbra{j} q^{H} \kket{i}_\omega
&=& \sum_{N,M} \langle j,N|V_{\omega}q^{L_0-\frac{c}{24}}
\ket{\omega^{-1}(i),M}
\langle j,N|q^{L_0-{c\over 24}}
\ket{\omega^{-1}(i),M}^{\dag} \\ \no
&=& \delta_{i,j}\delta_{i,\omega(i)} \ 
\tr_{{\cal H}_i} V_\omega q^{2L_0-\frac{c}{12}} = 
\delta_{i,j}\delta_{i,\omega(i)} \ \chi_j^{(0)} (2\tau).
\eeqa

Boundary and crosscap states are linear combinations of
the $\omega$-type Ishibashi-states.
For boundary states,
there is a long list of works that aim at determining the appropriate
linear combinations.
For crosscaps, the same amount of investigation has not been done.
Here, we do not attempt to determine the appropriate combinations
in full generality.
However, we will find that this can be done
in the case where an orbifold is ``mirror'' to the original (in the sense
described below).
The knowledge on the crosscaps for orbifolds turns out useful here.

\subsubsection{Mirror Symmetry}

\newcommand{\mapr}[1]{%
  \smash{\mathop{%
  \hbox to 1cm{\rightarrowfill}}\limits^{#1}}}
\newcommand{\mapdl}[1]{\Big\downarrow
  \llap{$\vcenter{\hbox{$\scriptstyle#1\,$}}$}}
\newcommand{\mapdr}[1]{\Big\downarrow
  \rlap{$\vcenter{\hbox{$\scriptstyle#1\,$}}$}}

An automorphism $\omega$ that conjugates the representations
$\omega(i)=\bi$ (any $i$) is called a mirror automorphism.
Two CFTs of symmetry algebra ${\mathcal A}\otimes {\mathcal A}$
are said to be mirror to each other when they are equivalent as 2d quantum
field theories and
the action of $W\otimes W'\in {\mathcal A}\otimes {\mathcal A}$
in one theory
is mapped to the action of $\omega(W)\otimes W'\in
{\mathcal A}\otimes  {\mathcal A}$ in the other.
Namely, if ${\mathcal H}_1$ and ${\mathcal H}_2$ are the Hilbert spaces
of states of the two theories and $\Psi$ is the isomorphism between them,
the following diagram commutes
\beq
\begin{array}{ccc}
{\mathcal H}_1&\mapr{\Psi}&{\mathcal H}_2\\
\mapdl{W\otimes W'\,\,}&&\mapdr{\omega(W)\otimes W'}\\
{\mathcal H}_1&\mapr{\Psi}&{\mathcal H}_2
\end{array}
\eeq
On each ${\mathcal A}\otimes{\mathcal A}$-irreducible subspace,
the isomorphism $\Psi$ acts as $V_{\omega}^{-1}\otimes {\rm id}$ times a
constant.
The two basic modular invariants
--- the charge conjugation modular invariant
${\mathcal H}^{\mathcal C}=\oplus_i\Scr{H}_i\otimes \Scr{H}_{\bi}$
and the diagonal modular invariant
${\mathcal H}^{\mathcal D}=\oplus_i\Scr{H}_i\otimes\Scr{H}_i$ ---
are mirror to each other.

A typical example of mirror symmetry is T-duality.
The sigma model on the circle of radius $R=\sqrt{k_1/k_2}$
and the model of radius $1/R=\sqrt{k_2/k_1}$, with $k_1, k_2$
integers, are both
RCFTs with chiral algebra $U(1)_{k_1k_2}$.
T-duality between them is a mirror symmetry.
Another example is the level $k$ $SU(2)/U(1)$ gauged WZW model,
which is the charge-conjugate modular invariant of the level $k$
parafermion algebra, and its orbifold by a $\Z_k$ symmetry group,
which is the diagonal modular invariant.
These examples will be studied in detail later in this paper.
A related example is the level $k$ $SU(2)/U(1)$ supersymmetric
gauged WZW model (Kazama-Suzuki model),
which is the charge-conjugate modular invariant of the
level $k$ superparafermion algebra, and its
orbifold by a $\Z_{k+2}\times \Z_2$ symmetry group,
which is the diagonal modular invariant.
This last example will be studied in detail in \cite{BH2}.
In fact, in this example, the two theories are mirror in the standard
sense: the isomorphism of the Hilbert spaces
acts on the $(2,2)$ supersymmetry algebra via
the standard mirror automorphism.

\subsubsection{A-branes/B-branes and A-parities/B-parities}

In what follows, D-branes and parities that preserve
the ordinary diagonal symmetry
${\mathcal A}\subset {\mathcal A}\otimes {\mathcal A}$
shall be referred to as
{\it A-branes} and {\it A-parities}.
Cardy branes and PSS parities are therefore A-branes and A-parities.
For a mirror automorphism $\omega$,
D-branes and parities that preserve
the $\omega$-diagonal
symmetry shall be referred to as
{\it B-branes} and {\it B-parities}.
A-branes and B-branes are exchanged under mirror symmetry,
and so are A- and B-parities.
Let $\Psi:{\mathcal H}_1\to{\mathcal H}_2$ be a mirror isomorphism as above.
If $|\Scr{B}\rangle_2$ and $|\Scr{C}\rangle_2$ are
the boundary and crosscap states corresponding
to an A-brane and an A-parity in ``theory 2'',
then $\Psi^{-1}|\Scr{B}\rangle_2$ and
$\Psi^{-1}|\Scr{C}\rangle_2$ correspond to
a B-brane and a B-parity in ``theory 1''.
(The terminology of ``A-type'' and ``B-type'' is the extension
of the one used for ${\mathcal N}=2$ supersymmetric theories \cite{OOY,BH2}.
Mirror symmetry for orientifolds is used in \cite{AAHV}
in that context.)

An RCFT ${\mathcal C}$ is sometimes
mirror to one of its orbifold models, ${\mathcal C}/G$, as in
the three examples mentioned above --- rational
$U(1)$, $SU(2)/U(1)$ coset model, and supersymmetric $SU(2)/U(1)$ model.
In such a case, one can construct B-branes/B-parities
in the model  ${\mathcal C}$
by applying the mirror isomorphism
$\Psi^{-1}$ to the A-branes/A-parities of the orbifold model,
which are in turn obtained by applying the orbifold technique
developed in the literature and in Section~\ref{subsec:orb}.

To be specific, let ${\mathcal C}$ be the charge-conjugate modular invariant
and ${\mathcal C}/G$ be the mirror diagonal modular invariant.
${\mathcal C}/G$
has A-parities $P_{g'}^{\theta}$ with crosscap (\ref{orbcross2}),
labelled by the solutions $\theta$ to (\ref{thetaEq}) and $g'\in\Scr{G}/G$.
Thus, ${\mathcal C}$ has B-parities $P_B^{\theta,g'}$
whose crosscap states are given by
\beq
|\Scr{C}_{P_B^{\theta,g'}}\rangle
=\Psi^{-1}|\Scr{C}_{P_{g'}^{\theta}}\rangle^{{\mathcal C}/G}.
\eeq
($\Psi^{-1}$
acts as $V_{\omega}\otimes 1$, up to a phase multiplication.)
B-parities obtained this way are not in general involutive.
We recall from (\ref{SqP}) that the square of $P_{g'}^{\theta}$
is the multiplication by
$\e^{2\pi i (\theta(g)-Q_{g'}(g))}$ on
${\mathcal H}_{g}=\oplus_{i}\Scr{H}_{i}\otimes\Scr{H}_{g(\bi)}$.
Since the orbifold model is the diagonal modular invariant,
only the subspaces with $g(\bi)=i$ remain in the spectrum
of ${\mathcal C}/G$.
Thus, $(P_{g'}^{\theta})^2=\e^{2\pi i (\theta(g)-Q_{g'}(g))}$
on $\Scr{H}_i\otimes \Scr{H}_i$ such that $g(\bi)=i$,
or on $\Scr{H}_{\bi}\otimes \Scr{H}_{\bi}$ such that $g(i)=\bi$.
Since the mirror isomorphism $\Psi^{-1}$ maps
$\Scr{H}_{\bi}\otimes \Scr{H}_{\bi}$ to
$\Scr{H}_i\otimes\Scr{H}_{\bi}$, we find
\beq
(P_B^{\theta,g'})^2
=\e^{2\pi i (\theta(g)-Q_{g'}(g))}\quad
\mbox{on $\Scr{H}_i\otimes\Scr{H}_{\bi}\subset {\mathcal H}^{\mathcal C}$
such that $g(i)=\bi$.}
\label{sqformula}
\eeq
Here we are assuming that $g(i)=\bi$ uniquely fixes $g$.
However, this is not always the case
if there are simple current fixed points.
In such a case, we need to trace back in order to see
from which twisted sector comes the subspace $\Scr{H}_{\bi}\otimes\Scr{H}_{\bi}$ in the orbifold theory.

\section{Circle of Radius $R$}\label{sec:arbR}

\newcommand{\talpha}{\widetilde{\alpha}}
\newcommand{\tilg}{\widetilde{g}}
\newcommand{\RP}{{\bf R}{\rm P}}
\newcommand{\mDx}{{\mit\Delta} x}
\newcommand{\mDa}{{\mit\Delta} a}

The sigma model whose target space is
$S^1$ of radius $R$ is described by a periodic scalar field
$X\equiv X+2\pi R$.
The algebra of oscillator modes
$\alpha_n$ and $\talpha_n$ of $X$ acts on the space of states
\beq
{\mathcal H}=\bigoplus_{l,m\in \Z}{\mathcal H}_{l,m},
\eeq
where the labels $l$ and $m$ on the Fock space ${\mathcal H}_{l,m}$ correspond
to the momentum and winding number, respectively. 
We denote by $|l,m\rangle\in {\mathcal H}_{l,m}$ the lowest energy state 
annihilated by the modes $\alpha_n$ and $\talpha_n$ with $n>0$. The energy of
this state is 
${1\over 2}(({l\over R})^2+(Rm)^2)
-{1\over 12}$.
There are two $U(1)$ symmetries
\beq
g_{{}_{\mDx} }:|l,m\rangle\mapsto \e^{-il\mDx/R}|l,m\rangle,
\quad
\tilg_{{}_{\mDa}}:|l,m\rangle\mapsto \e^{-imR\mDa}|l,m\rangle.
\label{actgg}
\eeq
We interpret $g_{{}_{\mDx}}$  as the rotation of the circle,
$X\to X+\mDx$.
Under T-duality, the sigma model on the circle of radius $R$
is mapped to the model on the circle of radius $1/R$.
The states and operators are mapped as follows
\beq
|l,m\rangle\to |m,l\rangle,\quad
\alpha_n\to -\alpha_n,\quad
\talpha_n\to\talpha_n.
\label{Tduality}
\eeq
The operation $\tilg_{{}_{\mDa}}$ is interpreted as the rotation of
the T-dual circle $X'\to X'+\mDa$.

\subsection{D-Branes}

There are two kinds of D-branes ---
D1-branes and D0-branes associated with the Neumann and Dirichlet
boundary conditions
on $X$ respectively.
We denote by $N_a$  the D1-brane with Wilson line $a$, and
by $D_{x_{}}$
the D0-brane located at $X=x_{}$.

The Heisenberg algebra $[\alpha_r,\alpha_{r'}]=r\delta_{r+r',0}$
acts on open string states, where $r\in \Z$ for N--N and D--D strings
and $r\in \Z+{1\over 2}$ for D--N and N--D strings.
 The $N_{a_1}$--$N_{a_2}$ string states are labelled
by the momentum $l\in \Z$ and the state $|l\rangle_{a_1,a_2}$ annihilated
by $\alpha_{n>0}$ has the lowest energy
$({l\over R}+{a_2-a_1\over 2\pi})^2-{1\over 24}$.
We assume the identification
$|l\rangle_{a_1,a_2}
=|l+p_1-p_2\rangle_{a_1+{2\pi\over R}p_1,a_2+{2\pi\over R}p_2}$
for integers $p_1,p_2$.
The $D_{x_1}$--$D_{x_2}$ string states are labelled by the winding number
$m\in\Z$ and the state
$|m\rangle_{x_1,x_2}$ annihilated by $\alpha_{n>0}$ has
the lowest energy $(Rm+{x_2-x_1\over 2\pi})^2-{1\over 24}$.
We assume the identification
$|m\rangle_{x_1,x_2}=|m+q_1-q_2\rangle_{x_1+2\pi Rq_1,x_2+2\pi Rq_2}$
for $q_1,q_2 \in \Z$.

Computing the partition function and performing the modular transform,
we find the boundary states for these branes:
\beqa
&&
|N_a\rangle=\sqrt{R\over \sqrt{2}}
\sum_{m\in\Z}\e^{-i Ram}\exp\left(-\sum_{n=1}^{\infty}
{1\over n}\alpha_{-n}\talpha_{-n}\right)|0,m\rangle,
\label{Nina}
\\
&&
|D_{x_{}}\rangle
=\sqrt{1\over R\sqrt{2}}\sum_{l\in \Z}\e^{-i{x_{}\over R}l}
\exp\left(\sum_{n=1}^{\infty}{1\over n}\alpha_{-n}\talpha_{-n}\right)
|l,0\rangle
\label{Din}
\eeqa

The D1-brane wrapped on a circle and the D0-brane in the dual circle
are mapped to
each other under T-duality (\ref{Tduality}),
where the Wilson line of a D1-brane is mapped to the location of the D0-brane.

The rotation symmetries (\ref{actgg}) act on the branes and
the open string states as
\beqa
&&g_{{}_{\mDx}}:\left\{
\begin{array}{l}
N_a\to N_a;\quad
|l\rangle_{a_1,a_2}\mapsto
\e^{-i\mDx({l\over R}+{a_2-a_1\over 2\pi})}|l\rangle_{a_1,a_2},\\
D_x\to D_{x+\mDx};\quad
|m\rangle_{x_1,x_2}\mapsto |m\rangle_{x_1+\mDx,x_2+\mDx},
\end{array}\right.
\\
&&\tilg_{{}_{\mDa}}:\left\{
\begin{array}{l}
N_a\to N_{a+\mDa};\quad
|l\rangle_{a_1,a_2}\mapsto
|l\rangle_{a_1+\mDa,a_2+\mDa},
\label{gdX}\\
D_x\to D_x;\quad
|m\rangle_{x_1,x_2}\mapsto
\e^{-i\mDa(Rm+{x_2-x_1\over 2\pi})}
|m\rangle_{x_1,x_2}.
\end{array}\right.
\eeqa
The extra phases, such as $\e^{-i{a_2-a_1\over 2\pi}\mDx}$ in
(\ref{gdX}), come from the parallel transport
of the open string boundary.
Note that $g_{2\pi R}$ and $\tilg_{2\pi\over R}$ act as the identity
on the closed string states, but not on the open string states
for generic values of Wilson lines and positions.
As a consequence, the symmetry group is no longer $U(1)\times U(1)$ but $\R\times\R$.

\subsection{$\Z_2$ Orientifolds}

Let us consider the parities
\beqa
&&\Omega:X(t,\s)\to X(t,-\s),\\
&&s\Omega:X(t,\s)\to X(t,-\s)+\pi R.
\eeqa
These act on the states and branes as
\beqa
&&\Omega:\left\{
\begin{array}{l}
|l,m\rangle\to |l,-m\rangle\\
N_a\to N_{-a};\quad |l\rangle_{a_1,a_2}\to |l\rangle_{-a_2,-a_1}\\
D_{x_{}}\to D_{x_{}};
\quad
|m\rangle_{x_1,x_2}\to |-m\rangle_{x_2,x_1}
\end{array}\right.
\label{acOm}
\\
&&s\Omega:\left\{
\begin{array}{l}
|l,m\rangle\to (-1)^l|l,-m\rangle\\
N_a\to N_{-a};\quad |l\rangle_{a_1,a_2}\to
\e^{-\pi i (l+{R(a_1+a_2)\over 2\pi})}|l\rangle_{-a_2,-a_1}\\
D_{x_{}}\to D_{x_{}+\pi R};
\quad
|m\rangle_{x_1,x_2}\to |-m\rangle_{x_2+\pi R,x_1+\pi R}
\end{array}\right.
\label{acsOm}
\eeqa
The parity $s\Omega$ acts in the same way as $g_{\pi R}\Omega$
on the closed string states as well as the DD string states,
but differs from it by an overall phase $\e^{-iRa_1}$
in the action on the NN string states.
This is to make $s\Omega$ involutive
(note that $(g_{\pi R}\Omega)^2=g_{2\pi R}\ne 1$ on NN string states).
Computing the partition functions and making the modular transform,
we find the following expressions for the crosscap states
\beqa
&&|\Scr{C}_{\Omega}\rangle
=\sqrt{R\sqrt{2}}\sum_{m'\in\Z}\exp\left(-\sum_{n=1}^{\infty}
{(-1)^n\over n}\alpha_{-n}\talpha_{-n}\right)|0,2m'\rangle,
\label{Om}
\\
&&|\Scr{C}_{s\Omega}\rangle
=\sqrt{R\sqrt{2}}\sum_{m'\in\Z}\exp\left(-\sum_{n=1}^{\infty}
{(-1)^n\over n}\alpha_{-n}\talpha_{-n}\right)|0,2m'+1\rangle.
\label{sOm}
\eeqa

Applying T-duality to (\ref{acOm}),(\ref{acsOm}) and
(\ref{Om}),(\ref{sOm})
in the system of radius $1/R$,
we find two other parity symmetries
\beqa
&&I\Omega:\left\{
\begin{array}{l}
|l,m\rangle\to |-l,m\rangle\\
N_a\to N_{a};\quad |l\rangle_{a_1,a_2}\to |-l\rangle_{a_2,a_1}\\
D_{x_{}}\to D_{-x_{}};
\quad
|m\rangle_{x_1,x_2}\to |m\rangle_{-x_2,-x_1}
\end{array}\right.
\\
&&I'\Omega:\left\{
\begin{array}{l}
|l,m\rangle\to (-1)^m|-l,m\rangle\\
N_a\to N_{a+{\pi\over R}};
\quad |l\rangle_{a_1,a_2}\to |-l\rangle_{a_2+{\pi\over R},a_1+{\pi\over R}}\\
D_{x_{}}\to D_{-x_{}};
\quad
|m\rangle_{x_1,x_2}\to
\e^{-\pi i(m+{x_1+x_2\over 2\pi R})}|m\rangle_{-x_2,-x_1}
\end{array}\right.
\eeqa
with the crosscap states
\beqa
&&|\Scr{C}_{I\Omega}\rangle
=\sqrt{\sqrt{2}\over R}\sum_{l'\in\Z}\exp\left(\sum_{n=1}^{\infty}
{(-1)^n\over n}\alpha_{-n}\talpha_{-n}\right)|2l',0\rangle,
\\
&&|\Scr{C}_{I'\Omega}\rangle
=\sqrt{\sqrt{2}\over R}\sum_{l'\in\Z}\exp\left(\sum_{n=1}^{\infty}
{(-1)^n\over n}\alpha_{-n}\talpha_{-n}\right)|2l'+1,0\rangle.
\eeqa
We see that they both correspond to the involution
\beq
X(t,\s)\to -X(t,-\s).
\eeq
There are two
orientifold fixed points of $X\to -X$; one at $X=0$ and
another one at $X=\pi R$.
The difference between $I\Omega $ and $I'\Omega$ arises, for instance,
in the $\RP^2$ diagram:
$\langle 0|\Scr{C}_{I\Omega}\rangle=\sqrt{R\sqrt{2}}$
whereas $\langle 0|\Scr{C}_{I'\Omega}\rangle=0$.
In a full string model that contains the circle as one of the
compactified dimensions, taking this overlap corresponds to
determining the  total tension of the orientifold planes. 
The fact that the overlap for $I'\Omega$ vanishes means that the
tensions of the orientifold planes located at the two fixed points
have opposite signs and cancel out. In string theory language,
the orientifold located at one of the fixed points 
is a ${\cal O}^+$-plane, the one at the other a
${\cal O}^-$-plane. For $I\Omega$ the tensions add up and the
two orientifold planes are of the same type.

This can be confirmed by comparing
the action on the open strings stretched between D0-branes.
We consider D0-branes at the fixed  points. 
The state $|0\rangle_{0,0}$ is a state of the
string ending on the D0-brane at $X\equiv 0$ without winding,
whereas $|1\rangle_{\pi R,-\pi R}$ is a state of
the string ending on the D0-brane at $X\equiv \pi R$ without winding.
The action of the two parities on these open string states are
\beq
I\Omega:\left\{\begin{array}{l}
|0\rangle_{0,0}\to |0\rangle_{0,0},\\
|1\rangle_{\pi R,-\pi R}\to
|1\rangle_{\pi R,-\pi R}
\end{array}\right.
\quad
I'\Omega:\left\{\begin{array}{l}
|0\rangle_{0,0}\to |0\rangle_{0,0},\\
|1\rangle_{\pi R,-\pi R}\to
-|1\rangle_{\pi R,-\pi R}
\end{array}\right.
\eeq
We see that the action of $I\Omega$ on the two string states is identical,
while $I'\Omega$ acts on them with opposite signs.
For $I\Omega$ orientifold, the orientifold planes at $X=0$ and $X=\pi R$
are both of $SO$-type.
For the $I'\Omega$ orientifold,
the orientifold plane at $X=0$ is of $SO$-type and the one at $X=\pi R$
is of $Sp$-type.

Since $I'\Omega$
is obtained from T-duality applied to $s\Omega$,
we have proved that the 
$s\Omega$ orientifold is T-dual to the orientifold with $Sp/SO$ mixture.
This was observed in the context of superstring theory in \cite{nvs}.

\subsection{Combination with Rotations I: Involutive Parities}

One can combine  the parity symmetries considered above
and the rotation symmetries $g_{{}_{\mDx}}$ and $\tilg_{{}_{\mDa}}$.
We first consider the combinations of the form
$gPg^{-1}$.
They are involutive
so that the crosscap states belong to the ordinary space
of states ${\mathcal H}$.
In fact, in order to find the crosscap state we can use the recipe
given in Appendix~\ref{app:some}:
$g|\Scr{C}_P\rangle=|\Scr{C}_{gPg^{-1}}\rangle$.

Let us first consider the
parity $g_{{}_{\mDx\over 2}}I\Omega g_{{}_{\mDx\over 2}}^{-1}$,
which is actually the same as $g_{{}_{\mDx}}I\Omega$.
The crosscap state is obtained by applying
$g_{{}_{\mDx/2}}$ to $|\Scr{C}_{I\Omega}\rangle$:
\beq
|\Scr{C}_{g_{{}_{\mDx}}I\Omega}\rangle
=\sqrt{\sqrt{2}\over R}\sum_{l'\in\Z}
\e^{-il'\mDx /R}\exp\left(\sum_{n=1}^{\infty}
{(-1)^n\over n}\alpha_{-n}\talpha_{-n}\right)|2l',0\rangle.
\label{CrdI}
\eeq
One can also consider $g_{{}_{\mDx\over 2}}I'\Omega g_{{}_{\mDx\over 2}}^{-1}$.
It is the same as $g_{{}_{\mDx}}I'\Omega$
in the action on the closed string and N--N strings,
but differs from it in the action on the D--D strings.
We therefore denote it  as
$\widetilde{g_{{}_{\mDx}}I'\Omega}$.
\beq
|\Scr{C}_{\widetilde{g_{{}_{\mDx}}I'\Omega}}\rangle
=\sqrt{\sqrt{2}\over R}\sum_{l'\in\Z}
\e^{-i(l'+{1\over 2})\mDx /R}\exp\left(\sum_{n=1}^{\infty}
{(-1)^n\over n}\alpha_{-n}\talpha_{-n}\right)|2l'+1,0\rangle,
\label{CrdIp}
\eeq
Both $g_{{}_{\mDx}}I\Omega$
and 
$\widetilde{g_{{}_{\mDx}}I'\Omega}$
act on the free boson as
$$
X(t,\s)\to -X(t,-\s)+\mDx.
$$
This action has two fixed points
at $X={\mDx\over 2}$ and $X={\mDx\over 2}+\pi R$.
If we move $\mDx$ from $0$ to $2\pi R$,
under which the two fixed points are exchanged,
the crosscap
for $I\Omega$ comes back to itself but the one for $I'\Omega$ comes back
with a sign flip.
This is because the two orientifold planes are of the same type
for $I\Omega$ (both $SO$-type),
while they are of the opposite type for $I'\Omega$ (one is $SO$-type and
the other is $Sp$-type).

We next consider the parity symmetries
$\tilg_{{}_{\mDa}}\Omega=\tilg_{{}_{\mDa\over 2}}\Omega
\tilg_{{}_{\mDa\over 2}}^{-1}$
and $\widetilde{\tilg_{{}_{\mDa}}s\Omega}:=
\tilg_{{}_{\mDa\over 2}}s\Omega
\tilg_{{}_{\mDa\over 2}}^{-1}$.
The crosscap states for these parities are
\beqa
&&
|\Scr{C}_{\tilg_{{}_{\mDa}}\Omega}\rangle
=\sqrt{R\sqrt{2}}\sum_{m'\in\Z}\e^{-im'R\mDa}
\exp\left(-\sum_{n=1}^{\infty}
{(-1)^n\over n}\alpha_{-n}\talpha_{-n}\right)|0,2m'\rangle,
\\
&&
|\Scr{C}_{\widetilde{\tilg_{{}_{\mDa}}s\Omega}}\rangle
=\sqrt{R\sqrt{2}}\sum_{m'\in\Z}\e^{-i(m'+{1\over 2})R\mDa}
\exp\left(-\sum_{n=1}^{\infty}
{(-1)^n\over n}\alpha_{-n}\talpha_{-n}\right)|0,2m'+1\rangle.\qquad
\eeqa
We note that $\tilg_{{}_{\mDa}}\Omega$ as well as
$\widetilde{\tilg_{{}_{\mDa}}s\Omega}$ change the Wilson line as
$a\to -a+\mDa$.

The symmetry $\tilg_{{}_{\mDa}}$ commutes with the parities
$I\Omega$ and $I'\Omega$, whereas
$g_{{}_{\mDx}}$ commutes with
$\Omega$ and $s\Omega$.
Thus, there is no dressing of the form $gPg^{-1}$
other than the ones considered above.

\subsection{Combination with Rotation II: Non-involutive Parities}

We can also consider parities that are not involutive.
An example is $g_{{}_{\mDx}}\Omega: X(t,\s)\to X(t,\s)+\mDx$,
where $(g_{{}_{\mDx}}\Omega)^2=g_{{}_{2\mDx}}$.
The crosscap states for such a parity $P$ are not in the ordinary
space of states but in the space of states with twisted boundary condition
determined by $P^2$.
We note that the effect of the twisting by
$g_{{}_{\mDx}}$ and $\tilg_{{}_{\mDa}}$
is just to shift the momentum and winding number
by $-{R\mDa\over 2\pi}$ and $-{\mDx\over 2\pi R}$:
\beq
{\mathcal H}_{g_{{}_{\mDx}}\tilg_{{}_{\mDa}}}
=\bigoplus_{l\in \Z-
{R\mDa\over 2\pi},
m\in\Z-{\mDx\over 2\pi R}}
{\mathcal H}_{l,m}.
\eeq

The Wilson line of a D1-brane is preserved
under the rotation $g_{{}_{\mDx}}$, while
the position of a D0-brane is preserved by
$\tilg_{{}_{\mDa}}$. Thus one can consider the $g_{{}_{\mDx}}$-twisted
boundary state for $N_a$ and the $\tilg_{{}_{\mDa}}$-twisted
boundary state for $D_x$.
Computing the twisted partition function and making the modular transform,
we obtain the following expressions for these twisted boundary states:
\beqa
&&
|N_a\rangle_{g_{{}_{\mDx}}}
=\sqrt{R\over \sqrt{2}}
\sum_{m\in\Z}
\e^{-i Ram}\exp\left(-\sum_{n=1}^{\infty}
{1\over n}\alpha_{-n}\talpha_{-n}\right)
|0,m-\mbox{$\mDx\over 2\pi R$}\rangle,
\\
&&
|D_{x_{}}\rangle_{\tilg_{{}_{\mDa}}}
=\sqrt{1\over R\sqrt{2}}\sum_{l\in\Z}
\e^{-i{x_{}\over R}l}
\exp\left(\sum_{n=1}^{\infty}{1\over n}\alpha_{-n}\talpha_{-n}\right)
|l-\mbox{$R\mDa\over 2\pi$},0\rangle.
\eeqa
Note that these states change by a phase under $\mDx\to\mDx+2\pi R$
and $\mDa\to\mDa+{2\pi\over R}$
because of the parallel transport involved in the action of
$g_{{}_{\mDx}}$ and $\tilg_{{}_{\mDa}}$.

The crosscap state for $g_{{}_{\mDx}}\Omega$ consists of states in
${\mathcal H}_{g_{{}_{2\mDx}}}$ and is given by
\beq
|\Scr{C}_{g_{{}_{\mDx}}\Omega}\rangle
=\sqrt{R\sqrt{2}}\sum_{m'\in\Z}
\exp\left(-\sum_{n=1}^{\infty}
{(-1)^n\over n}\alpha_{-n}\talpha_{-n}\right)|0,2m'-
\mbox{${\mDx\over \pi R}$}
\rangle.
\label{CgdX}
\eeq
The crosscap state for $\tilg_{{}_{\mDa}}I\Omega$
is a sum of states in ${\mathcal H}_{\tilg_{{}_{2\mDa}}}$:
\beq
|\Scr{C}_{\tilg_{{}_{\mDa}}I\Omega}\rangle
=\sqrt{\sqrt{2}\over R}\sum_{l'\in\Z}
\exp\left(\sum_{n=1}^{\infty}
{(-1)^n\over n}\alpha_{-n}\talpha_{-n}\right)
|2l'-\mbox{${R\mDa\over \pi}$},0\rangle.
\label{Ctgda}
\eeq
(\ref{CgdX}) interpolates between the ${\Omega}$ crosscap
and the $s\Omega$ crosscap. Recall that
$g_{\pi R}\Omega$ and $s\Omega$ are the same except for the difference
in the action on N--N string states by
an overall phase.
The same can be said of (\ref{Ctgda}).

One can also consider the parities
$g_{{}_{\mDx}}\tilg_{{}_{\mDa}}\Omega$
or $g_{{}_{\mDx}}\tilg_{{}_{\mDa}}I\Omega$.
The crosscap states are
given by
\beqa
&&|\Scr{C}_{g_{{}_{\mDx}}\tilg_{{}_{\mDa}}\Omega}\rangle
=\sqrt{R\sqrt{2}}\sum_{m'\in\Z}
\e^{-iR\mDa m'}
\exp\left(-\sum_{n=1}^{\infty}
{(-1)^n\over n}\alpha_{-n}\talpha_{-n}\right)|0,2m'-
\mbox{${\mDx\over \pi R}$}
\rangle,\qquad
\label{CxaO}\\
&&|\Scr{C}_{g_{{}_{\mDx}}\tilg_{{}_{\mDa}}I\Omega}\rangle
=\sqrt{\sqrt{2}\over R}\sum_{l'\in\Z}
\e^{-i{\mDx\over R}l'}
\exp\left(\sum_{n=1}^{\infty}
{(-1)^n\over n}\alpha_{-n}\talpha_{-n}\right)
|2l'-\mbox{${R\mDa\over \pi}$},0\rangle.
\label{CxaIO}
\eeqa
The crosscap for $g_{{}_{\mDx}}\tilg_{{}_{\mDa}}\Omega$
is obtained by applying
$\tilg_{{}_{\mDa\over 2}}$ to
$|\Scr{C}_{g_{{}_{\mDx}}\Omega}\rangle$
and multiplying by the phase $\e^{-i{\mDa\mDx\over 2\pi}}$.
The application of
$\tilg_{{}_{\mDa\over 2}}$ is because
$g_{{}_{\mDx}}\tilg_{{}_{\mDa}}\Omega
=\tilg_{{}_{\mDa\over 2}}g_{{}_{\mDx}}\Omega
\tilg_{{}_{\mDa\over 2}}^{-1}$.
The extra phase
arises because 
$\tilg_{{}_{\mDa}}|N_a\rangle_{g_{{}_{\mDx}}}$
is not just $|N_{a+\mDa}\rangle_{{g_{{}_{\mDx}}}}$
but has an  extra phase
$\e^{i{\mDa\mDx\over 2\pi}}\,\,$
\footnote{In fact,
${}_{g_{{}_{2\mDx}}}\langle N_a|q^{H_c}
|\Scr{C}_{g_{{}_{\mDx}}\tilg_{{}_{\mDa}}\Omega}\rangle$
should be the same as
${}_{g_{{}_{2\mDx}}}\langle N_{a-{\mDa\over 2}}|q^{H_c}
|\Scr{C}_{g_{{}_{\mDx}}\Omega}\rangle$.
The claimed construction
of $|\Scr{C}_{g_{{}_{\mDx}}\tilg_{{}_{\mDa}}\Omega}\rangle$
follows because
${}_{g_{{}_{2\mDx}}}\langle N_{a-{\mDa\over 2}}|
=\e^{-i{\mDa\mDx\over 2\pi}}
{}_{g_{{}_{2\mDx}}}\langle N_{a}|\tilg_{{}_{\mDa\over 2}}$.}.
The same can be said of
$|\Scr{C}_{g_{{}_{\mDx}}\tilg_{{}_{\mDa}}I\Omega}\rangle$.

\section{Rational $U(1)$}\label{sec:ratR}

\newcommand{\wtn}{\widetilde{n}}

Let us consider the case $R=\sqrt{k}$ for a positive integer $k$.
We can now use two approaches to construct D-branes and orientifolds:
on the one hand, we can insert the special value of $R$ in the formulae
worked out in the previous section.
On the other hand, the boson at this
particular radius is described by a rational conformal theory, so
that  the methods developed in Section \ref{general} can be applied. 
Needless to say,
the two approaches lead to the same results.

Let us briefly review the basic structure of the rational conformal field
theory description, and in particular collect the ingredients for the
construction of Section \ref{general}. 
At the radius $R=\sqrt{k}$,
the system has two copies of chiral algebra ${\mathcal A}=U(1)_k$.
One copy is generated by
the spin 1 and spin $k$ currents 
$$J=\sqrt{k}(\partial_t-\partial_{\s})X\,\,\,\,
{\rm and}\,\,\,\,
W_{\pm}=\e^{\pm 2i\sqrt{k} X_R},$$
while the other copy
is generated by 
$$\widetilde{J}=-\sqrt{k}(\partial_t+\partial_{\s})X\,\,\,\,
{\rm and}\,\,\,\, \widetilde{W}_{\pm}=\e^{\mp 2i\sqrt{k}X_L}.$$
Since $W_{\pm}$ has $J$-charge $\pm 2k$, the representation of $U(1)_k$
is labelled by a modulo $2k$ integer, $n$, and the representation space
is denoted by $\Scr{H}_n$.
Note that the state $|l,m\rangle$ of momentum $l$ and winding number $m$
has $(J,\widetilde{J})$-charge $(l-km,-l-km)$.
Thus, one may relabel the states as
$$
|l,m\rangle=|l-km\rangle\otimes |-l-km\rangle
=|n+2kp\rangle\otimes |-n-2k\tilde{p}\rangle,
$$
where we have made the
reparametrization
$l=n+k(p+\tilde{p})$ and $m=-p+\tilde{p}$.
If $l$ and $m$ run over integers,
then $(n,p,\tilde{p})$ runs over $\Z_{2k}\oplus\Z\oplus\Z$.
Thus, the space of states is given by
$$
{\mathcal H}=\bigoplus_{l,m}
{\mathcal H}_{l,m}
=\bigoplus_{n\in\Z_{2k}}\Scr{H}_n\otimes\Scr{H}_{-n}.
$$
The primary states are labelled by a mod $2k$ integer, $n\in\Z_{2k}$,
and the fusion rules are simply given by the addition modulo $2k$.
We choose the range $n=-k+1,\dots, k$
as a fundamental domain for $\Z_{2k}$.
For any integer $n$, we denote by $\hat{n}$ the representative
in this fundamental domain.
We also denote the addition of
two labels $n,m$ mod $2k$ by $\hat{+}$, such that
$n\hat{+}m \in \{-k+1,\dots, k\}$.
The conformal weight of the primary field with label $n$ is
$$
h_n=\frac{\hat{n}^2}{4k}.
$$

\subsubsection*{\it Modular Matrices}

The modular $T$ and $S$ matrices are
\beq
T_{nn'} = \delta_{n,n'}\e^{\frac{\pi i n^2}{2k}- \frac{i\pi}{12}},\qquad
S_{nn'} = \frac{1}{\sqrt{2k}} e^{-\frac{i\pi nn'}{k}}.
\eeq
For orientifold constructions one needs in addition the
$P=\sqrt{T}ST^2S\sqrt{T}$ and
$Y$ matrices.
In order to compute them, it is convenient to introduce related matrices
that do not involve $\sqrt{T}$:
\beq
Q=ST^2S
\label{defSTS}
\eeq
and
\beq
\tilde{Y}_{ab}^c = \sum_d \frac{S_{ab} Q_{bd}Q^*_{cd}}{S_{0d}}
=\sqrt{\frac{T_c}{T_b}} \ Y_{ab}^c.
\label{defYtilde}
\eeq
The absence of $\sqrt{T}$ makes the computation easier, and
we find
\beqa
&&Q_{nn'} =  \frac{1}{\sqrt{k}} \e^{\pi i\over 12}
\e^{-\frac{i\pi}{4k}(n+n')^2} \ \delta^{(2)}_{n+n'+k},
\nn\\
&&\tilde{Y}_{nn'}^{n''} = \e^{-\frac{\pi i}{4k}(-{n''}^2+{n'}^2)} \ 
\delta^{(2)}_{n'+n''} \
\left( \delta^{(2k)}_{n+\frac{n'-n''}{2}} + (-1)^{n'+k} \
\delta^{(2k)}_{n+\frac{n'-n''}{2}+k} \right).
\nn
\eeqa
The $P$-matrix and $Y$-tensor are then found to be
\beqa
\label{u1p}
&&P_{nn'}=\delta_{k+n+n'}^{(2)}{1\over \sqrt{k}}
\e^{-\pi i {\hat{n}\hat{n}'\over 2k}},\\
&&Y_{nn'}^{n''} = 
\delta^{(2)}_{n'+n''} \
\left( \delta^{(2k)}_{n+\frac{\hat{n}'-\hat{n}''}{2}} + (-1)^{n'+k} \
\delta^{(2k)}_{n+\frac{\hat{n}'-\hat{n}''}{2}+k} \right).
\label{Yt}
\eeqa

\subsubsection*{\it Discrete Symmetry}

The group of simple currents is the group of primaries itself,
$\Scr{G}=\Z_{2k}$. The charge $Q_n(n')$ is given by
$$
Q_n(n')={n^2\over 4k}+{(n')^2\over 4k}-{(n+n')^2\over 4k}
=-{nn'\over 2k}\qquad\mbox{mod 1}.
$$
Thus, we find a discrete symmetry group $\Z_{2k}$ generated by an element
$g$ that acts on $\Scr{H}_n\otimes\Scr{H}_{-n}$ by phase multiplication
$\e^{-\pi i {n\over k}}\times$.
In terms of the symmetries $g_{\mDx}$ and $\widetilde{g}_{\mDa}$,
this generator can be expressed as
\beq
g=g_{\pi R\over k}\widetilde{g}_{\pi\over R},
\eeq
where $R=\sqrt{k}$ is understood.
One can also show that among $g_{\mDx}\widetilde{g}_{\mDa}$
the symmetries that commute with the algebra
$U(1)_k\otimes U(1)_k$ are of the form
$g_{\pi Rn\over k}\widetilde{g}_{\pi n\over R}=g^n$ for some $n\in\Z_{2k}$.

\subsubsection*{\it T-duality}

The T-dual model has radius $1/R=1/\sqrt{k}=R/k$, and can actually be
regarded as the orbifold by the group
$G=\Z_k$ generated by $g_{2\pi R/k}=g^2$.
As a representation of the
$U(1)_k\otimes U(1)_k$ algebra, the space of states
is given by the diagonal modular invariant
$$
{\mathcal H}^{\rm T-dual}
=\bigoplus_n\Scr{H}_n\otimes\Scr{H}_n.
$$
T-duality induces the mirror automorphism $M:\alpha_n\to -\alpha_n$, and
the map of states $\Psi:{\mathcal H}\to
{\mathcal H}^{\rm T-dual}$ is given by $|l,m\rangle_R\mapsto
|m,l\rangle_{1\over R}$, which reads in the RCFT language as
$$
\Psi=V_M\otimes 1:|q\rangle\otimes |\widetilde{q}\rangle
\mapsto |-q\rangle\otimes |\widetilde{q}\rangle.
$$

\subsection{A-parities}

A-branes and A-parities correspond to 
the Cardy states and the PSS crosscaps
\beqa
&&
|\Scr{B}_n\rangle
={1\over (2k)^{1\over 4}}
\sum_{n'\in\Z_{2k}}\e^{-\pi i{nn'\over k}}\kket{n'},\\
&&
|\Scr{C}_n\rangle
={(2k)^{1\over 4}\over\sqrt{k}}\sum_{n'\in\Z_{2k}}
\e^{-\pi i{\hat{n}\hat{n}'\over 2k}}\delta^{(2)}_{n+n'+k}\cket{n'}.
\eeqa

To find out the geometrical meaning of these branes and parities,
we express these states in
terms of the basis $|l,m\rangle$ labelled by momentum
and winding number.
We first re-express Ishibashi states:
\beqa
&&\kket{n}=
\sum_{p\in\Z}\e^{\sum{1\over m}\alpha_{-m}\talpha_{-m}}
|n+2kp\rangle\otimes |-n-2kp\rangle
=\sum_{p\in\Z}\e^{\sum{1\over m}\alpha_{-m}\talpha_{-m}}
|n+2kp,0\rangle,\quad
\nn\\
&&\cket{n}=\e^{\pi i (L_0-h_n)}\kket{n}
=\e^{-\pi i h_n}\sum_{p\in\Z}
\e^{\sum{(-1)^m\over m}\alpha_{-m}\talpha_{-m}}\e^{\pi i {(n+2kp)^2\over 4k}}
|n+2kp,0\rangle.
\nn
\eeqa
Then, the Cardy states are expressed as
\beqa
|\Scr{B}_n\rangle
&=&
{1\over (2k)^{1\over 4}}
\sum_{n'\in\Z_{2k}}\e^{-\pi i{nn'\over k}}\sum_{p\in\Z}
\e^{\sum{1\over m}\alpha_{-m}\talpha_{-m}}|n'+2kp,0\rangle
\nn\\
&=&{1\over (2k)^{1\over 4}}
\sum_{l\in\Z}\e^{-\pi i {nl\over k}}\e^{\sum{1\over m}\alpha_{-m}\talpha_{-m}}
|l,0\rangle
\nn\\
&=&|D_{\pi R{n\over k}}\rangle.
\eeqa
Thus, the $n$-th Cardy state is identified as
the D0-brane located at the point $X=2\pi R{n\over 2k}$
of the circle.
To express the PSS crosscaps, it is convenient to use the
$Q$-matrices, 
\beqa\label{PSSU1}
|\Scr{C}_n\rangle
&=&
\e^{\pi i (h_n-{1\over 12})}\sum_{n'}{Q_{nn'}\over\sqrt{S_{0n'}}}
\e^{\pi i L_0}\kket{n'}
\nn\\
&=&
\e^{\pi i (h_n-{n^2\over 4k})}
\left({2\over k}\right)^{1\over 4}
\sum_{l''\in\Z}\e^{-\pi i {n\over k}(l''+{n+k\over 2})}
\e^{\sum{(-1)^m\over m}\alpha_{-m}\talpha_{-m}}
|2l''+n+k,0\rangle
\nn\\
&=&
\left\{
\begin{array}{ll}
|\Scr{C}_{g_{\pi Rn\over k}I\Omega}\rangle&\mbox{$(n+k)$ even}
\\
|\Scr{C}_{\widetilde{g_{\pi R\hat{n}\over k}I'\Omega}}\rangle
&\mbox{$(n+k)$ odd}
\end{array}\right.
\eeqa
In the last step, we used the crosscap formulae
(\ref{CrdI}) and (\ref{CrdIp}) for the parity symmetries
studied
in Section~\ref{sec:arbR}.
We see that the PSS parities are associated with reflections of the circle.
The $n$-th PSS parity has orientifold fixed points
at the diametrically opposite points,
$X=2\pi R\frac{n}{4k}$ and  $X=2\pi R\frac{n}{4k}+\pi R$.
For $n$ even, the location of the orientifold points 
coincides with the location of the Cardy branes
$\Scr{B}_{n/2}$ and $\Scr{B}_{n/2+k}$. 
For $n$ odd, the fixed points are halfway
between possible locations of Cardy branes. 
Furthermore, the two
orientifold points are both of the same type for $n+k$ even,
whereas they are of the opposite type for $n+k$ odd. 

We will now see how this information is encoded in the $Y$ tensor of the RCFT.
Using (\ref{Yt}) it is straightforward to write down the 
A-type M\"obius strips:
\beqa
\label{MAU1}
\langle\Scr{C}_n|q_t^{H}|\Scr{B}_m\rangle
&=&\sum_{n'} Y_{mn'}^n \widehat\chi_{n'}(\tau)
=\pm\widehat{\chi}_{n-2m}(\tau);\nn\\
\pm &=&\left\{\begin{array}{ll}
1&\mbox{if $\widehat{n-2m}=\hat{n}-2m$ mod $4k$}\\
(-1)^{n+k}&
\mbox{if $\widehat{n-2m}=\hat{n}-2m+2k$ mod $4k$}.
\end{array}
\right.
\eeqa
Since $\langle \Scr{B}_{m'}|q_t^H|\Scr{B}_{m}\rangle
=\sum_{n'}N_{n'm}^{m'}\chi_{n'}(\tau)=\chi_{m'-m}(\tau)$,
we see that the $P_n$-image of the Cardy brane
$\Scr{B}_{m}$ is $\Scr{B}_{n-m}$.
(Actually, this also follows from the general rule
(\ref{Pgi}).)
In particular, for $n$ even, the Cardy branes $\Scr{B}_{n/2}$ 
and $\Scr{B}_{n/2+k}$ are
left invariant, confirming that these branes are located at the orientifold
fixed points.
The two cases in
(\ref{MAU1}) are interchanged under the shift
$m\to m+k$.
In particular,
the amplitude
flips its sign under the exchange $\Scr{B}_m\leftrightarrow \Scr{B}_{m+k}$
if and only if $n+k$ is odd.
We note that the Cardy branes $\Scr{B}_m$ and $\Scr{B}_{m+k}$
are located at diametrically opposite points.
In this way, the RCFT data encode the fact that the crosscaps
with $n+k$ odd lead to orientifold projections of different types
at the two fixed points, whereas crosscaps for $n+k$ even give rise
to the same projection.

For completeness, let us also write the Klein bottles, which are
\beq
\langle\Scr{C}_n\ket{\Scr{C}_l}
= \sum_m \delta^{(2)}_{n+l} \
\left( \delta^{(2k)}_{m+\frac{n-l}{2}} + (-1)^{k+n} \ 
\delta^{(2k)}_{m+\frac{n-l}{2} +k}
\right) \chi_m
\eeq

\subsection{B-parities}

We next study B-parities.
To find B-crosscaps in our model, we first find A-crosscaps
in the mirror
$\Z_k$-orbifold model, and then bring them back by the mirror map.

To find the A-crosscaps in the orbifold model, we apply
the method of Section~\ref{subsec:orb}.
The bilinear form $q$ of the group $G=\Z_k$
is uniquely fixed by the requirement
$q(n,n)=-h_n=-{n^2\over 4k}$ (mod 1)
and is given by
\beq
q(n,m)=-{nm\over 4k}, \quad
n,m\,\,\mbox{even}.
\eeq
Note that it is well-defined, namely
invariant (mod 1) under $2k$ shifts of $n$ and $m$ 
since both of them are even.
Note also that
$2q(n,m)=-nm/2k=Q_n(m)$ (mod 1) as required.
To write down the Eq.~(\ref{thetaEq}) for $\theta$,
we first note that
\beqa
-\hat{Q}_n(m)+2q(n,m)
&=&-{\hat{n}^2\over 4k}-{\hat{m}^2\over 4k}+{(n\hat{+}m)^2\over 4k}
-{nm\over 2k}
\nn\\
&=&{(n\hat{+}m)^2\over 4k}-{(\hat{n}+\hat{m})^2\over 4k}
={n\hat{+}m\over 2}-{\hat{n}+\hat{m}\over 2}\quad\,\, \mbox{mod 2}.
\,\,\,\,\,\,\,
\label{Qcomp}
\eeqa
Thus, the equation is $\theta(n+m)=\theta(n)+\theta(m)+{n\hat{+}m\over 2}
-{\hat{n}+\hat{m}\over 2}$ and the solutions are
\beq
\theta_l(n)
={\hat{n}\over 2}+{nl\over k},\qquad
l\in \Z/k\Z.
\eeq
Then, the crosscap states (\ref{orbcross1}) and (\ref{orbcross2})
are given by
\beqa
&&|\Scr{C}_{P^{\theta_l}_{0}}\rangle
={\e^{i\omega_0}\over \sqrt{k}}\sum_{n:{\rm even}}
\e^{-\pi i ({\hat{n}\over 2}+{nl\over k})}
|\Scr{C}_n\rangle
=\e^{i\omega_0}(2k)^{1\over 4}\cket{-2l-k}.
\nn\\
&&|\Scr{C}_{P^{\theta_l}_{1}}\rangle
={\e^{i\omega_1}\over \sqrt{k}}\sum_{n:{\rm even}}
\e^{-\pi i ({\hat{n}\over 2}+{nl\over k}-Q_1(n))}
|\Scr{C}_{n+1}\rangle
=\e^{i\omega_1+\pi i{\widehat{2l+1+k}\over 2k}}(2k)^{1\over 4}\cket{-2l-1-k}.
\nn
\eeqa
In the latter we have chosen $n'=1$ as the representative of the
non-trivial element of $\Scr{G}/G=\Z_{2k}/\Z_k=\Z_2$.
The phases $\e^{i\omega_0}$ and $\e^{i\omega_1}$ can be tuned
so that no phases appear in front of the Ishibashi states.

B-parities in the original model are obtained from these by the mirror map.
We denote the mirror images of
$P_0^{\theta_l}$ and $P_1^{\theta_l}$
by $P_B^{2l+k}$ and $P_B^{2l+1+k}$ respectively.
Since the mirror map
$\Psi^{-1}$ sends the Ishibashi states $\cket{-n}$ to
B-type Ishibashi states $\cket{n}_B$,
we find that the crosscap states are given by
\beq
|\Scr{C}_{P_B^n}\rangle
=(2k)^{1\over 4}\cket{n}_B.
\label{CPB}
\eeq
These parities are not necessarily involutive.
Applying  formula (\ref{sqformula})
we find that
\beq
(P_B^{n})^2=g^{2n}.
\label{PBsqu}
\eeq
The crosscap state (\ref{CPB}) belongs to
the space $\Scr{H}_{n}\otimes\Scr{H}_{n}$.
Since $n=2n+(-n)=g^{2n}(\bar n)$,
$\Scr{H}_{n}\otimes\Scr{H}_{n}
=\Scr{H}_{n}\otimes\Scr{H}_{g^{2n}(\bar n)}$
is a space of states with $g^{2n}$-twisted boundary condition.
Namely, $|\Scr{C}_{P_B^{n}}\rangle$ is a $g^{2n}$-twisted
state, which is consistent with (\ref{PBsqu}).

Next, let us  examine the geometrical interpretation of these
parity symmetries.
To this end, we express the Ishibashi states in terms of the
$|l,m\rangle$ basis.
\beqa
&&\kket{n}_B=(V_M\otimes 1)\kket{-n}
=(V_M\otimes 1)\sum_{p\in\Z}\e^{\sum{1\over m}\alpha_{-m}\talpha_{-m}}
|-n+2kp\rangle\otimes |n-2kp\rangle
\nn\\
&&\qquad\qquad
=\sum_{p\in\Z}\e^{-\sum{1\over m}\alpha_{-m}\talpha_{-m}}
|n-2kp\rangle\otimes |n-2kp\rangle
=\sum_{p\in \Z}\e^{-\sum{1\over m}\alpha_{-m}\talpha_{-m}}
|0,2p-\mbox{$n\over k$}\rangle,
\nn\\
&&\cket{n}_B
=\e^{\pi i (L_0-h_n)}\kket{n}_B
=\e^{\pi i({n^2\over 4k}-h_n)}
\sum_{p\in\Z}
\e^{-\pi i(n+k)p}
\e^{-\sum{(-1)^m\over m}\alpha_{-m}\talpha_{-m}}
|0,2p-\mbox{$n\over k$}\rangle.
\nn
\eeqa
Comparing the latter with the formula (\ref{CxaO}), we find that
\beq
|\Scr{C}_{\tilg_{{\pi(n+k)\over R}}g_{\pi Rn\over k}\Omega}\rangle
=\e^{-\pi i({n^2\over 4k}-h_n)}
(2k)^{1\over 4}\cket{n}_B.
\label{Ccc}
\eeq
The crosscap state (\ref{CPB}) is equal to
(\ref{Ccc}), up to
an overall sign, which is $+$ if we choose $n=\hat{n}$.
Thus, we conclude that the RCFT parities are interpreted as
\beq
P_B^{n}
=g_{\pi R\hat{n}\over k}\tilde{g}_{\pi(\hat{n}+k)\over R}\Omega.
\eeq
If $n+k$ is even, $P_B^n$ is equal to $g_{\pi R\hat{n}\over k}\Omega$,
which is simply the worldsheet orientation reversal
$\Omega$ followed by the $n/2k$ rotation of the circle.
If $n+k$ is odd, $\widetilde{g}_{\pi (n+k)\over R}$
is non-trivial, and $P_B$ is not just
$\Omega$ followed by the $n/2k$ rotation, but it acts by extra sign
multiplication on odd-winding states.
Note that 
$g_{\pi Rn\over k}\tilde{g}_{\pi(n+k)\over R}
=g^n\tilde{g}_{\pi k\over R}$ and hence
$P_B^n=g^{\hat{n}}\tilde{g}_{\pi k\over R}\Omega$.
Since $(\tilde{g}_{\pi k\over R}\Omega)^2=1$,
this also explains (\ref{PBsqu}).

\subsubsection{Klein Bottles}

We record here the Klein bottle amplitudes:
\beqa
\Tr\!\mathop{}_{{\mathcal H}}\!
P_B^n q^H&=&
\langle\Scr{C}_{P_B^n}|\e^{-{\pi i\over 2\tau}H}|\Scr{C}_{P_B^n}\rangle
=\sqrt{2k}\chi_n(-1/2\tau)
\nn\\
&=&\sum_m\e^{-\pi i {mn\over k}}\chi_m(2\tau)
=\sum_m\e^{2\pi i Q_n(m)}\chi_m(2\tau).
\eeqa
This indeed shows that $P_B^{n}=g^nP_B^0$ on the closed string states.
One could also consider
$\langle\Scr{C}_{P_B^{n+k}}|q_t^H|\Scr{C}_{P_B^{n}}\rangle$,
which is interpreted as
$\Tr_{{\mathcal H}_{g^k}}P_B^nq^H$. This vanishes.

\subsubsection{M\"obius Strips}

Let us compute the M\"obius strip amplitudes.
Since $P_B^n$ are not in general involutive (\ref{PBsqu}),
we need to find the boundary states on a circle with $g^{2n}$-twisted 
boundary condition.
They are obtained via mirror symmetry from
the twisted boundary states for the A-branes in the orbifold model
${\mathcal C}/G$.

To find the A-brane boundary states in the orbifold model, we use the
method developed in Section~\ref{subsubsec:Borb}.
The symmetry $g^{2n}$ in the original model is mapped to the quantum symmetry
in the orbifold model associated with the character
$\rho_n$ of $G$ defined by $\rho_n(m)=mn/2k$ ($m$ even).
Applying  formula (\ref{bir}) with $i=n'=0,1$, we find
\beq
|\Scr{B}_{[n']}\rangle^{{\mathcal C}/G}_{\rho_n}
=\e^{i\lambda+{\pi inn'\over k}}\left({k\over 2}\right)^{1\over 4}
(\kket{-n}+(-1)^{n'}\kket{-n-k}).
\eeq
We choose the phase $\lambda$ so that $\e^{i\lambda+{\pi inn'\over k}}=1$.
The (twisted) boundary states
for B-branes in the original model
are obtained by applying the mirror map
$\Psi^{-1}$ to these states:
\beq
|\Scr{B}^B_{[n']}\rangle_{g^{2n}}
=
\left({k\over 2}\right)^{1\over 4}
(\kket{n}_B+(-1)^{n'}\kket{n+k}_B),\qquad
n'=0,1.
\eeq

To find the geometrical meaning, we express them
in terms of the $|l,m\rangle$ basis.
Using the expression for $\kket{n}_B$ obtained above, we find
\beqa
&&|\Scr{B}^B_{[0]}\rangle_{g^{2n}}
=\sqrt{\sqrt{k}\over\sqrt{2}}
\sum_{r\in \Z}\e^{-\sum{1\over m}\alpha_{-m}\talpha_{-m}}
|0,r-\mbox{${n\over k}$}\rangle
=|N_0\rangle_{g_{2\pi Rn\over k}},\\
&&|\Scr{B}^B_{[1]}\rangle_{g^{2n}}
=\sqrt{\sqrt{k}\over\sqrt{2}}\sum_{r\in \Z}(-1)^r
\e^{-\sum{1\over m}\alpha_{-m}\talpha_{-m}}
|0,r-\mbox{${n\over k}$}\rangle
=|N_{\pi\over R}\rangle_{g_{2\pi Rn\over k}}.
\eeqa
Thus, $\Scr{B}^B_{[0]}$ is the D1-brane wrapped on $S^1$ with trivial
Wilson line, while $\Scr{B}^B_{[1]}$ is the
D1-brane with Wilson line $\pi$ along the $S^1$.
Using the action of $\Omega$, $g_{\mDx}$ and $\widetilde{g}_{\mDa}$
on the D-branes studied in Section~\ref{sec:arbR},
we see that
$P_B^n=g_{\pi R\hat{n}\over k}\tilde{g}_{\pi(\hat{n}+k)\over R}\Omega$
maps the brane $N_a$ to $N_{-a+{\pi (\hat{n}+k)\over R}}$. In particular,
the B-branes $\Scr{B}^B_{[0]}$ and $\Scr{B}^B_{[1]}$ are invariant
under $P_B^n$ with even $n+k$,
while they are exchanged with each other under
$P_B^n$ with odd $n+k$.

Let us see how the RCFT data encode this information.
It is straightforward to compute the open string partition function
$\Tr_{ab}(g^{2n}q^H)
={}_{g^{2n}}\langle \Scr{B}_{[a]}^B|q_t^H|\Scr{B}_{[b]}^B\rangle_{g^{2n}}$:
\beqa
&&
\Tr_{00}(g^{2n}q^H)
=\Tr_{11}(g^{2n}q^H)
=\sum_{m:\,{\rm even}}\e^{-\pi i {mn\over k}}\chi_m(\tau)
\nn\\
&&
\Tr_{01}(g^{2n}q^H)
=\Tr_{10}(g^{2n}q^H)
=\sum_{m:\,{\rm odd}}\e^{-\pi i {mn\over k}}\chi_m(\tau).
\nn
\eeqa
This shows that 0--0 and 1--1 string states have even charges under $U(1)_k$
and 1-0 and 0-1 string states have odd charges.
(Also, $g^{2n}$ acts on the charge $m$ representation as
the phase multiplication by $\e^{-\pi i {nm\over k}}$.)
On the other hand, the M\"obius strip amplitudes are
\beq
{}_{g^{2n}}\langle \Scr{B}_{[a]}^B|q_t^H|\Scr{C}_{P_B^n}\rangle
=\left\{\begin{array}{ll}
{\displaystyle \sum_{m:\,{\rm even}}\e^{-\pi i {\hat{n}\hat{m}\over 2k}}
\widehat{\chi}_m(\tau)}&
\mbox{$n+k$ even},\\[0.5cm]
{\displaystyle \sum_{m:\,{\rm odd}}\e^{-\pi i {\hat{n}\hat{m}\over 2k}}
\widehat{\chi}_m(\tau)}&
\mbox{$n+k$ odd},
\end{array}
\right.
\eeq
for both $a=0,1$.
They are to be identified with the twisted open string partition functions
$\Tr_{[a],P_B^n[a]}(P_B^nq^H)$.
For $n+k$ even, only even charge states propagate,
which means that the brane $\Scr{B}_{[a]}^B$ is preserved under
the parity $P_B^n$.
For odd $n+k$, only odd charge states propagate,
which means that $\Scr{B}^B_{[0]}$ is exchanged with $\Scr{B}^B_{[1]}$.

\section{Parity Symmetry of (gauged) WZW Models}\label{sec:gWZW}

In this section, we study Parity symmetry of
the WZW model on a group manifold $G$,
and the $G$ mod $H$ gauged WZW model in which 
the vectorial rotation $g\mapsto h^{-1}gh$
is gauged.
We will focus on the case $G=SU(2)$, the group of $2\times 2$
unitary matrices of determinant 1, and its
diagonal subgroup $H=U(1)$.
We denote the Lie algebras of these groups as
$\fg$ and $\fh$.

\subsection{The Models}\label{subsec:Model}

Let $\Sigma$ be a $1+1$ dimensional worldsheet.
The level $k$ WZW action \cite{Wnab}
for a map $g:\Sigma\to SU(2)$
 in a background gauge field $A\in \Omega^1(\Sigma,\fg)$ is given by
\beqa
S_k(A,g)&=&{k\over 8\pi}\int_{\Sigma}
\tr\left(g^{-1}D^{\mu}g\,g^{-1}D_{\mu}g\right)\dd^2x
\nn\\
&&+{k\over 12\pi}\int_B\tr\left(\tilde{g}^{-1}\dd \tilde{g}\right)^3
-{k\over 4\pi}\int_{\Sigma}\tr\left\{A(g^{-1}\dd g+\dd g g^{-1})
+Ag^{-1}Ag\right\},
\label{WZWaction}
\eeqa
where $B$ is a 3-manifold bounded by the worldsheet,
$\partial B=\Sigma$, $\tilde{g}$ is an extension to it,
and
$g^{-1}D_{\mu}g:=g^{-1}\partial_{\mu}g+g^{-1}A_{\mu}g-A_{\mu}$.
Let us consider the {\it chiral}
gauge transformation
\beq
\begin{array}{l}
g\to h_1^{-1}g h_2,\\
A_+\to h_1^{-1}A_+h_1+h_1^{-1}\partial_+h_1,\\
A_-\to h_2^{-1}A_-h_2+h_2^{-1}\partial_-h_2,\\
\end{array}
\eeq
for any $h_i$ with values in $G$ or its complexification $G_c$. 
(Here we used the light-cone coordinate $x^{\pm}=t\pm \s$, with
$\partial_{\pm}={\partial\over \partial x^{\pm}}
={1\over 2}(\partial_t\pm\partial_{\s})$.)
This changes the action according to
the Polyakov--Wiegmann (PW) identity \cite{PW,Gaw}:
\beq
S_k(A,g)\to S_k(A^{h_1,h_2},h_1^{-1}gh_2)=S_k(A,g)-S_k(A,h_1h_2^{-1}).
\label{PWid}
\eeq
In particular, it is invariant under
the vectorial transformations
$g\to h^{-1}gh,
A\to h^{-1}Ah+h^{-1}\dd h$.
The action $S_k(A,g)$ can also be defined
when $A$ is a connection of a topologically
non-trivial $G/Z_G$-bundle on $\Sigma$ (where $Z_G$ is the center
of $G$) and $g$ is a section of the
associated adjoint bundle, so that the PW-identity still holds \cite{FIH}.

The level $k$ WZW model is the theory of the variable $g$
with the action $S_k(g)=S_k(0,g)$.
As a consequence of the PW identity (\ref{PWid}),
the action $S_k(g)$ is invariant under
\beq
g\to h_1(x^-)^{-1}gh_2(x^+).
\eeq
The corresponding currents
($X\in\fg$),
\beqa
&&J_n(X)={-k\over 2\pi i}\int_0^{2\pi}
\tr\left(\partial_- g g^{-1}X\right)\e^{in(t-\s)}\dd\s,
\label{Jint}\\
&&\wtJ_n(X)={k\over 2\pi i}\int_0^{2\pi}
\tr\left(g^{-1}\partial_+ g X\right)
\e^{in(t+\s)}\dd\s,
\label{wtJint}
\eeqa
obey the $SU(2)\times SU(2)$ current algebra relations at level $k$.
The Hilbert space of states decomposes into the irreducible representations
of this algebra $\whfg\oplus\whfg$.
Only the integrable representations
$\V_j=\V_j^{G,k}$ appear, where the spin $j$ ranges
over the set ${\rm P}_k=\{0,{1\over 2},1,...,{k\over 2}\}$:
\beq
{\cal H}^{G,k}=\bigoplus_{j\in {\rm P}_k}
\V_{j}\otimes \wtV_{j}.
\label{HilbG}
\eeq
The system is a conformal field theory with
$c={3k\over k+2}$.
The spin $j\in {\rm P}_k$ representation $V_j$ of $SU(2)$ is included in $\V_j$
as the space of Virasoro primary states
with $h_j={j(j+1)\over k+2}$, and the matrix elements
of $g$ in $V_j$ are the Virasoro primary fields corresponding to
the states in
$V_j\otimes V_j\subset\V_{j}\otimes \wtV_{j}$.
In particular, for spin ${1\over 2}$ representation,
we have the relation
\beq
g=\left(\begin{array}{cc}
g_{11}&g_{12}\\
g_{21}&g_{22}
\end{array}
\right)
\leftrightarrow
\left(\begin{array}{cc}
-|{1\over 2},-\half\rangle\otimes |{1\over 2},\half\rangle &
-|{1\over 2},-\half\rangle\otimes |{1\over 2},-\half\rangle\\
|{1\over 2},\half\rangle\otimes |{1\over 2},\half\rangle&
|{1\over 2},\half\rangle\otimes |{1\over 2},-\half\rangle
\end{array}
\right),
\label{FS}
\eeq
where $|j,m\rangle$ is the basis of $V_j$ with
$\sigma_3/2=m$.
The minus signs in (\ref{FS})
originate in the relation $v^a=\epsilon^{ab}v_{b}$
defining the isomorphism
$V_{1\over 2}\cong V_{1\over 2}^{\vee}$.
The relation for the higher spin representations can be obtained from this
by using the realization $V_j={\rm Sym}^{2j}V_{1\over 2}$.

\subsubsection*{\it Gauged WZW Models}

We next consider the $SU(2)$ mod $U(1)$ gauged WZW model.
This is the model with the action $S_k(A,g)$ where
$A$ also varies over $U(1)$ gauge fields.
To be precise, the gauge group $H\cong U(1)$ is the diagonal
subgroup of $SU(2)$ divided by the center $\Z_2=\{\pm 1\}$.
The model is a conformal field theory with the central charge
$c={3k\over k+2}-1$, which is known as the $\Z_k$ parafermion system.
Let us decompose the representation $\V_{j}$ of $\whfg$
into the irreducible representations of the subalgebra
$\whfh$ generated by $J_n(\sigma_3)$;
\beq
\V_{j}=\bigoplus_{n} \B_{j,n}\otimes \V_{-n}.
\eeq
The sum is over the eigenvalue of $-J_0(\sigma_3)$,
which are integers such that $2j+n$ is even.
The space $\B_{j,n}$ can be identified as the subspace of
$\V_j$ consisting of states obeying $J_m(\sigma_3)=0$, $m\geq 1$, and
$J_0(\sigma_3)=-n$.
The physical states of the $G$ mod $H$ model are
the gauge invariant states of the WZW model,
which satisfy
\beqa
&&
J_m(v)|{\rm phys}\rangle=\wtJ_m(v)|{\rm phys}\rangle=0,
\,\,\,m\geq 1,\,v\in \fh,
\label{Hinv1}\\
&&
\bigl(J_0(v)+\wtJ_0(v)\bigr)|{\rm phys}\rangle=0,
\,\,\,v\in \fh.
\label{Hinv2}
\eeqa
The subspace of $\V_{j}\otimes\wtV_{j}$
obeying this condition can be identified as
$\oplus_{n}\B_{j,n}\otimes
\wtB_{j,-n}$.
However, the space of states is not the whole sum of these spaces.
Since the gauge group has a non-trivial fundamental group
$\pi_1(U(1))=\Z$, there are
large gauge transformations which relate the physical states, acting on
the labels as $(j,n)\to ({k\over 2}-j,n+k)\to (j,n+2k)\to\cdots$.
The space of states is found by
selecting one member from each orbit,
\beq\label{parahilbert}
{\cal H}^{G/H,k}=\bigoplus_{(j,n)\in {\rm PF}_k}\,
\B_{j,n}\otimes
\wtB_{j,-n},
\eeq
where the sum is over 
$$
{\rm PF}_k={\{(j,n)\in {\rm P}_k\times \Z;2j+n\,{\rm even}\}\over\pi_1(U(1))}
={\{(j,n)\in {\rm P}_k\times\Z_{2k};
2j+n\,{\rm even}\}\over (j,n)\equiv ({k\over 2}-j,n+k)}.
$$

The $H$-valued chiral rotations $g\to h_1(x^-)^{-1}gh_2(x^+)$
commute with the gauge group and 
shift the action according to the PW-identity (\ref{PWid}):
\beq
S_k(A,h_1^{-1}(x^-)gh_2(x^+))
=S_k(A,g)+{k\over 2\pi}\int_{\Sigma}\tr(F_A\log(h_1h_2^{-1})),
\eeq
where $F_A=\dd A$ is the curvature of the gauge potential $A$.
For the constant $h_1,h_2$ with $h_1^{-1}=h_2=\exp(i\alpha\sigma_3/2)$,
the shift is $-k\alpha\times {i\over 2\pi}\int\tr(F_A\sigma_3)$.
Since $\tr((\sigma_3/2)\sigma_3)=1$,
the integral ${i\over 2\pi}\int\tr(F_A\sigma_3)$ is an integer
on a compact space. Thus, the path-integral weight $\e^{iS_k(A,g)}$
is invariant if $k\alpha\in 2\pi \Z$.
Therefore, the system has a symmetry generated by
\beq\label{Defa}
a:g\to \e^{\pi i\sigma_3/k}g\e^{\pi i\sigma_3/k}.
\label{genZk}
\eeq
This is an order $k$ symmetry since $(\e^{\pi i\sigma_3/k})^k=
\e^{\pi i\sigma_3}=-1$ acts trivially on $g$. 
Thus the system has an axial $\Z_k$ symmetry.
The ``axial anomaly'' $U(1)\to\Z_k$ can also be seen in the operator
formulation.
The axial rotation $g\to \e^{i\alpha\sigma_3/2}g\e^{i\alpha\sigma_3/2}$
acts on the $\B_{j,n}\otimes\wtB_{j,-n}$ subspace
as a multiplication by
$\e^{-i\alpha (-n)/2}\times\e^{i\alpha n/2}=\e^{i \alpha n}$.
However, it is consistent with the field identification
$(j,n)\equiv ({k\over 2}-j,n+k)$ only if $\alpha k\in 2\pi \Z$.
The rotation (\ref{genZk}) acts on this subspace
by multiplication by $\e^{2\pi i n/k}$, which is indeed well-defined.
The two reasonings are of course related since the field identification
originates from large gauge transformations that produce
topologically non-trivial gauge field configurations.

A geometric picture can be given to the model.
We parametrize the group element and the gauge field by
\beq
g=\e^{i(\phi+t)\sigma_3/2}\e^{i\theta \sigma_1}\e^{i(\phi-t)\sigma_3/2}
=\left(\begin{array}{cc}
\e^{i\phi}\cos\theta&i\e^{it}\sin\theta,\\
i\e^{-it}\sin\theta&\e^{-i\phi}\cos\theta
\end{array}\right),
\label{parametrization}
\eeq
and $A={i\over 2}\sigma_3a_{\mu}\dd x^{\mu}$.
The gauge transformation $h=\e^{i\lambda\sigma_3/2}$
acts on these variables as $t\to t-\lambda$,
$a_{\mu}\to a_{\mu}+\partial_{\mu}\lambda$.
Integrating out the gauge field $a_{\mu}$, we obtain the sigma model
on the space with metric
\beq
\dd s^2=k\Bigl[\,(\dd\theta)^2+\cot^2\theta(\dd\phi)^2\,\Bigr].
\label{discmetric}
\eeq
In terms of the complex coordinate $z=\e^{i\phi}\cos\theta$,
it is the disk $|z|\leq 1$ with the metric $\dd s^2=k|\dd z|^2/(1-|z|^2)$.
As discussed in \cite{MMS}, the string coupling appears to diverge
at the boundary $|z|=1$, but it is simply because of the choice of
variables.
The $\Z_k$ symmetry (\ref{genZk}) acts on the coordinates as the
shift $\phi\to \phi+2\pi/k$, or equivalently $z\to \e^{2\pi i/k}z$
--- the rotation of the disk with angle
$2\pi/k$.

\subsection{Parity Symmetry of WZW Models}
\label{subsec:PbWZW}

The WZW action $S_k(g)$
is not invariant under the simple Parity transformation
\beq
\Omega: t\to t,\,\,\,\s\to -\s,
\eeq
since the WZ term $\int_B\tr(\tilde{g}^{-1}\dd\tilde{g})^3$ flips its sign.
However, if $\Omega$ is combined with the transformation
\beq
{\cal I}: g\to g^{-1},
\eeq
it is invariant because $g^{-1}\dd g\to g\dd g^{-1}\to-g(g^{-1}\dd g)g^{-1}$
yields an extra minus sign to the WZ term.
The kinetic term is of course invariant under both $\Omega$ and ${\cal I}$.
Thus, WZW model has a Parity symmetry $P={\cal I}\Omega$.
Under this symmetry,
the currents (\ref{Jint}) and (\ref{wtJint})
transform as
\beq
\begin{array}{l}
J_n(X)\to \wtJ_n(X),\\
\wtJ_n(X)\to J_n(X).
\end{array}
\label{Pcurr}
\eeq
In particular, the right-moving highest weight state of spin $j$
is mapped to a left-moving highest weight state of spin $j$,
and vice-versa.
This shows that the Parity symmetry acts on the states
as the right--left exchange
$P:u\otimes v \mapsto \pm v\otimes u$,
up to the sign that may depend on the spin $j$ of the state.
To fix the sign, we recall the field-state correspondence
(\ref{FS}). Since $g\to g^{-1}$ sends the spin ${1\over 2}$
matrix elements as $g_{11}\leftrightarrow g_{22},
g_{12}\to -g_{12}$ and $g_{21}\to -g_{21}$, we find that the sign is $-1$ for
$j={1\over 2}$.
The sign for higher $j$ is $(-1)^{2j}$, since $V_j$ is realized as
the symmetric product of $2j$ copies of $V_{1\over 2}$.
Thus, we find that the action of $P$ is given by
\beq
P:u\otimes v\in \V_j\otimes\wtV_j\longmapsto
(-1)^{2j}v\otimes u\in \V_j\otimes\wtV_j.
\eeq
The partition function with $P$-twist is
\beq
\Tr^{}_{\cal H}(P\e^{2\pi i\tau H})
=\sum_{j\in {\rm P}_k}(-1)^{2j}\chi^{}_{j}(2\tau),
\label{Ppf}
\eeq
where $\tau$ is a positive imaginary number.

Variants of the above involution ${\cal I}$ can be considered.
One is an involution
\beq
{\cal I}^{-}:g\to -g^{-1}.
\eeq
The model is invariant under $P^{-}={\cal I}^{-}\Omega$, and
the currents transform in the same way as (\ref{Pcurr}).
Since $P^{-}$ is the composition of $P$ and multiplication by
the center $-{\bf 1}_2$, which is represented as $(-1)^{2j}$ on
$\V_j$,
it acts on the states as
\beq
P^{-}:
u\otimes v\in \V_j\otimes\wtV_j\longmapsto
v\otimes u\in \V_j\otimes\wtV_j.
\eeq
The $P^{-}$-twisted partition function is given by
\beq
\Tr^{}_{\cal H}(P^{-}\e^{2\pi i\tau H})
=\sum_{j\in{\rm P}_k} \chi^{}_{j}(2\tau).
\label{zPpf}
\eeq
More general involutions are
\beq
{\cal I}^{\pm}_{g_0}:g\to \pm g_0g^{-1}g_0,
\eeq
for any element $g_0$ of $G$.
$P_{g_0}={\cal I}_{g_0}^{\pm}\Omega$ is also a Parity symmetry of the model.
The currents transform as
\beq
\begin{array}{l}
J_n(X)\to \wtJ_n(g_0Xg_0^{-1}),\\
\wtJ_n(X)\to J_n(g_0^{-1}Xg_0).
\end{array}
\eeq
Since $P_{g_0}$ is the composition of $P^{\pm}$
and  $g\to g_0gg_0$ (where $P^+:=P$), it acts on the states as
\beq
P_{g_0}^{\pm}:u\otimes v\in \V_j\otimes\wtV_j\longmapsto
(\mp 1)^{2j} g_0^{-1}v\otimes g_0u\in
\V_j\otimes\wtV_j.
\eeq
The twisted partition function is independent of $g_0$
and reduces to (\ref{Ppf}) for $P^+_{g_0}$
and (\ref{zPpf}) for $P^-_{g_0}$.

\subsection{Parity Symmetry of Gauged WZW Models}
\label{subsec:PbgWZW}

Now, we study Parity symmetries of
the $SU(2)$ mod $U(1)$ gauged WZW model.

Under the involution $g\to g^{-1}$,
the covariant derivative is transformed
as $g^{-1}D_{\mu}g\to -g (g^{-1}D_{\mu}g)g^{-1}$ with
the gauge field $A$ fixed.
Thus, the kinetic term is invariant under
\beq
{\cal I}_A:(A,g)\to (A,g^{-1}).
\eeq
On the other hand,
the WZ term --- second line of (\ref{WZWaction}) --- flips its sign
under ${\cal I}_A$.
Thus, $P_A={\cal I}_A\Omega$
is a Parity symmetry of the gauged WZW model.
Note that $\Omega$ exchanges the $\pm$ components of the gauge
field: $(\Omega A)_{\pm}(t,\s)=A_{\mp}(t,-\s)$.
Another Parity is $P_B={\cal I}_B\Omega$
where 
\beq
{\cal I}_B:(A,g)\to(g_*^{-1}Ag_*,g_*^{-1}g^{-1}g_*),
\eeq 
\beq
g_*:=i\sigma_1=\left(\begin{array}{cc}
0&i\\
i&0
\end{array}\right).
\eeq
Conjugation by $g_*$ preserves the $U(1)$ subgroup,
acting as $g_*^{-1}hg_*=h^{-1}$.
Thus, a $U(1)$ bundle with connection $A$ is mapped to
another $U(1)$ bundle with connection $-A$.
By PW-identity (\ref{PWid}), we have $kS(g_*^{-1}Ag_*,g_*^{-1}gg_*)=kS(A,g)$.
Thus, $g_*$-conjugation is a symmetry of the system.
The Parity $P_B$ is obtained by combining $P_A$ with this symmetry.
Note that $g_*$ can be replaced by $g_*h_1$ for any $h_1\in H$
by a gauge transformation.
The two involutions ${\cal I}_A$ and ${\cal I}_B$ act on the
disk coordinate $z=\e^{i\phi}\cos\theta$ as
\beqa
&&{\cal I}_A:z\to\bz,
\\
&&{\cal I}_B:z\to z.
\eeqa
In fact, up to the $\Z_k$ axial rotations, these are the only
Parity symmetry obtained by starting from the involutions of $SU(2)$
of the type $g\to\pm g_0g^{-1}g_0$.

These Parity symmetries act on the states as
\beqa
&&P_A:u\otimes v\in \B_{j,n}\otimes\wtB_{j,-n}\longmapsto
(-1)^{2j}v\otimes u\in \B_{j,-n}\otimes \wtB_{j,n},
\label{bPA}\\[0.1cm]
&&P_B:u\otimes v\in \B_{j,n}\otimes\wtB_{j,-n}\longmapsto
(-1)^{2j}g_*v\otimes g_*u\in \B_{j,n}\otimes\wtB_{j,-n}.
\label{bPB}
\eeqa
Here $g_*u\in \B_{j,-n}$ for $u\in \B_{j,n}$ is defined by considering
$u$ as an element of
$\V_j$: if $u$ is a charge $n$ highest weight state with respect to
$\whfh$, then $g_*u$ is a charge $-n$ highest weight state
which can be regarded as an element of $\B_{j,-n}$.

One can check that (\ref{bPA}) and (\ref{bPB})
are consistent with the field identification.
Let us start with $P_A$. The problem is trivial if
$(j,n)$ is not equivalent to $(j,-n)$ since one can choose the phase of the
states so that
$P_A$ is compatible with the field identification.
The cases with
$(j,n)\equiv (j,-n)$ consist of
$(j,n)=(j,0)$ and $(j,n)=({k\over 4},\pm{k\over 2})$.
The case $(j,0)$ is trivial for an obvious reason.
Non-trivial is the latter case.
Let $u_{\pm}=|{k\over 4},\mp{k\over 4}\rangle\in V_{k\over 4}\subset
\V_{k\over 4}$ be the vector representing the primary state of
$\B_{{k\over 4},\pm {k\over 2}}$.
The field identification identifies the states
$u_+\otimes u_-
\in\B_{{k\over 4},{k\over 2}}\otimes \B_{{k\over 4},-{k\over 2}}$
and
$u_-\otimes u_+
\in \B_{{k\over 4},-{k\over 2}}\otimes \B_{{k\over 4},{k\over 2}}$,
up to a constant;
\beq
u_+\otimes u_-=\epsilon u_-\otimes u_+.
\label{FIeg}
\eeq
Then, $P_A$ maps these states as
$$
\begin{array}{l}
u_+\otimes u_-\mapsto (-1)^{k\over 2} u_-\otimes u_+
=(-1)^{k\over 2}\epsilon^{-1}u_+\otimes u_-,\\
u_-\otimes u_+\mapsto (-1)^{k\over 2} u_+\otimes u_-
=(-1)^{k\over 2}\epsilon u_-\otimes u_+.
\end{array}
$$
These are indeed the same map, provided
$\epsilon^{-1}=\epsilon$, or $\epsilon=\pm 1$.
Let us next consider the action of $P_B$
on the ground state $|0,0\rangle\otimes |0,0\rangle$
in $\B_{0,0}\otimes \wtB_{0,0}$
which is identified with the state
$|{k\over 2},{k\over 2}\rangle\otimes |{k\over 2},-{k\over 2}\rangle$
in $\B_{{k\over 2},-k}\otimes \wtB_{{k\over 2},k}$,
up to some phase,
say $c$.
Now, $g_*$ sends $|j,m\rangle\to i^{2j}|j,-m\rangle$.
Thus, $P_B$ maps these states as
\beqa
|0,0\rangle\otimes |0,0\rangle&\mapsto&
|0,0\rangle\otimes |0,0\rangle,
\nn\\
\mbox{$c|{k\over 2},{k\over 2}\rangle\otimes
|{k\over 2},-{k\over 2}\rangle$}
&\mapsto&\mbox{$c(-1)^{k}i^{k}|{k\over 2},{k\over 2}\rangle\otimes
i^{k}|{k\over 2},-{k\over 2}\rangle$}.\nn
\eeqa
Since $(-1)^{k}i^{k}i^{k}=1$, it is indeed compatible with
the field identification.

Using (\ref{bPA})-(\ref{bPB}),
one can compute  the twisted partition function.
For $P_A$, only  representations with $(j,-n)\equiv
(j,n)$ contribute. As we have seen above, these are
$(j,0)$ with even $j$ and $({k\over 4},{k\over 2})$
(the latter is possible only when $k$ is even).
It is then easy to see that
\beq
\Tr(P_A\e^{2\pi i \tau H})=
\sum_{j:{\rm integer}}
\chi^{}_{j,0}(2\tau)+\delta_k^{(2)}\epsilon(-1)^{k\over 2}
\chi_{{k\over 4},{k\over 2}}(2\tau),
\eeq
where $\epsilon$ is the constant that appears in the field identification
(\ref{FIeg}), which we learned to be a sign $\pm 1$.
For $P_B$, all the representations contribute.
The trace on the subspace $\Scr{H}_{j,n}\otimes\Scr{H}_{j,-n}$ is
\beqa
\Tr_{j,n}(P_B\e^{2\pi i\tau H})
&=&\sum_{N,M}
\langle N|\otimes \langle M|q^H(-1)^{2j}g_*|M\rangle
\otimes g_*|N\rangle
\nn\\
&=&\sum_{N,M}q^{E_N}(-1)^{2j}\langle N|g_*|M\rangle
\langle M|g_*|N\rangle
=\Tr\!\!\mathop{}_{\Scr{H}_{j,n}}q^{2(L_0-{c\over 24})}(-1)^{2j}g_*^2,
\nn
\eeqa
where $\{|N\rangle\}$ and $\{|M\rangle\}$ are the basis vectors of 
$\Scr{H}_{j,n}$ and $\Scr{H}_{j,-n}$.
We note here that $g_*^2$ is equal to $-{\bf 1}_2$ and thus acts on
the spin $j$ representation as $(-1)^{2j}$.
Thus, this contribution is just $\chi_{j,n}(2\tau)$
and the total trace is the sum over $(j,n)\in {\rm PF}_k$.

One could also consider parities combined with the axial rotation symmetry
$a^{\ell}$.
Such parities map the state
$u\otimes v\in \Scr{H}_{j,n}\otimes \Scr{H}_{j,-n}$
as $a^{\ell}P_A:
u\otimes v\mapsto (-1)^{2j}\e^{-{2\pi i \ell n\over k}}v\otimes u$
and $a^{\ell}P_B:u\otimes v\mapsto (-1)^{2j}\e^{2\pi i\ell n\over k}
g_*v\otimes g_*u$.
Note that all $a^{\ell}P_A$ are involutive but
$a^{\ell}P_B$ are not;
\beq
(a^{\ell}P_A)^2=1,\qquad
(a^{\ell}P_B)^2=a^{2\ell}.
\eeq
For $a^{\ell}P_B$, only the one with $\ell=0$ and $\ell={k\over 2}$
are involutive (the latter applies only when $k$ is even).
This can also be understood from the geometrical point of view,
$a^{\ell}{\mathcal I}_A:z\to \e^{2\pi i\ell\over k}\bz$
and $a^{\ell}{\mathcal I}_B:z\to \e^{2\pi i\ell \over k}z$. 
The twisted partition functions are
\beqa
&&
\Tr(a^{\ell}P_A\e^{2\pi i \tau H})=
\sum_{j:{\rm integer}}
\chi^{}_{j,0}(2\tau)+\delta_k^{(2)}\epsilon(-1)^{k\over 2}
(-1)^{\ell}\chi_{{k\over 4},{k\over 2}}(2\tau),
\label{TrPAe}\\
&&
\Tr(a^{\ell}P_B\e^{2\pi i\tau H})
=\sum_{(j,n)\in {\rm PF}_k}\e^{2\pi i\ell n\over k}\chi^{}_{j,n}(2\tau).
\label{TrPBe}
\eeqa
We recall that $\epsilon$ is some sign which has not been determined yet.


\section{Parafermions: RCFT versus geometry}\label{sec:paraf}

In this section, we describe the crosscap states
of the $SU(2)/U(1)$ coset model
following the general procedure given in Section~\ref{general}.
Comparison with some of the results in Section~\ref{sec:gWZW}
will provide the geometric interpretation of the PSS and other
parity symmetries. This is also confirmed by using localized wave packets.

As a warm-up and for later use, we briefly review orientifolds of
$SU(2)_k$ \cite{PSS,PSSII,HOcs,BiSt} and their geometrical interpretation.

\subsection{Orientifolds of $SU(2)$: Summary of the RCFT}

Following Section \ref{general}, we collect 
the basic RCFT data. The $S$-matrix of the
$SU(2)$ theory is given by the well known expression
\beq
S_{jj'} = \sqrt{\frac{2}{k+2}} \ \sin \pi \frac{(2j+1)(2j'+1)}{k+2}.
\eeq
From this and $\Delta_j=\frac{j(j+1)}{k+2}$ one computes the
$P$-matrix
\beq\label{su2p}
P_{jj'} = \frac{2}{\sqrt{k+2}} \ \sin \pi \frac{(2j+1)(2j'+1)}{2(k+2)}.   \
\delta^{(2)}_{2j+2j'+k}
\eeq
With this, $k+1$ Cardy boundary states can be constructed,
labelled by the integral highest weight representations. 
It is by now well known \cite{AlSc,ASR,FFFS} 
that the Cardy boundary state $\ket{\Scr{B}_J}$
corresponds geometrically to a brane wrapping the conjugacy class $C_J$ 
of $SU(2)$
containing the group element $e^{2\pi i J\sigma_3/k}$. 

The simple current
group of the model is $\Z_2$, and is generated by the sector labelled $k/2$.
Fusing $k/2$ with itself yields the identity representation. Hence, one
can construct two different crosscap states, the standard PSS state
$\ket{\Scr{C}}$ and a simple current induced state
$|\Scr{C}_\frac{k}{2}\rangle$.
The geometrical interpretation
of those crosscap states has been given in \cite{HSSc,BCW,ib}: the standard
PSS crosscap corresponds to the involution ${\cal I}: g \to g^{-1}$,
whereas the simple current induced crosscap corresponds to 
${\cal I}^- : g\to - g^{-1}$. 

In terms of the RCFT data, the Klein bottle  amplitudes are
obtained as
\beqa
\Tr^{}_{\cal H}(P\e^{2\pi i\tau H})
&=&  \sum_{j\in {\rm P}_k} Y_{j0}^0 \ \chi_j(2\tau) =
\sum_{j\in {\rm P}_k}(-1)^{2j}\chi^{}_{j}(2\tau) \\ \no
\Tr^{}_{\cal H}(P^-\e^{2\pi i\tau H})
&=& \sum_{j\in {\rm P}_k} Y_{j\frac{k}{2}}^{\frac{k}{2}}
\chi^{}_{j}(2\tau) 
=\sum_{j\in {\rm P}_k}\chi^{}_{j}(2\tau)
\eeqa
where we have used the fact that $Y_{j0}^0 = (-1)^{2j}$
and that $Y_{j\frac{k}{2}}^{\frac{k}{2}}=1$ for all $j$.

To construct and interpret the M\"obius strips, one needs to know that
$Y_{J 0}^j= (-1)^j (-1)^{2J} N_{JJ}^j $ 
and $Y_{J \frac{k}{2}}^j = N_{J \frac{k}{2}-J}^j$.
With the help of these identities, one obtains
\beqa
\langle \Scr{C} |\e^{-{\pi i\over 4\tau}H}\ket{\Scr{B}_{J}} 
&=& \sum_j Y_{J0}^{j} \ \widehat\chi_j(\tau)
= \sum_j N_{JJ}^j \ (-1)^{2J} \ (-1)^j \ \widehat\chi_j(\tau) \\ \no
\langle \Scr{C}_{k\over 2} | \e^{-{\pi i\over 4\tau}H}\ket{\Scr{B}_{J}} 
&=& \sum_j Y_{J\frac{k}{2}}^{j} \ \widehat\chi_j(\tau)
= \sum_j N_{J \frac{k}{2}-J}^j \ \widehat\chi_j (\tau)
\eeqa
To make the connection to geometry for the first line, recall that
${\cal I}: g\to g^{-1}$ acts as the anti-podal map on the $S^2$ wrapped
by the brane. In the classical limit, the primary fields living on the
brane become functions on $S^2$. More precisely, the algebra of functions
on $S^2$ is spanned by the spherical harmonics $Y_{j,m}$, $j \in \Z$. Under
reflection they transform as $Y_{j,m} \to (-1)^j Y_{j,m}$. This is
exactly the action of $P$ that one reads off from the M\"obius amplitude.

For the second line, observe that the spectrum of open string states
in the M\"obius amplitude is exactly that of a brane $J$ and its
image $\frac{k}{2} - J $ under ${\cal I}^-$. 

In this way, the CFT data encodes the geometry of the orientifold.

\subsection{Parafermions}

We first review some basic facts on the $SU(2)/U(1)$ coset model, as
a rational conformal field theory.
The Hilbert space
$$
{\mathcal H}=\bigoplus_{(j,n)\in PF_k}\Scr{H}_{j,n}\otimes\Scr{H}_{j,-n},
$$
decomposes into  irreducible representations of
the parafermion algebra.
The conformal weight of the primary field with label
$(j,n)$
is given by
\beq\label{paraweight}
h_{j,m} = \frac{j(j+1)}{(k+2)} - \frac{n^2}{4k}
\eeq
if $(j,n)$ is in the range
$j=0,\dots, k/2$ and $-2j \leq n \leq 2j$.
We shall call the latter
{\it the standard range} (abbreviated by S.R.).
Any label $(j,n)$ can be reflected to the standard range by
field identification $(j,n)\to ({k\over 2}-j,n+k)$.

\subsubsection*{\it Modular matrices}

The $S$ and $T$-matrices of the coset model have the factorized form
\beq
S_{(j,n)(j',n')} = 2 \ S_{jj'} \ S^*_{nn'},~~~
T_{(j,n),(j',n')}=T_{jj'}T^*_{nn'},
\label{factorPF}
\eeq
where it is understood that the matrices with
pure $j$ labels are those of the $SU(2)_k$
WZW model, and matrices with pure $n$ labels are
those of $U(1)_k$.
Using this factorization property, we find
\beq
N_{(j,n)(j',n')}^{\,\,\,(j'',n'')}
=N_{jj'}^{\,\,j''}\delta_{n+n',n''}^{(2k)}
+N_{jj'}^{\,\,{k\over 2}-j''}\delta_{n+n',n''+k}^{(2k)}.
\label{paraN}
\eeq
We also need to determine $P$ and $Y$.
As a first step, it is useful to consider instead the quantities
$Q=ST^2S$ and $\tilde{Y}$ defined by (\ref{defSTS}) and (\ref{defYtilde}).
The computation is easy since $\sqrt{T}$ is not involved
and one can use the factorization (\ref{factorPF}).
The result is
\beqa
&&
Q_{(j,n)(j',n')} = Q_{jj'} Q_{nn'}^* + Q_{\frac{k}{2}-j, j'} Q_{n+k,n'}^*,
\\
&&
\tilde{Y}_{(j,n)(j',n')}^{(j'',n'')} = \tilde{Y}_{jj'}^j \
\overline{\tilde{Y}}_{nn'}^{n''} + \tilde{Y}_{jj'}^{\frac{k}{2}-j} \
\overline{\tilde{Y}}_{nn'}^{n''+k}.
\eeqa
From this, one can compute $P=\sqrt{T}Q\sqrt{T}$ and
$Y_{ab}^{c}=\sqrt{T_b/T_c}\tilde{Y}_{ab}^c$.

\subsubsection*{\it Discrete Symmetries}

The group of simple currents of the model is $\Z_k$
generated by $(0,2)$.
The monodromy charge of the field in the representation $(j,n)$
under the simple current $(0,2\ell)$ is
\beq
Q_{(0,2\ell)}(j,n) = \frac{\ell n}{k}.
\eeq
Accordingly, there is a symmetry group $\Z_k$
acting on the states as
\beq
a^{\ell}: \psi_{(j,n)} \to
\e^{\frac{2\pi i \ell n}{k}} \ \psi_{(j,n)}
\quad {\rm for} \quad \psi_{(j,n)} \in \Scr{H}_{j,n} \otimes \Scr{H}_{j,-n},
\eeq
where we denote the generator of the group by $a$.
This is in fact equivalent to the $\Z_k$ axial rotation symmetry
of the gauged WZW model (\ref{Defa}).

\subsubsection*{\it Mirror symmetry}

The orbifold by the full symmetry group $\Z_k$
has the Hilbert space of states
\beq
{\cal H}^M = \bigoplus_{(j,n)\in PF_k}
\Scr{H}_{j,n} \otimes \Scr{H}_{j,n},
\eeq
and can be regarded as the mirror of the original model.
The mirror map $\Psi: {\cal H} \to {\cal H}^M$ 
acts on states as
$\Psi=V_M \otimes 1: 
\ket{j,n} \otimes \ket{j,-n} \to  \ket{j,-n} \otimes \ket{j,-n}$.

\subsection{A-type parities: RCFT and geometry}

According to Section~\ref{general},
there are Cardy branes (A-branes) $\Scr{B}_{j,n}$
labelled by the representation and PSS parities (A-parities) $P_{\ell}$
labelled by the simple currents.
The branes are transformed under the parities as (\ref{Pgi}),
which reads
\beq
P_{\ell}:\Scr{B}_{j,n}\to\Scr{B}_{j,2\ell-n}.
\label{paraBT}
\eeq
The crosscap state for the parity $P_{\ell}$ is
\beq
\ket{\Scr{C}_{\ell}} = \sum_{(j,n) \in PF_k} 
\frac{P_{(0,2\ell)(j,n)}}{\sqrt{S_{(0,0)(j,n)}}} \ \cket{j,n}.
\eeq
Explicit expressions for the A-type crosscap states can be found
in the appendix.
Of particular interest
is the coefficient of the identity $(0,0)$, since in a full string
model this coefficient would give rise to a contribution to the total tension
of the orientifold plane. It is given by
\beq
T_{O^A_{\ell}}=\left\{
\begin{array}{ll}
\frac{1}{[(k+2)k]^{\frac{1}{4}}} \ 
\cot^{\half}\!\left[\frac{\pi}{2(k+2)}\right]
& k \ {\rm odd} \\
\frac{1}{[(k+2)k]^{\frac{1}{4}}} \ 
\left( \cot^{\frac{1}{2}}\!\left[\frac{\pi}{2(k+2)}\right]
+ (-1)^{\ell}
\tan^{\frac{1}{2}}\!\left[\frac{\pi}{2(k+2)}\right] \right)
&k \ {\rm even}
\end{array}\right.
\eeq
For $k$ odd, it is manifestly independent of $\ell$
--- all the PSS orientifolds have the same tension.
For $k$ even, there is an additional term that depends on $\ell$ mod $2$.
The latter is consistent with the action of the
$\Z_k$ generator $a$  on the crosscap states
$$
a: \ket{\Scr{C}_{\ell}} \to \ket{\Scr{C}_{\ell+2}};$$
which implies that orientifolds related by symmetry operations 
have the same mass, as required.
In the geometric
limit of infinite $k$, the $\ell$ dependence drops out and  we get 
equal masses also for orientifolds that are not related by symmetry.

The Cardy formula yields the boundary states
\beq
\ket{\Scr{B}_{j,n} } = \sum_{(j',n')\in PF_k} 
\frac{S_{(j,n)(j',n')}}{\sqrt{S_{(0,0)(j',n')}}} \ \kket{(j',n')}.
\eeq
whose tension is
\beq
T_{\Scr{B}_{j,n}} = \frac{\sqrt{2}}{[k(k+2)]^{\frac{1}{4}}} \, 
\frac{\sin \frac{\pi(2j+1)}{k+2}}{\sqrt{\sin \pi \frac{1}{k+2}}},
\eeq
which is $n$-independent.

\subsubsection{The one-loop amplitudes}

We next compute the cylinder, MS and KB amplitudes.
Details are recorded in Appendix~\ref{app:detai}.
The cylinder amplitude are
\beq
\langle \Scr{B}_{j_1,n_1} | \e^{-{\pi i\over \tau}H}\ket{\Scr{B}_{j_2,n_2}}
= \sum_{2j+n \,{\rm even}}
N_{j_1j_2}^{j} \delta_{n_2-n_1+n}^{(2k)}\chi_{j,n}(\tau).
\label{AparaBB}
\eeq
The M\"obius strip with boundary condition $\Scr{B}_{j_2,n_2}$ is
\beq
\langle\Scr{C}_{\ell}|\e^{-{\pi i\over 4\tau}H} |\Scr{B}_{j_2,n_2}\rangle 
=\sum_{2j+n\,{\rm even}}
N_{j_2j_2}^j\delta_{2n_2-2\ell+n}^{(2k)}
\epsilon_{j,n}\widehat{\chi}_{j,n}(\tau),
\label{AparaBC}
\eeq
where
$\epsilon_{j,n}$ is a sign factor which is
$(-1)^{2j+n\over 2}, 1, (-1)^n$ if
$(j,n),({k\over 2}-j,n+k),({k\over 2}-j,n-k)$ is
in the standard range, respectively.
Comparison of (\ref{AparaBC}) with (\ref{AparaBB}) implies
that the image brane of $\Scr{B}_{j_2,n_2}$
is $\Scr{B}_{j_2,2\ell-n_2}$, which is indeed correct
 (\ref{paraBT}).
The Klein bottle amplitudes are
\beq
\langle \Scr{C}_{\ell_1}|\e^{-{\pi i\over 2\tau}H}
|\Scr{C}_{\ell_2}\rangle
=\sum_{2j+\ell_1-\ell_2\,{\rm even}}
\chi_{j,\ell_1-\ell_2}(2\tau)
+\delta_{k}^{(2)}\delta_{\ell_1,\ell_2}^{(2)}
(-1)^{\ell_1}
\chi_{{k\over 4},{k\over 2}+\ell_1-\ell_2}(2\tau).
\label{AparaCC}
\eeq
In particular,
for $\ell_1=\ell_2$ we have
\beq
\langle \Scr{C}_{\ell}|\e^{-{\pi i\over 2\tau}H}
|\Scr{C}_{\ell}\rangle
=\sum_{j\,{\rm integer}}
\chi_{j,0}(2\tau)
+\delta_{k}^{(2)}(-1)^{\ell}
\chi_{{k\over 4},{k\over 2}}(2\tau).
\label{AparaKB}
\eeq
For $k$ odd, this is independent of 
$\ell$, whereas for $k$ even there is an $\ell$ dependent term, which
plays only a role at finite $k$.
This behavior was observed before, when we computed the tensions
of the orientifolds.

\subsubsection{Geometrical interpretation}

As noted in Section~\ref{sec:gWZW},
the model has a $\sigma$-model interpretation,
where the target space is a disk $|z|\leq 1$.
A geometrical interpretation of the Cardy boundary states in that
geometry was provided by \cite{MMS}. 
It was found  that the Cardy states with $j=0$
correspond to D0-branes distributed at the $k$ symmetric points at
the boundary of the disk:
$\Scr{B}_{0,n}$ ($n$ even)
is the D0-brane at the boundary point
$z=\e^{\pi i n\over k}$.
The $\Z_k$ symmetry rotations act on the boundary states by shifting
$n$.
The branes
with higher $j$ are D1 branes
stretched between two special points separated by
the angle $4\pi j/k$:
$\Scr{B}_{j,n}$ ($2j+n$ even)
is the D1-brane along the straight line
connecting the points $z=\e^{{\pi i\over k}(n+2j)}$
and $z=\e^{{\pi i\over k}(n-2j)}$.
Branes of a given $j$ are related by $\Z_k$ symmetry, just as in
the case $j=0$.

The parity $P_A$ found in the gauged WZW model analysis
acts on the disk as $z\to\bz$, a reflection with respect to the
diameter ${\rm Im}(z)=0$.
Since it maps the special points as $z=\e^{{\pi i\over k}(n\pm 2j)}\mapsto
\e^{-{\pi i\over k}(n\pm 2j)}$, it transform the Cardy branes as
$P_A:\Scr{B}_{j,n}\to\Scr{B}_{j,-n}$.
Also, the combination with the axial rotation symmetry $a^{\ell}$
would act
on the branes as $a^{\ell}P_A:\Scr{B}_{j,n}\to\Scr{B}_{j,2\ell-n}$.
Comparing with the rule (\ref{paraBT}), we find the relation of gauged
WZW parities to the PSS parities:
\beq
P_{\ell}=a^{\ell}P_A.
\eeq
Indeed, under this identification,
the Klein bottle amplitude (\ref{AparaKB}) is consistent with
the partition function (\ref{TrPAe}) in the gauged WZW model.
The sign $\epsilon$ of the field identification
is now determined to be $\epsilon=(-1)^{k\over 2}$.
It would be interesting to examine this using the functional integral method.

\subsubsection{Sketch of a shape computation}

We now give an independent argument to determine the location of
the orientifold planes.
In \cite{MMS} the location and geometry of D-branes was tested by
scattering graviton wave packets from the
D-branes
and taking the classical limit. These computations have
been repeated  for orientifolds in \cite{HSS,hikida,ib}. Suitable
graviton wave packets are localized $\delta$-functions, which are
written down for parafermions in \cite{MMS} appendix D. Since closed string
states with $j\sim k$  are not well-localized, one only uses  states
with $j\ll k$ as part of the test wave function. This means that the
shape of the brane is encoded in the couplings of the brane
to bulk fields with $j \ll k$. From the above argument, it is expected 
that the orientifold planes in the parafermion theory are located
along diameters of the disk. On the other hand, it is known that
the D-branes  $\Scr{B}_{\frac{k}{4},n}$ are also located along diameters. To
compute the shape it is therefore not required to repeat the computation
of \cite{MMS}, but to merely compare the coefficients of the crosscap
state with the coefficients of the boundary states
$|\Scr{B}_{\frac{k}{4},n}\rangle$.
If the asymptotic behavior of the boundary and crosscap
coefficients is the same, it can  be concluded that their
locations coincide.
The D-brane  couplings of the Cardy state 
$|\Scr{B}_{\frac{k}{4},n}\rangle$ to the ground states $\ket{j,m}$ 
are given by
\beq
B_{(\frac{k}{4}, n)(j,m)} = \frac{\sqrt{2}}{[k(k+2)]^{\frac{1}{4}}} \
\frac{(-1)^{2j} \delta^{(2)}_{2j}}{\sqrt{\sin \pi\frac{2j+1}{k+2}}} \,
e^{\frac{\pi i n m}{k}}
\eeq
and the couplings of the PSS-crosscap state (for $k$ even) are
\beq
\Gamma_{j,m} = \frac{\sqrt{2}}{[k(k+2)]^{\frac{1}{4}}} \ \delta^{(2)}_{2j} \, 
\frac{ \sin \pi \frac{2j+1}{2(k+2)} \, + (-1)^{\frac{m+2j}{2}}
\cos \pi \frac{2j+1}{2(k+2)}}{\sqrt{\sin \pi\frac{2j+1}{k+2}}}.
\eeq
In the large $k$ limit, the contribution of the second term is
dominant. In that limit, the orientifold-couplings behave 
exactly like those of the
boundary state $\ket{\Scr{B}_{k/4, k/2}}$ and the conclusion is that
the PSS crosscap is located along
the same diameter as that brane. 
This matches with the earlier
conclusion
based on the brane transformation rule (\ref{paraBT}).
The other
crosscap states differ only in the
$m$-dependence and hence correspond to rotated diameters.

\subsection{B-type Parities: RCFT and geometry}

As before, we construct B-type crosscap states by constructing
A-type crosscaps in the mirror $\Z_k$ orbifold followed by an
application of the mirror map.

To construct
the crosscap states in the orbifold theory, we first need the
bilinear form $q$ for the group $G=\Z_k$.
It is uniquely fixed by the requirement $q(g,g)=-h_g$
and is given by
\beq
q((0,n),(0,m)) := \frac{nm}{4k},\qquad\mbox{$n$, $m$ even}.
\eeq
Note that it is well-defined (invariant under
$2k$ shifts of $n$ and $m$)
and obeys $2q(g_1,g_2) = Q_{g_1}(g_2)$. 
We also need $\hat{Q}_{g}(h)$ defined modulo 2.
For this, we note that
\beq
h_{(0,n)} = -\frac{n^2}{4k} +\frac{|n|}{2},\qquad\mbox{for $-k\leq n\leq k$}.
\eeq
Then, we have (for $n$, $m$ even)
\beqa
\hat{Q}_{(0,n)}((0,m)) 
&=& -\frac{n^2}{4k} -\frac{m^2}{4k} + \frac{(n\hat{+}m)^2}{4k}
+\frac{|n|}{2} +\frac{|m|}{2} - 
\frac{|n\hat{+}m|}{2}
\nn\\
&=& \frac{n\hat{+}m-|n\hat{+}m|}{2}
-\frac{n-|n|}{2} -\frac{m-|m|}{2} +\frac{nm}{2k}
=\frac{nm}{2k} \quad {\rm mod} \ 2.
\nn
\eeqa
In the second step we used (\ref{Qcomp}).
In the last step we have used that ${n-|n|\over 2}$ is
an even integer if $n$ is even.
Thus, we arrive at the conclusion that $\hat{Q}=2q$ mod $2$. 
Therefore, Eq.~(\ref{thetaEq}) is homogeneous,
$\theta(gh)=\theta(g)+\theta(h)$, and the general solution is given by
\beq
\theta_r(\ell)=-2\frac{r\ell}{k},\qquad r\in \Z/k\Z.
\eeq

Following the procedure given in
Section~\ref{general},
we find B-parities $P^r_B$ parametrized by a mod-$k$ integer $r$,
whose crosscaps are given by
\beq
|\Scr{C}^B_r\rangle
={1\over\sqrt{k}}\sum_{\ell}\e^{2\pi i r\ell\over k}(V_M\otimes 1)
|\Scr{C}_{\ell}\rangle,
\label{CBr}
\eeq
where $V_M:\Scr{H}_{j,n}\to \Scr{H}_{j,-n}$ is the map induced from the
mirror automorphism.
More explicitly, they are expressed as
\beqa
\ket{\Scr{C}_r^B}&=&
\frac{k^{\frac{1}{4}}}{(k+2)^{\frac{1}{4}}}\Biggl[
\sum_{j,(j,-2r)\in S.R.} 
(-1)^{j+r}
\sqrt{\cot\mbox{$\frac{\pi(2j+1)}{2(k+2)}$}}
\ \cket{(j,2r)}_B \nn\\
&& ~~+ \sum_{j,(j,-2r)\notin S.R.} 
\sqrt{\cot\mbox{$\frac{\pi(2j+1)}{2(k+2)}$}}
\cket{(j,2r)}_B \Biggr].
\eeqa
The details of the computation are summarized in Appendix~\ref{subapp:Bcc}.
The square of the B-type involutions $P^r_B$ can be computed to be
\beq
(P^r_B)^2 = a^{2r}.
\eeq
This is consistent with the crosscap
$|\Scr{C}_r^B\rangle$ being a $a^{2r}$-twisted state
(it belongs to
$\oplus_j\Scr{H}_{j,2r}\otimes\Scr{H}_{j,2r}$, which is a subspace of
$\oplus_{j,n}\Scr{H}_{j,n}\otimes\Scr{H}_{j,4r-n}={\mathcal H}_{a^{2r}}$).
As before, the tensions of the orientifold planes can  be determined as
overlaps with the ground state. They are only non-vanishing for the
involutive crosscaps $P_B^0$ and $P_B^{\frac{k}{2}}$ (the latter exists
only for $k$ even). The result is
\beqa
T_{\Scr{C}_0^B} &=& \frac{k^{\frac{1}{4}}}{(k+2)^{\frac{1}{4}}}
\cot^{\frac{1}{2}}\!\left[\frac{\pi}{2(k+2)}\right]
\label{TB0}\\
T_{\Scr{C}_{\frac{k}{2}}^B} &=& \frac{k^{\frac{1}{4}}}{(k+2)^{\frac{1}{4}}}
\tan^{\frac{1}{2}}\!\left[\frac{\pi}{2(k+2)}\right]
\label{TBk2}  
\eeqa
Note the different behavior of the tension as $k\to \infty$: the
tension of the orientifold plane for$P^0_B$ becomes infinite, whereas 
that for $P^{\frac{k}{2}}_B$ goes to zero.

\subsubsection*{\it Boundary states for B-branes}

In this section we write down the B-type boundary states, which are obtained
by taking the average over the $\Z_k$-orbit of the A-type boundary states,
followed by the action of the mirror map $\Psi^{-1}$.
Since there is only one orbit for each $j$, they are parametrized by
just $j$ (with the identification
$j\equiv {k\over 2}-j$):
\beq
\ket{\Scr{B}_j^B}
={1\over \sqrt{k}}\sum_{\ell}(V_M\otimes 1)\ket{\Scr{B}_{j,n+2\ell}}
=(2k)^{1\over 4}\sum_{j'\,{\rm integer}}{S_{jj'}\over\sqrt{S_{0j'}}}
\cket{(j',0)}_B,
\eeq
where $n$ is an arbitrary integer such that $2j+n$ is even.
For $j={k\over 4}$, which is possible only when $k$ is even,
the $\Z_k$ action has a fixed point,
$\ell={k\over 2}:({k\over 4},n)\mapsto ({k\over 4},n+k)\equiv ({k\over 4},n)$,
and special care is needed in the construction of the boundary states.
In fact, it splits into two B-branes distinguished by $\eta=\pm 1$
\cite{MMS}, with the boundary states
\beq
\ket{\Scr{B}_{\frac{k}{4},\eta}^B}= \frac{1}{2} \
(2k)^{\frac{1}{4}} \sum_{j\,{\rm integer}} 
\frac{S_{{k\over 4}j}}{\sqrt{S_{0j}}}
\kket{(j,2r)}_B + \frac{\eta}{2} \ [k(k+2)]^{\frac{1}{4}} \ 
\kket{\frac{k}{4}, \frac{k}{2}}_B
\eeq
Under the symmetry $a^{\ell}$, these boundary states
are transformed as $|\Scr{B}_j^B\rangle\to|\Scr{B}_j^B\rangle$
and $|\Scr{B}^B_{{k\over 4},\eta}\rangle\to
|\Scr{B}^B_{{k\over 4},(-1)^{\ell}\eta}\rangle$.
Thus, each of the ordinary B-branes $\Scr{B}_j^B$ ($j\not\equiv {k\over 4}$)
is invariant under $\Z_k$, but the special B-branes
$\Scr{B}^B_{{k\over 4},+}$ and $\Scr{B}^B_{{k\over 4},-}$ are exchanged
under odd elements of $\Z_k$.

One can also consider boundary states on a circle with  $\Z_k$
twisted boundary
conditions, which can be used, for example, when we compute
the M\"obius strip amplitudes with respect to non-involutive
parities.
The twist adds an appropriate
phase factor in the $\Z_k$-average, as explained
in Section~\ref{subsubsec:Borb}.
For an ordinary B-brane $\Scr{B}_j^B$, 
since it is invariant under any element of $\Z_k$,
one can consider the boundary state with any twist $a^r$.
The symmetry $a^r$ is mapped under mirror symmetry to
the quantum symmetry of the orbifold model, associated with the character
$a^{\ell}\to\e^{-{2\pi i r \ell\over k}}$. Therefore, the relevant average is
\beq
\ket{\Scr{B}_j^B}_{a^{r}}
={\e^{\pi i rn\over k}
\over \sqrt{k}}\sum_{\ell}\e^{2\pi i r\ell\over k}
(V_M\otimes 1)\ket{\Scr{B}_{j,n+2\ell}}
=(2k)^{1\over 4}\sum_{j'\,{\rm integer}}{S_{jj'}\over\sqrt{S_{0j'}}}
\cket{(j',r)}_B.
\label{BjB}
\eeq
The overall phase $\e^{\pi i rn\over k}$ is chosen so that $n$-dependence
disappears in the final expression.
The special ones $\Scr{B}_{{k\over 4},\pm}$ are invariant only under even
elements $a^{2r}$. Thus we can only consider even twists:
\beq
\ket{\Scr{B}_{\frac{k}{4},\eta}^B}_{a^{2r}}= \frac{1}{2} \
(2k)^{\frac{1}{4}} \sum_{j\,{\rm integer}} 
\frac{S_{{k\over 4}j}}{\sqrt{S_{0j}}}
\kket{(j,2r)}_B + \frac{\eta}{2} \ [k(k+2)]^{\frac{1}{4}} \ 
\kket{\mbox{$(\frac{k}{4}, \frac{k}{2}+2r)$}}_B
\eeq

\subsubsection{The one loop amplitudes}

We present here the cylinder, MS and KB amplitudes.
When the average formulae
(\ref{CBr}), (\ref{BjB}), etc, are available,
the computation is easily done using
the results (\ref{AparaBB}), (\ref{AparaBC}) and (\ref{AparaCC})
for the A-type amplitudes.
In this way we obtain
\beq
{}_{a^{r}}\langle \Scr{B}_{j_1}^B|\e^{-{\pi i\over\tau}H}
|\Scr{B}_{j_2}^B\rangle_{a^{r}}
=\sum_{2j+n\,{\rm even}}
N_{j_1j_2}^{\,\,j}\e^{\pi i rn\over k}\chi_{j,n}(\tau).
\label{BparaBB}
\eeq
\beq
\langle \Scr{C}_r^B|\e^{-{\pi i\over 4\tau}H}
|\Scr{B}_{j'}^B\rangle_{a^{2r}}
=\sum_{2j+n\,{\rm even}}
N_{j'j'}^{\,\,j}\e^{\pi i rn\over k}\epsilon_{j,-n}\widehat{\chi}_{j,n}(\tau).
\label{BparaBC}
\eeq
\beq
\langle \Scr{C}_r^B|\e^{-{\pi i\over 2\tau}H}
|\Scr{C}_r^B\rangle
=\sum_{(j,n)\in PF_k}\e^{2\pi i rn\over k}\chi_{j,n}(2\tau).
\label{BparaCC}
\eeq
Those involving the special B-branes may be computed independently.
\beqa
{}_{a^{2r}}\langle \Scr{B}^B_{\frac{k}{4},\eta} | \e^{-{\pi i\over \tau}H} 
|\Scr{B}^B_{\frac{k}{4},\eta'}\rangle_{a^{2r}}
&=& \frac{1}{4} \sum_{2j+n\,{\rm even}} \delta^{(2)}_{2j} \
(1+\eta\eta'(-1)^{\frac{2j-n}{2}}) \ \e^{\frac{2\pi i rn}{k}} \ \chi_{j,n}
\nn\\
&=& \sum_{(j,n)\in PF_k\atop j\,{\rm integer}}
{1\over 2}(1+\eta\eta'(-1)^{2j+n\over 2})
\e^{2\pi i rn\over k}\chi_{j,n}(\tau).
\label{BparaSBB}
\eeqa
The last step would fail if $2r$ were formally replaced by an odd integer,
which is consistent with
the boundary state for $\Scr{B}_{{k\over 4},\eta}$
not admitting odd twists.
The M\"obius strip with special boundary conditions is
\beqa
\langle \Scr{C}_r^B|\e^{-{\pi i\over 4\tau}H}
|\Scr{B}_{\frac{k}{4},\eta}^B\rangle_{a^{2r}}
&=&\frac{1}{2} \sum_{2j+n\,{\rm even}} \e^{\frac{\pi irn}{k}}
\epsilon_{j,-n}\widehat{\chi}_{j,n}(\tau)
\nn\\
&=&\sum_{(j,n)\not\in S.R.\atop j\,{\rm integer}}
{1\over 2}(1+(-1)^r(-1)^{2j+n\over 2})\e^{\pi i rn\over k}
\widehat{\chi}_{j,n}(\tau).
\label{BparaSBC}
\eeqa
$\epsilon_{j,-n}$ are evaluated in the last step.

Let us examine how the B-parities $P_B^r$ transform the B-branes.
Comparing (\ref{BparaBC}) and (\ref{BparaBB}), one realizes that the
$P_B^r$-image of $\Scr{B}_j^B$ is $\Scr{B}_j^B$ itself.
Comparing (\ref{BparaSBB}) and (\ref{BparaSBC}),
we see that the propagating modes in the loop channel are
the same if $\eta\eta'=(-1)^r$, namely if $r$ even and $\eta=\eta'$
or $r$ odd and $\eta=-\eta'$.
This means that $P_B^r$ with even $r$ preserves each of
the two special branes,
$\Scr{B}_{{k\over 4},+}^B$ and $\Scr{B}_{{k\over 4},-}^B$,
while $P_B^r$ with odd $r$ exchanges them.

\subsubsection{Geometrical interpretation}

To find the geometrical meaning of the parity symmetries,
we first look at the Klein bottle amplitudes (\ref{BparaCC}).
Comparing this with the result (\ref{TrPBe}) in the gauged WZW model,
we find that the B-parities $P_B^r$ are identified as
\beq
P_B^r=a^rP_B.
\label{PBint}
\eeq
This means that the RCFT parity $P_B^r$ is the orientation reversal 
$\Omega$ followed by the
rotation $z\to \e^{2\pi i r\over k} z$ of the disk.

Let us next see how the
M\"obius amplitudes fit with this interpretation.
The geometrical interpretation of B-type boundary states has already
been given in \cite{MMS}: the brane $\Scr{B}_{j=0}^B$
corresponds to a D0 sitting at the center of the disk,
and the branes of higher $j< {k\over 4}$ are $D2$ branes on a disk whose
radius depends on $j$.
Each of these are invariant under the rotation $z\to\e^{2\pi i r\over k}z$,
and hence also by $P_B^r$.
This agrees with the conclusion from the cylinder and MS amplitudes.
Moreover, the $r$-dependence of
the MS amplitudes (\ref{BparaBC})
resides only in the phase factor $\e^{\pi irn\over k}$,
which is identified as the effect of the $a^r$-action,
as can be seen from the
cylinder amplitude (\ref{BparaBB}).
This supports the structure of (\ref{PBint}).
The special B-branes $\Scr{B}_{{k\over 4},\pm}$ are interpreted
as D2-branes covering the whole disc. How the two are distinguished
is not yet understood in a geometric way, but the boundary states
show that they are exchanged
under the unit axial rotation $a$ (or other
odd rotations).
The interpretation (\ref{PBint}) is therefore consistent
with the conclusion from the cylinder/MS comparison
that the two are exchanged under $P_B^r$ with odd $r$
and preserved under $P_B^r$ with even $r$.

Finally we examine the tension formulae (\ref{TB0}) and (\ref{TBk2})
from the geometric point of view.
The orientifold fixed point set of the parity $P_B^0$ is the whole disk. 
This fits with the divergence of its tension in the geometric limit.
The orientifold fixed point set of the other involutive parity
$P_B^{k\over 2}$ (present only for even $k$)
is just one point, the center of the disk.
This is consistent with the fact that it becomes light in the geometric limit.

\subsubsection{Sketch of a shape computation}

The localized wave packets one uses to determine the shape of
branes are naturally  elements of the Hilbert space 
${\cal H}$ (\ref{parahilbert}). Hence, the scattering experiment
makes sense only for crosscap states containing Ishibashi states
in that space. These are the crosscap states leading to involutive
parities.

From our previous considerations, we already know that the orientifold
plane corresponding to $P_B^0$ extends over the whole disk. So does
the D-brane $\Scr{B}_{\frac{k}{4},\eta}$, which exists for $k$ even.
We can therefore independently confirm the location of $\Scr{C}_{P_B^0}$ by
comparing its couplings $\Gamma_{(j,m)}$ to the closed string
sector with the couplings of the brane $B_{\frac{k}{4},(j,m)}$.
Explicitly
\beqa
B_{\frac{k}{4}(j,m)} &=& \frac{\sqrt{2}k^{\frac{1}{4}}}{(k+2)^{\frac{1}{4}}} \
\delta^{(2k)}_m \ \delta^{(2)}_{2j} \
\frac{(-1)^j}{\sqrt{\sin \pi \frac{2j+1}{k+2}}} \\ 
\Gamma_{(j,m)} &=& \frac{\sqrt{2}k^{\frac{1}{4}}}{(k+2)^{\frac{1}{4}}} \
\delta^{(2k)}_m \ \delta^{(2)}_{2j} \ 
\frac{(-1)^j\, \cos \pi \frac{2j+1}{2(k+2)}}{\sqrt{\sin \pi \frac{2j+1}{k+2}}} .
\eeqa
In the large $k$ limit, these couplings do indeed agree, confirming that
the crosscap and boundary state are located at the same place.

Similarly, we know that $P_B^{\frac{k}{2}}$ corresponds to a fixed point
set consisting of the center of the disk. That is exactly the location of 
the boundary state with $J=0$. The closed string couplings of the
boundary and crosscap state are given by
\beqa
B_{0 ,(j,m)} &=& \frac{\sqrt{2}k^{\frac{1}{4}}}{(k+2)^{\frac{1}{4}}} \
\delta^{(2k)}_m \ \delta^{(2)}_{2j} \
\frac{\sin \pi \frac{2j+1}{k+2}}{\sqrt{\sin \pi \frac{2j+1}{k+2}}} \\ 
\Gamma_{(j,m)} &=&   \frac{\sqrt{2}k^{\frac{1}{4}}}{(k+2)^{\frac{1}{4}}} \
\delta^{(2k)}_m \ \delta^{(2)}_{2j} \
\frac{ \sin \pi \frac{2j+1}{2(k+2)}}{\sqrt{\sin \pi \frac{2j+1}{k+2}}} .
\eeqa
In the large $k$ limit we can (for small $j$) replace the $\sin$ by the angle,
and notice that the two couplings have the same large $k$ behaviour. This
confirms that in this case the orientifold is located at the center
of the disk, just as the boundary state.

\section*{Acknowledgements}

We would like to thank B.~Acharya, A.~Adem, M.~Aganagic,
C.~Angelantonj, J.~Maldacena, R.~Myers, 
A.~Sagnotti, E.~Sharpe and C.~Vafa.
KH is supported in part by NSF DMS 0074329 and NSF PHY 0070928.

\newpage

\appendix{Conventions on Modular Matrices}\label{app:PSY}

Here we collect the conventions on the modular matrices that are used
throughout the paper.
For a representation $\Scr{H}_i$ of the chiral algebra ${\mathcal A}$
we define the character by
\beq
\chi_i(\tau,u)
:=\Tr\!\mathop{}_{\Scr{H}_i}\left(\e^{2\pi i\tau (L_0-{c\over 24})}
\e^{2\pi i J_0(u)}\right),
\eeq
where $\tau$ is in the upper half-plane
and $J_0(u)$ is the zero mode of a spin 1 current
(or sum of commuting spin 1 currents) in ${\mathcal A}$.
We note that
\beq
\chi_i(\tau,u)=\chi_{\bi}(\tau,-u),
\label{Cc}
\eeq
and therefore, $i$ and its conjugate $\bi$ can be distinguished
by the introduction of $u$.
We also note
$\overline{\chi_i(\tau,u)}=\chi_i(-\overline{\tau},-\overline{u})
=\chi_{\bi}(-\overline{\tau},\overline{u})$.
The parameter $u$ is usually suppressed in the main text, but its presence
is always borne in mind.
We encountered its importance already in the discussion of
D-branes \cite{zuber}.

For an element $({a\,b\atop c\,d})$ of $SL(2,\Z)$, the characters
transform as
\beq
\chi_i\Bigl({b+d\tau\over a+c\tau},{u\over a+c\tau}\Bigr)
=\sum_j\chi_j(\tau,u)M_{ji}(\mbox{${a\,b\atop c\,d}$}).
\eeq
$M({a\, b\atop c\,d})$ is $\tau$-independent at $u=0$.
We define $S$, $T$, $C$ as $M({a\, b\atop c\,d})|_{u=0}$
for the $SL(2,\Z)$ elements
\beq
{\mathcal S}=\left(\begin{array}{cc}
0&-1\\
1&0
\end{array}\right),\qquad
{\mathcal T}=\left(\begin{array}{cc}
1&1\\
0&1
\end{array}\right),\qquad
{\mathcal C}=\left(\begin{array}{cc}
-1&0\\
0&-1
\end{array}\right),
\label{STC}
\eeq
respectively. 
$S,T$ and $C$ are all unitary and
obey the same algebraic relations as the above $SL(2,\Z)$ elements, such as
$[S,C]=[T,C]=0$,
$S^2=C$,
$(ST)^3=C$, $STS=T^{-1}ST^{-1}$.
$C$ is the charge conjugation matrix, $C_{ij}=\delta_{j,\bi}$,
because of relation (\ref{Cc}).
$T$ is a diagonal matrix $T_{ij}
=\delta_{i,j}\e^{2\pi i(h_i-{c\over 24})}=:\delta_{i,j}T_i$.
More non-trivial is the fact that
$S$ is a symmetric matrix, $S_{ij}=S_{ji}$ \cite{DV}.

The modular transformation of the M\"obius strip involves
\beqa
\widehat{\chi}_i(\mbox{$-{1\over 4\tau},{u\over 2\tau}$})
&=&\sqrt{T_i}^{-1}\chi_i(\mbox{$-{1\over 4\tau}+{1\over 2},{u\over 2\tau}$})
=\chi_i(\mbox{$-{1\over 4\tau}-{1\over 2},{u\over 2\tau}$})\sqrt{T_i}
\nn\\
&=&
\sum_k\chi_k(\mbox{${4\tau\over 1+2\tau},{2u\over 1+2\tau}$})S_{ki}\sqrt{T_i}
=
\sum_k\chi_k(\mbox{$-{2\over 1+2\tau},{2u\over 1+2\tau}$})T_k^2S_{ki}\sqrt{T_i}
\nn\\
&=&\sum_{k,l}\chi_l(\mbox{$\tau+{1\over 2},u$})
S_{lk}T_k^2S_{ki}\sqrt{T_i}
=
\sum_l\widehat{\chi}_l(\tau,u)(\sqrt{T}ST^2S\sqrt{T})_{li}.
\nn
\eeqa
This introduces the new matrix
\beq
P=\sqrt{T}ST^2S\sqrt{T}.
\eeq
Using the properties of $S,T$ and $C$, we find
that $P$ is a unitary, symmetric matrix such that $P^2=C$.

\appendix{Some properties of $Q_g(i)$}\label{app:Q}

A simple current $g$ is a representation of the chiral algebra
${\mathcal A}$ such that the fusion product of $g$ with any representation
$i$ is just one representation, which we denote by $g(i)$.
The set of simple currents forms an abelian group $\Scr{G}$.
Let us introduce the number (defined modulo $1$)
\beq
Q_g(i):=h_g+h_i-h_{g(i)}\qquad\mbox{mod 1}.
\eeq
They obey the following properties:

\noindent
(i) If $N_{ij}^{\,\,\,k}\ne 0$, then $Q_g(i)+Q_g(j)=Q_g(k)$.\\
(ii) $Q_{g_1}(i)+Q_{g_2}(i)=Q_{g_1g_2}(i)$.\\
(iii) $S_{g(i)j}=\e^{2\pi i Q_g(j)}S_{ij}$.\\
(iv) $g\mapsto \e^{2\pi i Q_g(i)}$
is a homomorphism $\Scr{G}\to U(1)$, as a consequence of (ii).\\
(v) $Q_{\bar g}(\bi)=Q_g(i)$.\\
(vi) $Q_g(\bi)=-Q_g(i)$, as a consequence of (ii) and (v).\\
(vii) $\e^{2\pi i Q_g(i)}=\pm 1$ if $i=\bi$, as a consequence of (vi).\\
(viii) $Q_g(g)=-2h_g$.\\
(ix) $Q_{g_1}(g_2(i))-Q_{g_1}(i)=Q_{g_1}(g_2)$, as a consequence of
(i).

\noindent
(i) follows from the operator product expansion of
$g$, $i$, and $j$. Since $N_{g_2i}^{\,\,g_2(i)}\ne 0$
we find $Q_{g_1}(g_2)+Q_{g_1}(i)=Q_{g_1}(g_2(i))$ by (i).
This is in fact equivalent to (ii).
(iii) is shown in \cite{ScYa,SYQ,Intri}.
(v) is because $h_{\bi}=h_i$.
(viii) is because $Q_g(g)=-Q_{g^{-1}}(g)=-h_{g^{-1}}-h_g+h_1=-2h_g$.

Let us consider $Q_{g_1}(g_2)=h_{g_1}+h_{g_2}-h_{g_1g_2}$,
which is symmetric under the exchange $g_1\leftrightarrow g_2$.
By the property (ii) above,
$(g_1,g_2)\mapsto Q_{g_1}(g_2)$ is a symmetric bilinear form
of $\Scr{G}$ with values in $\R/\Z$.
There is not always  a symmetric bilinear form
$q(g_1,g_2)$ of $\Scr{G}$ such that
\beq
Q_{g_1}(g_2)=2q(g_1,g_2)
\qquad
\mbox{mod $1$}.
\eeq
However, for some subgroup
$G$ of $\Scr{G}$, there can be such a symmetric bilinear form.
The existence of such a form is the condition for the
existence of an $G$-orbifold with modular invariant partition function
(see Appendix~\ref{app:Twist}).
For example let us consider a simple current $g$ of order $N$,
$g^N=1$.
Let us note from (ii) and (viii) that
$Q_{g^n}(g^m)=-2nmh_g$.
For the group $G$ generated by $g$,
a candidate bilinear form is thus
$q(g^n,g^m)=-nmh_g$. However, this is well-defined
(as a function with values in $\R/\Z$)
if and only if $Nh_g$ is an integer.

\subsubsection*{\it Formulae involving $S$, $T$ and $P$}

We record some formulae
involving $S,T,P$.
We first quote from \cite{ScYa,SYQ,Intri} that
\beq
S_{g(i),j} = e^{2\pi i Q_g(j)} \, S_{ij} \quad, \quad 
T_{g(i)} = e^{2\pi i (h_g-Q_g(i))} \, T_i.
\eeq
One can derive a similar relation for the $P$-matrix \cite{HSSc,FHSSW}:
\beq
P_{g^{2m}(i),j} = \phi (2m,i) \, e^{2\pi i m Q_g(j)} P_{ij},
\eeq
where 
\beq
\phi (2m,i) := e^{\pi i ( h_i-h_{g^{2m}i} - 2Q_{g^m}(g^m(i)))}.
\eeq
We are particularly interested in the case that $i=0$. In that case,
the expression in brackets in the exponent becomes
$-h_{g^{2m}}-2Q_{g^m}(g^m) = h_{g^{2m}}-4h_{g^m}$ mod $1$. This can be
rewritten as $-Q_{g^{-m}}(g^{-m}) - Q_{g^m}(g^{-m})$. Applying property
(ii) above, one obtains that this is $0$ mod $1$.
The conclusion is that $\phi(2m,0)$ is just a sign, and therefore
\beq\label{Pvac}
P_{g^{2m},j} = \pm e^{2\pi i m Q_g(l)} \, P_{0,j}
\eeq

\appendix{Orbifolds}\label{app:Twist}

We consider the model ${\mathcal C}$ with the Hilbert space of states
${\mathcal H}^{\mathcal C}=\oplus_i\Scr{H}_i\otimes\Scr{H}_{\bi}$.
Simple currents act on ${\mathcal H}^{\mathcal C}$ by
\beq
g:v\mapsto 
\e^{2\pi i Q_g(i)}v,\qquad v\in\Scr{H}_i\otimes\Scr{H}_{\bi}.
\eeq
This is a discrete symmetry of the system.
We record and explain some known facts on orbifold by a group of
simple currents.

\subsection{$g$-Twisted Hilbert Space}\label{subap:gtw}

Let us quantize the system on the circle $x\equiv x+1$
with $g$-twisted boundary condition.
Namely, we impose the boundary condition
$\phi(x)=g\phi(x+1)$ on the fields.
We are interested in the wavefunctions of such a system,
the $g$-twisted states.
We show that the space of such states is given by
\beq
{\mathcal H}_g=\bigoplus_i\Scr{H}_i\otimes \Scr{H}_{g(\bi)}.
\label{Hg}
\eeq

The partition function
$\Tr_{{\mathcal H}_g}q^{L_0-c/24}\overline{q}^{L_0-c/24}$
is realized as the path-integral on the torus
$(x,y)\equiv (x+1,y)\equiv (x,y+1)$ (with complex coordinate $z=x+\tau y$)
on which the fields are periodic in ``time'',
$\phi(x,y)=\phi(x,y+1)$, but obey the $g$-twisted boundary condition in
``space'' $\phi(x,y)=g\phi(x+1,y)$.
One can switch the role of time and space by the coordinate transformation
$(x,y)^t={\mathcal S}^{-1}(x',y')^t$ where
${\mathcal S}$ is the $SL(2,\Z)$ matrix (\ref{STC}).
Then, we have $\phi(x',y')=\phi(x'+1,y')=g\phi(x',y'+1)$.
Thus, the partition function can be written as
\beqa
\Tr\!\!\mathop{}_{{\mathcal H}_g}q^{L_0-c/24}\overline{q}^{L_0-c/24}
&=&\Tr\!\!\mathop{}_{{\mathcal H}}g{q'}^{L_0-c/24}\overline{q'}^{L_0-c/24}
=\sum_i\e^{2\pi i Q_g(i)}\chi_i(\tau')\overline{\chi_i(\tau')}
\nn\\
&=&\sum_{ijk}\e^{2\pi i Q_g(i)}\chi_j(\tau)S^{-1}_{ji}
\overline{\chi_k(\tau)S^{-1}_{ki}}
\nn\\
&=&\sum_{ijk}S^{-1}_{g^{-1}(j)i}\overline{S^{-1}_{ki}}
\chi_j(\tau)\overline{\chi_k(\tau)},
\nn
\eeqa
where we have used $S_{g^{-1}(j)i}=\e^{-2\pi i Q_g(i)}S_{ji}$.
Since $\sum_iS_{g^{-1}(j)i}^{-1}\overline{S^{-1}_{ki}}=\delta_{k,g^{-1}(j)}$
we find that
the partition function equals
\beq
\Tr_{{\mathcal H}_g}q^{L_0-c/24}\overline{q}^{L_0-c/24}
=\sum_{j}\chi_j(\tau)\overline{\chi_{g^{-1}(j)}(\tau)}.
\nonumber
\eeq
Since $\overline{g^{-1}(j)}=g(\bj)$, this proves Eq.~(\ref{Hg}).

\subsection{$g$ action on ${\mathcal H}_{g^n}$}

We next show that
the discrete symmetry $g^m$ acts on ${\mathcal H}_{g^n}$
by the phase multiplication
\beq
\e^{-2\pi i(Q_{g^m}(g^n(\bi))+nmh_g)}\times \quad
\mbox{on the subspace}\quad
\Scr{H}_i\otimes \Scr{H}_{g^n(\bi)}.
\label{actQgg}
\eeq

To find the action of $g$ on ${\mathcal H}_{g^n}$ we compute
the partition function
$\Tr_{{\mathcal H}_{g^n}}gq^{L_0-c/24}\overline{q}^{L_0-c/24}$.
This is realized as the path-integral on the torus
with boundary condition
$\phi(x,y)=g^n\phi(x+1,y)=g\phi(x,y+1)$.
Under the change of coordinates $(x,y)^t={\mathcal S}
{\mathcal T}^n{\mathcal S}^{-1}(x',y')^t$, the boundary
condition becomes $\phi(x',y')=\phi(x'+1,y')=g\phi(x',y'+1)$.
Thus the partition function
can be written as
\beqa
\Tr\!\!\mathop{}_{{\mathcal H}_{g^n}}gq^{L_0-c/24}\overline{q}^{L_0-c/24}
&=&\Tr\!\!\mathop{}_{{\mathcal H}}g{q'}^{L_0-c/24}\overline{q'}^{L_0-c/24}
=\sum_i\e^{2\pi i Q_g(i)}\chi_i(\tau')\overline{\chi_i(\tau')}
\nn\\
&=&\sum_{ijk}
\e^{2\pi i Q_g(i)}\chi_j(\tau)(ST^nS^{-1})_{ji}
\overline{\chi_k(\tau)(ST^nS^{-1})_{ki}}
\nn\\
&=&\sum_{jkl}\chi_j(\tau)S_{jg(l)}T_{g(l)}^n
\overline{\chi_k(\tau)S_{kl}T_l^n}
\eeqa
We note here that $T_{g(l)}^n\overline{T_l^n}=\e^{2\pi i n(h_{g(l)}-h_l)}
=\e^{2\pi i Q_{g^{-n}}(g(l))}\e^{-2\pi i nh_g}$.
Using $S_{jg(l)}\e^{2\pi i Q_{g^{-n}}(g(l))}=S_{g^{-n}(j)g(l)}
=\e^{2\pi i Q_g(g^{-n}(j))}S_{g^{-n}(j)l}$, we find that the partition
function is given by
\beq
\Tr\!\!\mathop{}_{{\mathcal H}_{g^n}}gq^{L_0-c/24}\overline{q}^{L_0-c/24}
=\sum_j\e^{2\pi i (Q_g(g^{-n}(j))-nh_g)}
\chi_j(\tau)\overline{\chi_{g^{-n}(j)}(\tau)}.
\eeq
This shows the action of $g$ on ${\mathcal H}_{g^n}$.
$g^m$ action (\ref{actQgg}) is obtained by iteration.


\noindent
{\bf Remark.}
If $g$ is an order $N$ simple current, we find $Q_{g^N}(-)=0$ mod 1,
by the property (ii) in Appendix~\ref{app:Q}.
Thus, $g^N$ acts on ${\mathcal H}_{g^n}$ by 
phase multiplication $\e^{-2\pi i nNh_g}\times$.
If we want $g$ to be order $N$ also in the action on the twisted states,
we need $Nh_g$ to be an integer.
This explains why we need $Nh_g\in\Z$ in order to have
the orbifold of ${\mathcal C}$
by the group $G=\{g^n\}_{n=0}^{N-1}$.

\subsection{$g_1$ action on ${\mathcal H}_{g_2}$}

Suppose a group $G$ of simple currents has
a symmetric bilinear form $q(g_1,g_2)\in\R/\Z$
obeying $Q_{g_1}(g_2)=2q(g_1,g_2)$ mod $1$
such that $q(g,g)=-h_g$.
Then, one can define a $G$-orbifold of ${\mathcal C}$
with the modular-invariant partition function
\beq
Z^{{\mathcal C}/G}
={1\over |G|}\sum_{i,g_1,g_2}
\e^{2\pi i(Q_{g_2}(i)-q(g_2,g_1))}
\chi_i(\tau)\overline{\chi_{g_1^{-1}(i)}(\tau)}.
\label{MIPF}
\eeq
(Note that the phase can also be written as
$\e^{-2\pi i(Q_{g_2}(g_1(\bi))-q(g_2,g_1))}$.)
This shows that a $g_2$ action on ${\mathcal H}_{g_1}$
(for $g_1,g_2\in G$)
is given by
\beq
\e^{-2\pi i(Q_{g_2}(g_1(\bi))-q(g_2,g_1))}\times
\quad\mbox{on the subspace}\quad
\Scr{H}_i\otimes\Scr{H}_{g_1(\bi)}.
\label{TTT}
\eeq

It is straightforward to show modular invariance of (\ref{MIPF}).
Under $\tau\to \tau+1$, the extra phase
$h_i-h_{g_1^{-1}(i)}=Q_{g_1^{-1}}(i)-h_{g_1}$ appears.
At this point, we use $h_{g_1}=-q(g_1,g_1)=q(g_1^{-1},g_1)$.
Then, we see that
the expression for $Z(\tau+1)$ is the same as (\ref{MIPF}) with $g_2$
replaced by $g_2g_1^{-1}$.
On the other hand
\beq
Z \left( -{1\over\tau} \right)
={1\over |G|}\sum\e^{2\pi i (Q_{g_2}(i)-q(g_2,g_1))}
\chi_j(\tau)S_{ji}\overline{\chi_k(\tau)}S^{-1}_{kg_1^{-1}(i)}.
\nn
\eeq
We now use
$\e^{2\pi iQ_{g_2}(i)}S_{ji}=S_{g_2(j)i}$ and
$S^{-1}_{kg_1^{-1}(i)}=\e^{2\pi i Q_{g_1}(k)}S^{-1}_{ki}$,
so that the sum over $i$ can be performed,
$\sum_iS_{g_2(j)i}S^{-1}_{ki}=\delta_{k,g_2(i)}$.
Noting $Q_{g_1}(g_2(j))-q(g_2,g_1)=Q_{g_1}(j)+q(g_1,g_2)$, we find that
it is the same as (\ref{MIPF}) with $(g_1,g_2)$
replaced by $(g_2^{-1},g_1)$.

The complete set of all simple current modular invariant partition functions
is obtained in \cite{KrSc}.
There are other modular invariants
associated with the ``discrete torsion''
\cite{Vafa:DT}.
Addition of a discrete torsion corresponds to
 shifting $q(g_2,g_1)$ in (\ref{TTT}) by
an antisymmetric bilinear form of $G$ with values in $\R/\Z$
that vanishes on the diagonals $(g,g)$.

\subsection{Quantum Symmetry}\label{subap:quantums}

For each character $g\mapsto \e^{2\pi i \rho(g)}$ of $G$,
there is a symmetry $g_{\rho}$ of
the orbifold theory ${\mathcal C}/G$
that acts on the $g$-twisted sector
states by the multiplication by the phase
$\e^{2\pi i\rho(g)}$. This is called a {\it quantum symmetry}.
The group of quantum symmetries is the character group $G^{\vee}$, which
is isomorphic to $G$ itself.

Let us find out what  the $g_{\rho}$-twisted states are
in the orbifold theory.
We recall that the untwisted
states of the orbifold model are $G$-invariant states in
$\oplus_{h\in G}{\mathcal H}_{h}$, the states obeying $g=1$,
$\forall g\in G$.
We claim that the $g_{\rho}$-twisted states
are the states in $\oplus_{h\in G}{\mathcal H}_{h}$
obeying $g=\e^{2\pi i\rho(g)}$, $\forall g\in G$.
Namely
\beq
({\mathcal H}^{{\mathcal C}/G})_{g_{\rho}}
=\bigoplus_{j,h}
\left.
\Scr{H}_i\otimes\Scr{H}_{h(\bi)}
\right|_{g=\e^{2\pi i \rho(g)},\forall g\in G}.
\eeq
Since the action of $g$ on $\Scr{H}_i\otimes\Scr{H}_{h(\bi)}$
is given in (\ref{TTT}), it is
the sum of $\Scr{H}_i\otimes\Scr{H}_{h(\bi)}$ over those $(i,h)$ such that
$\e^{-2\pi i (Q_g(h(\bi))-q(g,h))}=\e^{2\pi i \rho(g)}$ for any $g\in G$.

This is shown as in Appendix \ref{subap:gtw}.
The partition function on the $g_{\rho}$-twisted circle
is the same as the partition function on the untwisted circle,
but with a $g_{\rho}$ insertion in the evolution operator:
\beq
Z_{\rho}^{{\mathcal C}/G}
=
{1\over |G|}
\sum_{i,g_1,g_2}
\e^{2\pi i\rho(g_1)}\e^{2\pi i (Q_{g_2}(i)-q(g_2,g_1))}
\chi_i(\tau')\overline{\chi_{g_1^{-1}(i)}(\tau')}.
\eeq
After a manipulation similar to
Appendix~\ref{subap:gtw}, we find that it is equal to
\beq
{1\over |G|}\sum_{j,g_1,g_2}
\e^{2\pi i\rho(g_1)}
\e^{2\pi i (Q_{g_1}(g_2(\bj))-q(g_2,g_1))}\chi_j(\tau)
\overline{\chi_{g_2^{-1}(j)}(\tau)}.
\eeq
This shows the claim.

\appendix{Alternative Way of Dressing}\label{app:some}

For a crosscap state $|\Scr{C}_P\rangle$ and a symmetry
$g$, what is $g|\Scr{C}_P\rangle$?
We first note that
\beqa
\langle \Scr{B}_{\alpha}|\e^{-{L\over 2} H_c(2\beta)}g|\Scr{C}_P\rangle
&=&\langle \Scr{B}_{g^{-1}(\alpha)}|\e^{-{L\over 2} H_c(2\beta)}
|\Scr{C}_P\rangle
\nn\\
&=&\Tr\!\!\mathop{}_{{\mathcal H}_{g^{-1}(\alpha),Pg^{-1}(\alpha)}}
P\e^{-\beta H_o(L)}
\nn\\
&=&\Tr\!\!\mathop{}_{{\mathcal H}_{\alpha,gPg^{-1}(\alpha)}}
gPg^{-1}\e^{-\beta H_o(L)}.
\eeqa
This suggests that $g|\Scr{C}_P\rangle$ is the crosscap state
for the parity $gPg^{-1}$.
Under this interpretation,
$\langle\Scr{C}_{P'}|q^{H_c}g|\Scr{C}_P\rangle$ can be regarded as
$\langle\Scr{C}_{P'}|q^{H_c}|\Scr{C}_{gPg^{-1}}\rangle$ as well as
$\langle\Scr{C}_{g^{-1}P'g}|q^{H_c}|\Scr{C}_P\rangle$.
The two indeed agree because
\beq
\Tr\!\!\mathop{}_{{\mathcal H}_{P'gP^{-1}g^{-1}}}gPg^{-1}
\e^{-\beta H_c}
=\Tr\!\!\mathop{}_{{\mathcal H}_{g^{-1}P'gP^{-1}}}P
\e^{-\beta H_c},
\eeq
where we have used that
$g$ maps ${\mathcal H}_{h}$ to ${\mathcal H}_{ghg^{-1}}$
(we have in mind $h=g^{-1}P'gP^{-1}$).
Thus, we conclude that
\beq
g|\Scr{C}_P\rangle=|\Scr{C}_{gPg^{-1}}\rangle.
\label{gC}
\eeq
If $P$ and $gP$ are both involutive (or more weakly
if $(gP)^2=P^2$), then we find $gPg=P$ and hence
\beq
gPg^{-1}=g^2P.
\eeq

\newcommand{\wtP}{\widetilde{P}}
\newcommand{\wtg}{\widetilde{g}}

Let us apply this to the case of $P=P_0$ and $g$ the symmetry
associated with a simple current:
The crosscap state for the parity
$\wtP_g:=gP_0g^{-1}$ is $g|\Scr{C}_{P_0}\rangle$ with the coefficients
\beq
\widetilde{\gamma}_{gi}=\e^{2\pi i Q_g(i)}{P_{0i}\over\sqrt{S_{0i}}}.
\eeq
Since $(gP_0)^2=P_0^2=1$ on ${\mathcal H}$, we find
$\wtP_g=g^2P_0$ on ${\mathcal H}$.
Since the action of $\wtP_g$ and $P_{g^2}$ agree on ${\mathcal H}$,
it is a natural question to ask whether they are the same parity symmetry.
In fact, we found in Appendix~\ref{app:Q} that
\beq
P_{g^2i}=\pm \e^{2\pi i Q_g(i)}P_{0i},
\eeq
where the sign $\pm$ depends only on $g$ but not on $i$.
Thus, we indeed see that the crosscap states
$|\Scr{C}_{P_{g^2}}\rangle$ and $|\Scr{C}_{\wtP_{g}}\rangle$
agree up to a sign.


\appendix{Partition Functions of the Circle Sigma Model}

We record here the cylinder, Klein bottle and M\"obius strip amplitudes
of the circle sigma model.
This can be used to justify the formula for the boundary and crosscap states
used in Section~\ref{sec:arbR}.
To perform the modular transformation, we will make use of the 
Poisson resummation formula
$$
\sum_{n\in \Z}
\e^{-\pi \alpha n^2-2\pi i\beta n}
={1\over \sqrt{\alpha}}\sum_{m\in\Z}
\e^{-{\pi\over\alpha}(m+\beta)^2},
$$
as well as the relations
$$
f_1(\e^{-\pi/T})=\sqrt{T}f_1(\e^{-\pi T}),
\quad\,
f_3(\e^{-\pi/T})=f_3(\e^{-\pi T}),
\quad\,
f_2(\e^{-\pi/T})=f_4(\e^{-\pi T}),
$$
among the functions
\beqa
&&\mbox{$f_1(q)=q^{1\over 12}\prod_{n=1}^{\infty}(1-q^{2n}),\qquad
f_2(q)=\sqrt{2}q^{1\over 12}\prod_{n=1}^{\infty}(1+q^{2n})$,}
\nn\\
&&\mbox{$f_3(q)=q^{-{1\over 24}}\prod_{n=1}^{\infty}(1+q^{2n-1}),\qquad
f_4(q)=q^{-{1\over 24}}\prod_{n=1}^{\infty}(1-q^{2n-1})$.}
\nn
\eeqa
Note that 
$f_1(q)=\eta(q^2)$.

The cylinder amplitudes are ($q=\e^{-2\pi T}, q_t=\e^{-\pi/T}$):
\beqa
&&\Tr\!\!\mathop{}_{N_{a_1}N_{a_2}}(q^H)
={\sum_{l}q^{({l\over R}+{\mDa\over 2\pi})^2}
\over
\eta(q)}
={R\over\sqrt{2}}{\sum_m q_t^{{1\over 2}(Rm)^2}
\e^{-iR\mDa m}\over\eta(q_t^2)}
=\langle N_{a_1}|q_t^H|N_{a_2}\rangle,
\qquad
\\
&&
\Tr\!\!\mathop{}_{D_{x_1}D_{x_2}}(q^H)
={\sum_{m}q^{(Rm+{\mDx\over 2\pi})^2}
\over
\eta(q)}
={1\over R\sqrt{2}}{\sum_l q_t^{{1\over 2}({l\over R})^2}
\e^{-i{\mDx\over R} l}\over\eta(q_t^2)}
=\langle D_{x_1}|q_t^H|D_{x_2}\rangle,
\\
&&
\Tr\!\!\mathop{}_{DN}(q^H)={q^{1\over 48}\over
\prod_{n=1}^{\infty}(1-q^{n-{1\over 2}})}
={1\over
\sqrt{2}q_t^{1\over 12}\prod_{n=1}^{\infty}(1+q_t^{2n})}
=\langle D|q_t^H|N\rangle.
\eeqa
Here $\mDa=a_2-a_1$ and $\mDx=x_2-x_1$.

The Klein bottle amplitudes are ($q=\e^{-2\pi T}, q_t=\e^{-\pi/2T}$):
\beqa
&&
\Tr\!\!\mathop{}_{\mathcal H}(\Omega q^H)
={\sum_l q^{l^2\over 2R^2}\over \eta(q^2)}
=R\sqrt{2}{\sum_m^{\rm even} q_t^{R^2m^2\over 2}\over\eta(q_t^2)}
=\langle\Scr{C}_{\Omega}|q_t^H|\Scr{C}_{\Omega}\rangle,\\
&&
\Tr\!\!\mathop{}_{\mathcal H}(s\Omega q^H)
={\sum_l (-1)^lq^{l^2\over 2R^2}\over \eta(q^2)}
=R\sqrt{2}{\sum_m^{\rm odd} q_t^{R^2m^2\over 2}\over\eta(q_t^2)}
=\langle\Scr{C}_{s\Omega}|q_t^H|\Scr{C}_{s\Omega}\rangle.
\eeqa

The M\"obius strip amplitudes are:
($q=\e^{-2\pi T}, q_t=\e^{-\pi/4T}$)
\beqa
&&
\Tr\!\!\mathop{}_{N_{a}N_{-a}}(\Omega q^H)
={\sum_{l}q^{({l\over R}+{\mDa\over 2\pi})^2}
\over
q^{1\over 24}\prod (1-(-1)^nq^n)}
=R{\sum_{m}^{\rm even} q_t^{{1\over 2}(Rm)^2}
\e^{-iR\mDa m}
\over
q_t^{1\over 12}\prod (1-(-1)^nq_t^{2n})}
=\langle N_{a}|q_t^H|\Scr{C}_{\Omega}\rangle,
\qquad
\\
&&
\Tr\!\!\mathop{}_{N_{a}N_{-a}}(s\Omega q^H)
={\sum_{l}(-1)^lq^{({l\over R}+{\mDa\over 2\pi})^2}
\over
q^{1\over 24}\prod (1-(-1)^nq^n)}
=R{\sum_{m}^{\rm odd} q_t^{{1\over 2}(Rm)^2}
\e^{-iR\mDa m}
\over
q_t^{1\over 12}\prod (1-(-1)^nq_t^{2n})}
=\langle N_{a}|q_t^H|\Scr{C}_{s\Omega}\rangle,
\qquad\quad
\\
&&
\Tr\!\!\mathop{}_{D_xD_x}(\Omega q^H)
={1
\over
q^{1\over 24}\prod (1+(-1)^nq^n)}
={1
\over
q_t^{1\over 12}\prod (1+(-1)^nq_t^{2n})}
=\langle D_x|q_t^H|\Scr{C}_{\Omega}\rangle,
\qquad
\\
&&
\Tr\!\!\mathop{}_{D_xD_{x-\pi R}}(s\Omega q^H)=0
=\langle D_x|q_t^H|\Scr{C}_{s\Omega}\rangle.
\eeqa
Here $\mDa=(-a)-a=-2a$.
The last partition function vanishes because
$s\Omega$ maps $|m\rangle_{x,x-\pi R}$
to $|-m\rangle_{x,x+\pi R}=|1-m\rangle_{x,x-\pi R}$, which cannot be the same
as $|m\rangle_{x,x-\pi R}$ for integer $m$.

The cylinder with $g_{\mDx}$-twist is ($q=\e^{-2\pi T},q_t=\e^{-\pi/T}$)
\beqa
&&
\Tr\!\!\mathop{}_{N_{a_1}N_{a_2}}(g_{\mDx}q^H)
={\sum_{l}q^{({l\over R}+{\mDa\over 2\pi})^2}
\e^{-i\mDx({l\over R}+{\mDa\over 2\pi})}
\over
\eta(q)}
={R\over\sqrt{2}}
{\sum_m q_t^{{1\over 2}(Rm-{\mDx\over 2\pi})^2}
\e^{-iR\mDa m}\over\eta(q_t^2)}
\nn\\
&&\qquad\qquad\qquad\qquad
={}_{g_{\mDx}}\langle N_{a_1}|q_t^H|N_{a_2}\rangle_{g_{\mDx}},
\eeqa
where $\mDa=a_2-a_1$.
M\"obius strip with $g_{\mDx}$-twist is
($q=\e^{-2\pi T},q_t=\e^{-\pi/4T}$)
\beqa
&&
\Tr\!\!\mathop{}_{N_{a}N_{-a}}(g_{\mDx}q^H)
={\sum_{l}q^{({l\over R}+{(-2a)\over 2\pi})^2}
\e^{-i\mDx({l\over R}+{(-2a)\over 2\pi})}
\over
q^{1\over 24}\prod_{n=1}^{\infty}(1-(-1)^nq^n)}
=R
{\sum_m^{\rm even} q_t^{{1\over 2}(Rm-{\mDx\over \pi})^2}
\e^{iRa m}\over
q_t^{1\over 12}\prod_{n=1}^{\infty}(1-(-1)^nq_t^{2n})}
\nn\\
&&\qquad\qquad\qquad\qquad
={}_{g_{2\mDx}}\langle N_{a}|q_t^H|\Scr{C}_{g_{\mDx}\Omega}\rangle.
\eeqa

\appendix{ Formulae for $SU(2)/U(1)$}

\subsection{A-type crosscaps}

We compute the explicit coefficients of the A-type crosscaps
\beq
\ket{\Scr{C}_{\ell}} = \sum_{(j,n) \in PF_k} 
\frac{P_{(0,2\ell)(j,n)}}{\sqrt{S_{(0,0)(j,n)}}} \ \cket{(j,n)}
\eeq
using the formula
\beq
P_{(j,n)(j',n')} = T^{\half}_{j,n} \left(
Q_{jj'} Q_{nn'}^* + Q_{\frac{k}{2}-j, j'} Q_{n+k,k'}^* \right)
T^{\half}_{j',n'}.
\eeq
The subtlety is that $T_{j,n}^{\half}$ does not usually
factorize as $T_j^{\half}T_n^{-\half}$ except in the standard range
(henceforth $S.R.$) where (\ref{paraweight}) holds.
Using
\beqa
&&
T^{\half}_{j,n} = T^{\half}_j \ T^{-\half}_n,\qquad (j,n)\in S.R.,
\nn\\
&&
T^{\half}_{j,n} = T^{\half}_{\frac{k}{2}-j} \ T^{-\half}_{n+k} 
= T^{\half}_j \ T^{-\half}_n (-1)^{\frac{2j+n}{2}},
\qquad (\mbox{${k\over 2}$}-j,n+k)\in S.R.,
\nn\\
&&
T^{\half}_{j,n} = T^{\half}_{\frac{k}{2}-j} \ T^{-\half}_{n-k} 
= T^{\half}_j \ T^{-\half}_n (-1)^{\frac{2j-n}{2}},
\qquad (\mbox{${k\over 2}$}-j,n-k)\in S.R.,
\eeqa
the required $P$-matrix elements are computed to be
\beq
P_{(0,2\ell)(j,n)} = (-1)^{\ell} P^{SU(2)}_{0,j} \
\left( P^{U(1)}_{2\ell,n} \right)^* +  P^{SU(2)}_{\frac{k}{2},j} \ 
\left( P^{U(1)}_{2\ell+k,n}\right)^*,
\eeq
for $(j,n)$ in the standard range and $\ell$ in the range
$-k\leq 2\ell\leq k$. If $(k/2-j,n\pm k)$ is in the standard range,
one needs the extra sign factor $(-1)^{\frac{2j\pm n}{2}}$.
The explicit expression for the crosscap is
\beqa
\lefteqn{\ket{\Scr{C}_{\ell}} = {1\over[k(k+2)]^{\frac{1}{4}}}\times}
\nn\\
&& \sum_{(j,n)\in S.R.}
\e^{\frac{\pi i \ell n}{k}}
\left(\delta^{(2)}_{n+k}
(-1)^{\ell} \sqrt{\tan\mbox{$\frac{\pi(2j+1)}{2(k+2)}$}}
+\delta^{(2)}_{n}
 (-1)^{\frac{2j+n}{2}} 
\sqrt{\cot\mbox{$\frac{\pi(2j+1)}{2(k+2)}$}}
\right) \cket{(j,n)},
\nn\\
&&
\label{Aexpl}
\eeqa
where, on the RHS, we need to bring $\ell$ in the range
$-k\leq 2\ell\leq k$.

\subsection{B-type crosscaps}\label{subapp:Bcc}

We first construct A-type crosscaps in the orbifold, and then
apply the mirror map. The crosscaps of the orbifold are
\beq
\ket{\Scr{C}_{P^{\theta_r}}} = \frac{1}{\sqrt{k}} \ \sum_{\ell}
\e^{-\pi i \theta_r (\ell)} \ \ket{\Scr{C}_{\ell}},
\eeq
where $\theta_r(\ell)= -2r\ell/k$, as explained in the main text.
We also have set $\omega_1=0$.
Inserting the states (\ref{Aexpl}), one sees that the following
summations over $\ell$ are relevant:
$$
\sum_{\ell}(-1)^{\ell} \e^{\frac{\pi i \ell}{k}(2r+n)}
= k \delta^{(2k)}_{2r+n+k},\qquad
\sum_{\ell} \e^{\frac{\pi i \ell}{k}(2r+n)}
= k \delta^{(2k)}_{2r+n}.
$$
The first sum plays a role when one sums up $\ell$ in the first term
($\sim \sqrt{\tan}$)
in the parenthesis in (\ref{Aexpl}), and the second sum is relevant
for the second term ($\sim \sqrt{\cot}$) in the parenthesis. 
$n$ is projected on either $n=-2r-k$ or $n=-2r$. Only one term
gives a contribution, since in (\ref{Aexpl}) we are summing over
$(j,m)$ in the standard range.
This leads to the following expressions
\beqa
\ket{\Scr{C}_{P^{\theta_r}}}&=&
\frac{k^{\frac{1}{4}}}{(k+2)^{\frac{1}{4}}}\Biggl[
\sum_{j,(j,-2r)\in S.R.} 
(-1)^{j} \ (-1)^r \ 
\sqrt{\cot\mbox{$\frac{\pi(2j+1)}{2(k+2)}$}}
\ \cket{(j,-2r)} \nn\\
&& ~~+ \sum_{j,(j,-2r)\notin S.R.} 
\sqrt{\cot\mbox{$\frac{\pi(2j+1)}{2(k+2)}$}}
\cket{(j,-2r)} \Biggr].
\eeqa
Applying the mirror map, one obtains the B-type
crosscaps
\beqa
\ket{\Scr{C}_r^B}&=&
\frac{k^{\frac{1}{4}}}{(k+2)^{\frac{1}{4}}}\Biggl[
\sum_{j,(j,-2r)\in S.R.} 
(-1)^{j} \ (-1)^r \ 
\sqrt{\cot\mbox{$\frac{\pi(2j+1)}{2(k+2)}$}}
\ \cket{(j,2r)}_B \nn\\
&& ~~+ \sum_{j,(j,-2r)\notin S.R.} 
\sqrt{\cot\mbox{$\frac{\pi(2j+1)}{2(k+2)}$}}
\cket{(j,2r)}_B \Biggr]
\eeqa
For the standard crosscap state with $r=0$, only the first term contributes,
since all states $(j,0)$ are in the standard range. For the state
with $r=k/2$ ($k$ even), only the second term contributes.
The respective crosscap states can be rewritten as
\beqa
&&
\ket{\Scr{C}_{0}^B}
={k^{1\over 4}\over (k+2)^{1\over 4}}\sum_{j\,{\rm integer}}
(-1)^j\sqrt{\cot\mbox{${\pi(2j+1)\over 2(k+2)}$}}
\cket{(j,0)}_B,
\\
&&
\ket{\Scr{C}_{\frac{k}{2}}^B}
={k^{1\over 4}\over (k+2)^{1\over 4}}\sum_{j\,{\rm integer}}
\sqrt{\tan\mbox{${\pi(2j+1)\over 2(k+2)}$}}
\cket{(j,0)}_B.
\eeqa

\subsection{Computation of one-loop amplitudes}\label{app:detai}

Here we record some detail of the computation of
the one-loop amplitudes
(\ref{AparaBB}), (\ref{AparaBC}), and (\ref{AparaCC}).
For (\ref{AparaBB}):
\beqa
\langle \Scr{B}_{J,M} | q_t^H\ket{\Scr{B}_{J'M'}}
&=& \sum_{(j,m)\in PF_k} N_{(J,-M)(J',M')}^{(j,-m)}  \chi_{j,m}(\tau)
\nn\\
&=&{1\over 2}\sum_{2j+m \,{\rm even}}N_{(J,-M)(J',M')}^{(j,-m)}
\chi_{j,m}(\tau)
=\sum_{2j+m \,{\rm even}}
N_{JJ'}^{j} \delta^{(2k)}_{M'-M+m} \chi_{j,m}(\tau),
\nn
\eeqa
where we have used (\ref{paraN}) in the last step.
For (\ref{AparaBC}), we first note
$$
\langle\Scr{C}_{\ell}|q_t^H\ket{\Scr{B}_{(J,M)}} 
=\sum_{(j,m)\in PF_k} Y_{(J,M) \ (j,m)}^{(0,2\ell)} 
\widehat{\chi}_{j,m}(\tau) 
= \sum_{2j+m\,{\rm even}} \tilde{Y}_{Jj}^{\frac{k}{2}} 
\overline{\tilde{Y}}_{Mm}^{2\ell+k} 
T^{-\frac{1}{2}}_{0,2\ell} \,\chi_{j,m}(\tau+\frac{1}{2}).
$$
Inserting the known $Y$-tensors from the $U(1)$ theory,
we see that this is equal to
\beqa
&&\sum_{2j+m\,{\rm even}} Y_{Jj}^{\frac{k}{2}} \ \delta^{(2)}_{m+k} \ 
\left( \delta^{(2k)}_{M-\ell+\frac{m-k}{2}}+ (-1)^{m+k} \ 
\delta^{(2k)}_{M-\ell+\frac{m+k}{2}} \right)
e^{-\pi i (h_j-\frac{m^2}{4k} - \frac{c}{24})} \ \chi_{j,m} (\tau+\half)
\nn\\
&&=
\sum_{2j+m\,{\rm even}}
Y_{Jj}^{\frac{k}{2}} \ \delta^{(2k)}_{2M-2\ell-k+m} \ 
e^{-\pi i (h_j-\frac{m^2}{4k} - \frac{c}{24})} \chi_{j,m}(\tau+\half).
\nn
\eeqa
Replacing $(j,m)\to ({k\over 2}-j,m+k)$ in the sum, and using
$Y_{J,{k\over 2}-j}^{k\over 2}=N_{JJ}^j$, we find
this to be equal to
$$
\sum_{2j+m\,{\rm even}}
N_{JJ}^j \delta^{(2k)}_{2M-2\ell+m} \ 
e^{\pi i (h_{j,m}-h_{{k\over 2}-j}+\frac{(m+k)^2}{4k})}
\widehat{\chi}_{j,m}(\tau).
$$
It is straightforward to see that
$$
\epsilon_{j,m}:=
\e^{\pi i (h_{j,m}-h_{{k\over 2}-j}+{(m+k)^2\over 4k})}
=\left\{\begin{array}{ll}
1&({k\over 2}-j,m+k)\in S.R.\\
(-1)^{2j+m\over 2}&(j,m)\in S.R.\\
(-1)^m&({k\over 2}-j,m-k)\in S.R.
\end{array}\right.
$$
This shows (\ref{AparaBC}).
Computation of (\ref{AparaCC}) is similarly straightforward.
It is convenient to use
$Y_{j0}^0=(-1)^{2j}$ and $Y_{j0}^{k\over 2}=N_{j,{{k\over 2}-j}}^{\,\,0}
=\delta_{j,{k\over 2}-j}=\delta_{j,{k\over 4}}$ (the latter is possible only
for $k$ even).

\end{document}